\documentclass[longauth]{aa}
\usepackage{graphicx}
\usepackage[varg]{txfonts}
\usepackage[colorlinks=true,     linkcolor=blue, citecolor=blue, filecolor=blue, urlcolor=blue]{hyperref}
\usepackage{silence}
\begin{document} 

   \title{Massive stars exploding in a He-rich circumstellar medium}

   \subtitle{XI. Diverse evolution of five Ibn SNe\,2020nxt, 2020taz, 2021bbv, 2023utc and 2024aej}

   \authorrunning{Z.-Y. Wang et al.} 
   \titlerunning{Five SNe Ibn}
   
   \author{
    Z.-Y. Wang\inst{\ref{inst1},\ref{inst2}}\and
    A. Pastorello\inst{\ref{inst3}} \and
    Y.-Z. Cai\inst{\ref{inst6},\ref{inst7},\ref{inst8}}\thanks{Corresponding authors: caiyongzhi@ynao.ac.cn (CYZ)}\and
    M. Fraser\inst{\ref{inst45}}\and 
    A. Reguitti\inst{\ref{inst5},\ref{inst3}} \and 
    W.-L. Lin\inst{\ref{inst46},\ref{inst36}}\thanks{linwl@xmu.edu.cn (LWL)} \and
    L. Tartaglia\inst{\ref{inst30}}\and \newline
    D. Andrew Howell\inst{\ref{inst9},\ref{inst10}}\and 
    S. Benetti\inst{\ref{inst3}} \and
    E. Cappellaro\inst{\ref{inst3}} \and
    Z.-H. Chen \inst{\ref{inst36}} \and
    N. Elias-Rosa\inst{\ref{inst3},\ref{inst15}} \and
    J. Farah\inst{\ref{inst9},\ref{inst10}} \and
    A. Fiore\inst{\ref{inst55},\ref{inst3}} \and \newline
    D. Hiramatsu\inst{\ref{inst19},\ref{inst20}} \and 
    E. Kankare\inst{\ref{inst51}} \and
    Z.-T. Li \inst{\ref{inst43}} \and
    P. Lundqvist\inst{\ref{inst17}} \and
    P. A. Mazzali\inst{\ref{inst52},\ref{inst53}} \and
    C. McCully\inst{\ref{inst9}} \and
    J. Mo\inst{\ref{inst36}} \and 
    S. Moran\inst{\ref{inst50}} \and \newline
    M. Newsome\inst{\ref{inst9},\ref{inst10}} \and
    E. Padilla Gonzalez\inst{\ref{inst10},\ref{inst9}} \and
    C. Pellegrino\inst{\ref{inst9},\ref{inst10}} \and
    Z.-H. Peng\inst{\ref{inst58}}\and
    S. J. Smartt\inst{\ref{inst48}} \and 
    S. Srivastav\inst{\ref{inst47}} \and \newline
    M. D. Stritzinger\inst{\ref{inst49}} \and
    G. Terreran\inst{\ref{inst61}} \and
    L. Tomasella\inst{\ref{inst3}}  \and
    G. Valerin\inst{\ref{inst3}} \and
    G.-J. Wang\inst{\ref{inst41},\ref{inst42}}\and 
    X.-F. Wang\inst{\ref{inst36},\ref{inst62}}\and 
    T. de Boer\inst{\ref{inst60}}\and \newline
    K. C. Chambers\inst{\ref{inst60}}\and
    H. Gao\inst{\ref{inst60}}\and
    F.-Z. Guo\inst{\ref{inst36}} \and 
    C. P. Guti\'errez\inst{\ref{inst33},\ref{inst15}}\and
    T. Kangas\inst{\ref{inst54},\ref{inst51}}\and 
    E. Karamehmetoglu\inst{\ref{inst17},\ref{inst49}}\and 
    G.-C. Li\inst{\ref{inst36}} \and \newline
    C.-C. Lin\inst{\ref{inst60}}\and 
    T. B. Lowe\inst{\ref{inst60}}\and
    X.-R. Ma\inst{\ref{inst36}} \and 
    E. A. Magnier\inst{\ref{inst60}}\and
    P. Minguez\inst{\ref{inst60}}\and 
    S.-P. Pei\inst{\ref{inst101}}\and 
    T. M. Reynolds\inst{\ref{inst51},\ref{inst56},\ref{inst57}}\and \newline 
    R. J. Wainscoat\inst{\ref{inst60}}\and 
    B. Wang\inst{\ref{inst6},\ref{inst7},\ref{inst8}}\and 
    S. Williams\inst{\ref{inst51},\ref{inst54}}\and 
    C.-Y. Wu\inst{\ref{inst6},\ref{inst7},\ref{inst8}} \and 
    J.-J. Zhang\inst{\ref{inst6},\ref{inst7},\ref{inst8}} \and
    X.-H. Zhang\inst{\ref{inst40}} \and \newline 
    X.-J. Zhu\inst{\ref{inst2},\ref{inst37},\ref{inst38}}\thanks{zhuxj@bnu.edu.cn (ZXJ)}
    }

   \institute{
    \label{inst1}School of Physics and Astronomy, Beijing Normal University, Beijing 100875, China \and
    \label{inst2}Department of Physics, Faculty of Arts and Sciences, Beijing Normal University, Zhuhai 519087, China \and
    \label{inst3}INAF - Osservatorio Astronomico di Padova, Vicolo dell'Osservatorio 5, 35122 Padova, Italy \and
    \label{inst6}Yunnan Observatories, Chinese Academy of Sciences, Kunming 650216, China \and
    \label{inst7}International Centre of Supernovae, Yunnan Key Laboratory, Kunming 650216, China \and
    \label{inst8}Key Laboratory for the Structure and Evolution of Celestial Objects, Chinese Academy of Sciences, Kunming 650216, China \newpage \and
    \label{inst45}School of Physics, O'Brien Centre for Science North, University College Dublin, Belfield, Dublin 4, Ireland \and
    \label{inst5}INAF - Osservatorio Astronomico di Brera, Via E. Bianchi 46, 23807 Merate (LC), Italy \and
    \label{inst46}Department of Astronomy, Xiamen University, Xiamen, Fujian 361005, China \and
    \label{inst36}Department of Physics, Tsinghua University, Beijing 100084, China \and
    \label{inst30}INAF - Osservatorio Astronomico d'Abruzzo, Via M. Maggini snc, 64100 Teramo, Italy \and
    \label{inst9}Las Cumbres Observatory, 6740 Cortona Drive, Suite 102, Goleta, CA 93117-5575, USA \and
    \label{inst10}Department of Physics, University of California, Santa Barbara, CA 93106-9530, USA \and
    \label{inst15}Institute of Space Sciences (ICE, CSIC), Campus UAB, Carrer de Can Magrans, s/n, E-08193 Barcelona, Spain \and
    \label{inst55}Institut f\"ur Theoretische Physik, Goethe Universit\"at, Max-von-Laue-Str. 1, 60438 Frankfurt am Main, Germany \and
    \label{inst19}Center for Astrophysics | Harvard \& Smithsonian, 60 Garden Street, Cambridge, MA 02138-1516, USA  \and
    \label{inst20}The NSF AI Institute for Artificial Intelligence and Fundamental Interactions, USA \and
    \label{inst51}Tuorla Observatory, Department of Physics and Astronomy, University of Turku, FI-20014 Turku, Finland \and
    \label{inst43}Key Laboratory of Optical Astronomy, National Astronomical Observatories, Chinese Academy of Sciences, Beijing 100101, China \and
    \label{inst17}The Oskar Klein Centre, Department of Astronomy, Stockholm University, AlbaNova, SE-10691 Stockholm, Sweden \and
    \label{inst52}Astrophysics Research Institute, Liverpool John Moores University, IC2, Liverpool Science Park, 146 Brownlow Hill, Liverpool L3 5RF, UK \and
    \label{inst53}Max-Planck-Institut f\"ur Astrophysik, Karl-Schwarzschild Str. 1, D-85748 Garching, Germany \and
    \label{inst50}
    School of Physics and Astronomy, University of Leicester, University Road, Leicester LE1 7RH, UK \and
    \label{inst58}College of Science, Chongqing University of Posts and Telecommunications, Chongqing 400065, China \and
    \label{inst48}Astrophysics Research Centre, School of Mathematics and Physics, Queen's University Belfast, BT7 1NN, UK \and
    \label{inst47}Astrophysics sub-Department, Department of Physics, University of Oxford, Keble Road, Oxford, OX1 3RH, UK \and
    \label{inst49}Department of Physics and Astronomy, Aarhus University, Ny Munkegade 120, DK-8000 Aarhus C, Denmark \and
    \label{inst61}Adler Planetarium, 1300 S. DuSable Lake Shore Drive, Chicago, IL 60605, USA \and
    \label{inst41}Department of Physics, Stellenbosch University, Matieland 7602, South Africa \and
    \label{inst42}National Institute for Theoretical and Computational Sciences (NITheCS), South Africa \and
    \label{inst62}Purple Mountain Observatory, Chinese Academy of Sciences, Nanjing, 210023, China \and
    \label{inst60}Institute for Astronomy, University of Hawai'i, 2680 Woodlawn Drive, Honolulu, HI 96822, USA \and
    \label{inst33}Institut d'Estudis Espacials de Catalunya (IEEC), 08860 Castelldefels (Barcelona), Spain \and
    \label{inst54}Finnish Centre for Astronomy with ESO (FINCA), Quantum, Vesilinnantie 5, University of Turku, FI-20014 Turku, Finland \and
    \label{inst101}School of Physics and Electrical Engineering, Liupanshui Normal University, Liupanshui, Guizhou, 553004, China \and
    \label{inst56}Cosmic Dawn Center (DAWN) \and
    \label{inst57}Niels Bohr Institute, University of Copenhagen, Jagtvej 128, 2200 København N, Denmark \and
    \label{inst40}School of Physics and Electronic Information, Jiangsu Second Normal University, Nanjing, Jiangsu 211200, China \and
    \label{inst37}Institute for Frontier in Astronomy and Astrophysics, Beijing Normal University, Beijing 102206, China \and 
    \label{inst38}Advanced Institute of Natural Sciences, Beijing Normal University, Zhuhai 519087, China
    }

   \date{Received March 26, 2025; accepted June 16, 2025}
 
  \abstract
  {
    We present the photometric and spectroscopic analysis of five Type Ibn supernovae (SNe): SN\,2020nxt, SN\,2020taz, SN\,2021bbv, SN\,2023utc, and SN\,2024aej. These events share key observational features and belong to a family of objects similar to the prototypical Type Ibn SN\,2006jc. The SNe exhibit rise times of approximately 10 days and peak absolute magnitudes ranging from $-$16.5 to $-$19 mag. Notably, SN\,2023utc is the faintest Type Ibn supernova discovered to date, with an exceptionally low $r$-band absolute magnitude of $-16.4$ mag. The pseudo-bolometric light curves peak at $(1-10) \times 10^{42}$ erg s$^{-1}$, with total radiated energies on the order of $(1-10) \times 10^{48}$ erg. Spectroscopically, these SNe display relatively slow spectral evolution; the early spectra are characterised by a hot blue continuum and prominent He~\textsc{i} emission lines. Early spectra show blackbody temperatures exceeding $10000~\mathrm{K}$, with a subsequent decline in temperature during later phases. Narrow He~\textsc{i} lines, indicative of unshocked circumstellar material (CSM), show velocities of approximately $1000~\mathrm{km~s^{-1}}$. The spectra suggest that the progenitors of these SNe underwent significant mass loss prior to the explosion, resulting in a He-rich CSM. Light curve modelling yields estimates for the ejecta mass ($M_{\rm ej}$) in the range $1-3~M_{\odot}$, with kinetic energies ($E_{\rm Kin}$) of $(0.1-1) \times 10^{50}$ erg. The inferred CSM mass ranges from $0.2$ to $1~M_{\odot}$. These findings are consistent with expectations for core-collapse events arising from relatively massive, envelope-stripped progenitors.
  }
 

   \maketitle
\nolinenumbers  
\section{Introduction}

Type~Ibn supernovae (SNe~Ibn) are a subclass of stellar explosions characterised by narrow ($\sim$1000 km s$^{-1}$) helium emission lines in their spectra, indicating the presence of He-rich circumstellar material \citep[CSM,][]{Smith2017hsn..book..403S, Gal-Yam2017hsn..book..195G}. The first SN~Ibn, SN~1999cq, was identified by \citet{Matheson2000AJ....119.2303M}, but the formal designation of the new SN type was introduced later by \citet{2008MNRAS.389..113P}, after the publications of the first studies on SN~2006jc, the prototypical SN of this class \citep[e.g.,][]{Foley2007ApJ...657L.105F, Pastorello2007Natur.447..829P, Anupama2009MNRAS.392..894A}. 

SNe~Ibn are rare, with only 73 confirmed events to date.\footnote{Data based on a query from the Transient Name Server (\url{https://www.wis-tns.org/}) on 26 March 2025.} \citet{Maeda2022ApJ...927...25M} estimated their volumetric rate to be $\sim$1\% of all core-collapse supernovae (CC~SNe), while \citet{Perley2020ApJ...904...35P} reported a detection rate of 0.66\% within the Zwicky Transient Facility (ZTF) transient sample. Furthermore, \citet{Ma2025arXiv250404507M,Ma2025arXiv250404393M} analysed a nearby SN sample within 40\,Mpc — compiled primarily from wide-field surveys conducted between 2016 and 2023 — and found that SNe~Ibn comprise around 1\% of the total sample. Despite significant progress in the discovery and characterisation of SNe~Ibn, their progenitor systems remain enigmatic due to the limited sample size and the diversity in their observed properties \citep{Maund2016ApJ...833..128M,Maeda2022ApJ...927...25M,Dessart2022A&A...658A.130D}.

\cite{Hosseinzadeh2017ApJ...836..158H} suggest that Type~Ibn SNe exhibit an overall photometric homogeneity but show significant spectral diversity around the maximum light. From a spectroscopic point of view, SNe~Ibn are characterised by narrow He~\textsc{i} emission lines, with full-width at half maximum (FWHM) velocities ranging from a hundred to a few thousand $\rm{km\,s^{-1}}$ \citep{Pastorello2016MNRAS.456..853P, Hosseinzadeh2017ApJ...836..158H}. In some cases, weak hydrogen lines have been also detected in the spectra, suggesting the presence of a residual amount of H in the CSM \citep{Pastorello2008MNRAS.389..131P,Smith2012MNRAS.426.1905S,Pastorello2015MNRAS.449.1921P,Reguitti2022A&A...662L..10R,Wang2024A&A...691A.156W}. SNe~Ibn light curves usually exhibit fast rise times ($\leq 15$ days), rapid post-peak declines (0.05--0.15~mag~$\rm{day^{-1}}$), and a peak absolute magnitude of $M \sim -19$~mag. 

Despite the findings of photometric homogeneity \citep{Hosseinzadeh2017ApJ...836..158H}, there are a few outliers, including the superluminous SN ASASSN-14ms ($M_V \sim -20.5$~mag; \citealt{Vallely2018MNRAS.475.2344V, Wang2021ApJ...917...97W}), long-lasting transients such as OGLE2012-SN-006 \citep{Pastorello2015MNRAS.449.1941P}, the double-peaked iPTF13beo \citep{Gorbikov2014MNRAS.443..671G}, and the slow-rising OGLE-2014-SN-131 \citep{Karamehmetoglu2017AA...602A..93K}. This could be due to a variety of progenitor systems and explosion mechanisms for SNe~Ibn, although in most cases they are considered stripped-envelope CC~SNe interacting with He-rich environments \citep{Chugai2009MNRAS.400..866C}.\footnote{But see, e.g., \citet{Sanders2013ApJ...769...39S} and \citet{Kool2023Natur.617..477K}.} 

The progenitor systems of SNe~Ibn are an area of active investigation, with multiple channels proposed to explain their diverse observational characteristics. Initially, the progenitors were identified as massive ($M_{\rm ZAMS}$~$\ge$~25~$M_\odot$) hydrogen-poor Wolf-Rayet (WR) stars embedded in He-rich CSM \citep{Pastorello2007Natur.447..829P}. These WR stars undergo substantial mass loss prior to core collapse, resulting in the formation of a dense CSM. When the fast-moving SN ejecta collide with this slow CSM, shocks are generated that heat and ionize the He-rich material in the CSM, leading to the formation of the narrow He emission lines characteristic of SNe~Ibn. The frequent association of SNe~Ibn with active star-forming regions supports this scenario \citep{Taddia2015A&A...580A.131T, Pastorello2015MNRAS.449.1921P}. However, the Type Ibn SN PS1-12sk occurred in the outskirts of an elliptical galaxy with low star formation activity, challenging the massive star progenitor scenario as the sole channel producing Type Ibn SNe \citep{Sanders2013ApJ...769...39S}.

An alternative progenitor scenario involves lower-mass (final masses $\lesssim$5~M$_{\odot}$; \citealp{Dessart2022A&A...658A.130D}) helium stars in binary systems, where binary interaction drives episodic mass losses gathering the CSM prior to the core collapse \citep{Maund2016ApJ...833..128M}. Late-time Hubble Space Telescope (HST) images of SNe Ibn explosion sites indicate that at least some SNe Ibn were in relatively low-mass binary systems \citep{Sun2020MNRAS.491.6000S}. Binary systems may account for SNe~Ibn in older stellar populations, with \citet{Dessart2022A&A...658A.130D} providing evidence for low-mass binary progenitors through spectral modelling. Furthermore, some type~IIb SNe, which explode within dense He-rich CSM and evolve into SNe Ibn features, suggest that multiple progenitor pathways might lead to SN~Ibn \citep{Prentice2020MNRAS.499.1450P}. Thus, while SNe~Ibn are unified by their spectroscopic signatures, their progenitor systems likely encompass a range of evolutionary scenarios, reflecting both single and binary star channels. \citet{Metzger2022ApJ...932...84M} proposed a novel mechanism for type~Ibn SNe, suggesting that their emission could be powered by disk outflows resulting from hyper-accretion onto a compact object, such as a neutron star or black hole, located in the vicinity of a helium star, rather than being driven by stellar mass loss. This mechanism also predicts a relatively low yield of $^{56}\mathrm{Ni}$.

X-ray observations, though only occasionally employed, provide a direct means of probing the mass-loss history of the SN progenitor. The X-ray emission is produced by forward and reverse shocks arising from the ejecta-CSM interaction \citep{Chevalier1994ApJ...420..268C}. It allows to constrain the density of CSM and the mass distribution, allowing estimates of progenitor mass-loss rates in the last years before the explosion \citep[e.g.][]{Immler2001ApJ...561L.107I, Tsuna2021ApJ...914...64T, Margalit2022ApJ...928..122M}. To our knowledge, only two Type~Ibn SNe, SN~2006jc \citep{Immler2008ApJ...674L..85I} and SN~2022ablq \citep{Pellegrino2024ApJ...977....2P}, have well-sampled X-ray light curves. In SN~2006jc, the X-ray flux peaked at $\sim$100~days after the explosion and was attributed to the shock encountering a dense shell ejected two years earlier, consistent with a previously observed optical outburst \citep{Immler2008ApJ...674L..85I}. The inferred CSM mass exceeded $0.01~M_{\odot}$. For SN~2022ablq, enhanced mass-loss rates ranging from 0.05 to $0.5~M_{\odot}$~yr$^{-1}$ was noticed from 2 to 0.5~years before the explosion, suggesting an eruptive event from a lower-mass progenitor rather than steady winds from a Wolf-Rayet star \citep{Pellegrino2024ApJ...977....2P}. More recently, \citet{Inoue2025ApJ...980...86I} developed broadband X-ray light-curve models for SNe~Ibn/Icn, providing theoretical predictions to guide future high-cadence X-ray observations of interacting transients.

In this paper, we present a detailed analysis of the photometric and spectroscopic observations of five Type~Ibn supernovae, SNe~2020nxt, 2020taz, 2021bbv, 2023utc, and 2024aej, to investigate their observational properties and compare them with previously studied events. The basic discovery details, including distance and extinction estimates, are outlined in Section~\ref{sec:information}. In Section~\ref{sec:photometry}, we examine their light curves and fit bolometric light curve models to derive key physical parameters. A comprehensive analysis of the spectral properties of these SNe~Ibn is presented in Section~\ref{sec:spectroscopy}. Finally, we discuss and summarize our findings in Section~\ref{sec:Discussion}.

\section{Basic sample information} 
\label{sec:information}

\begin{figure*}
\begin{center}
\includegraphics[width=0.33\linewidth]{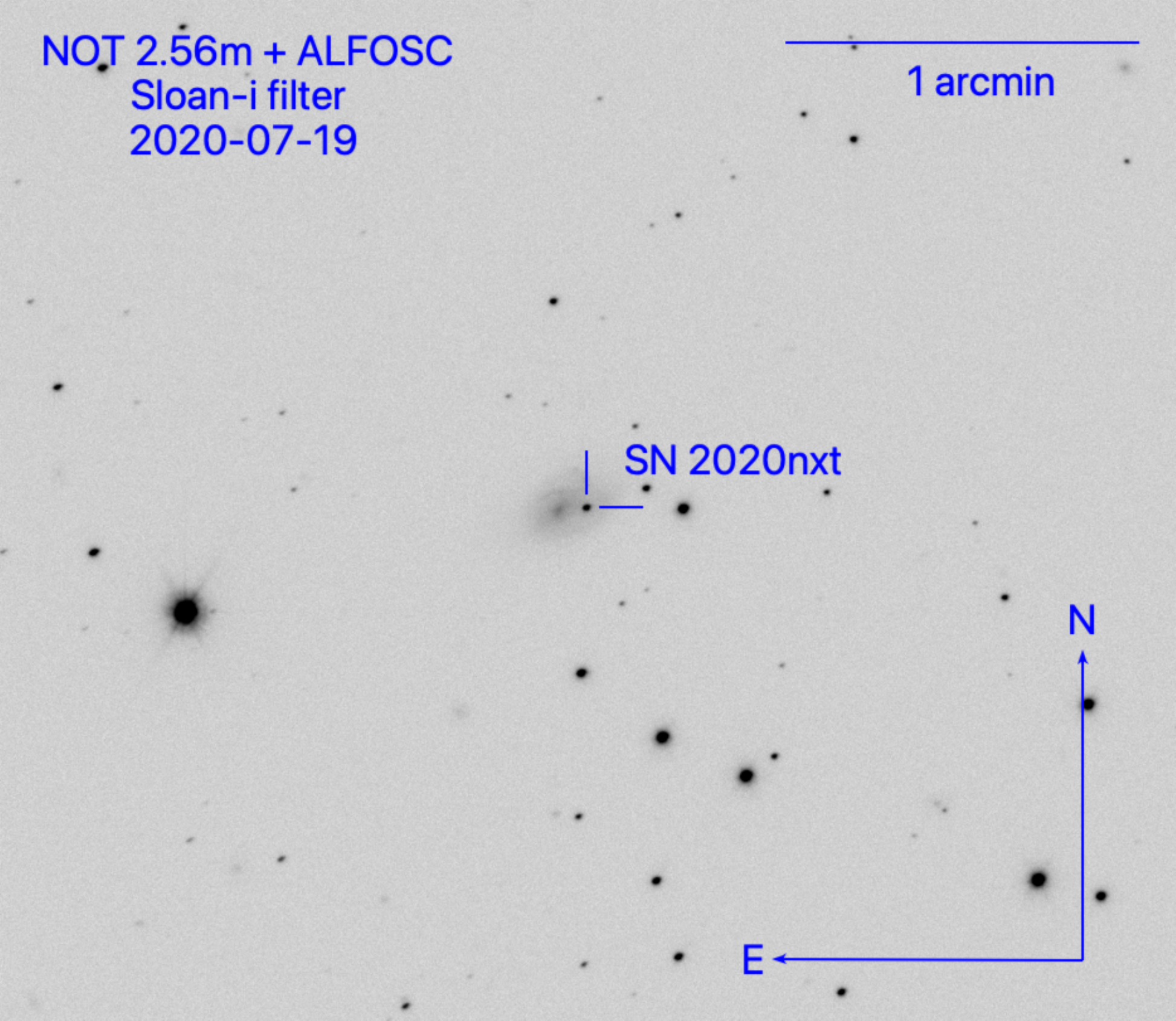}
\includegraphics[width=0.33\linewidth]{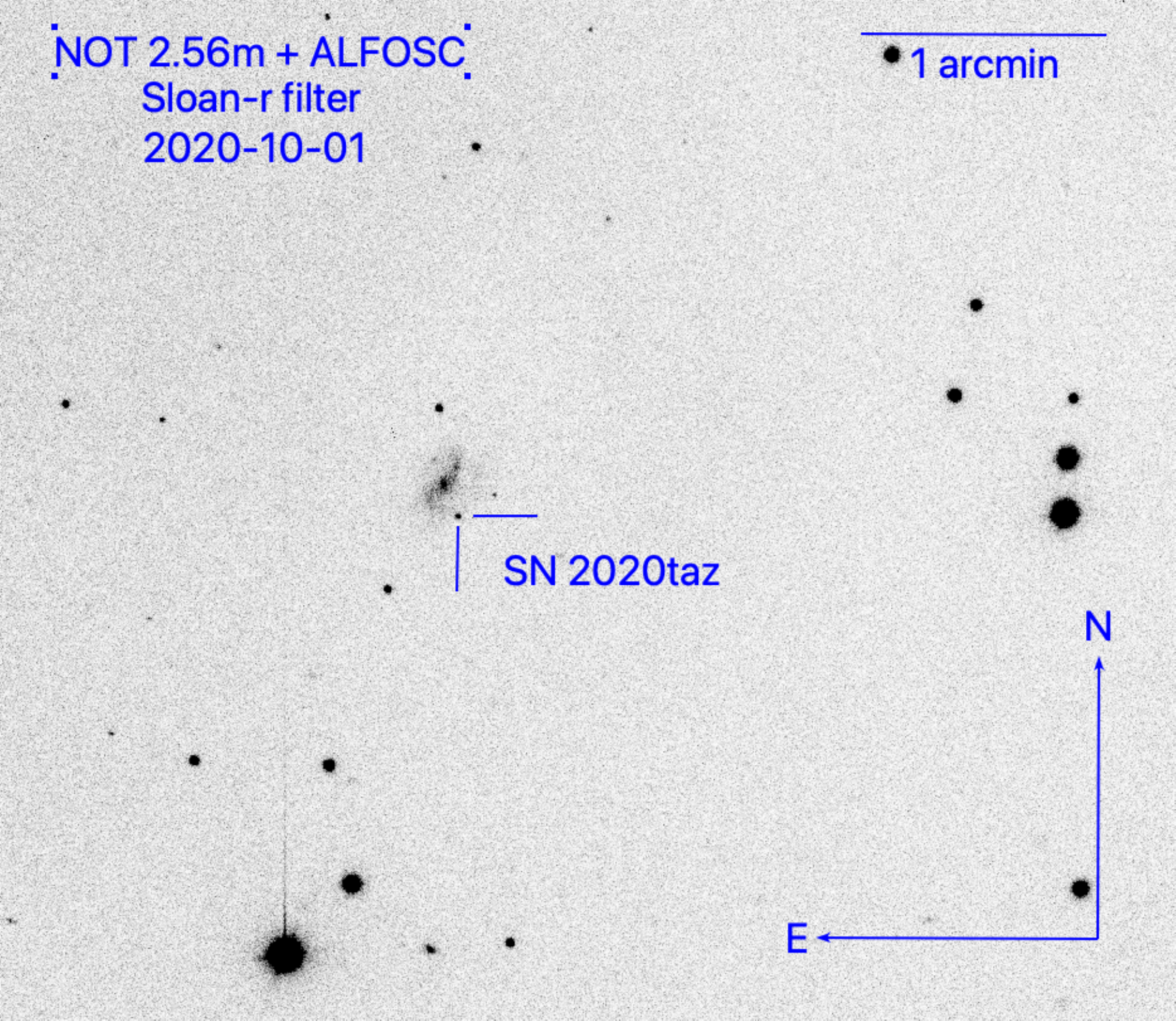}
\includegraphics[width=0.33\linewidth]{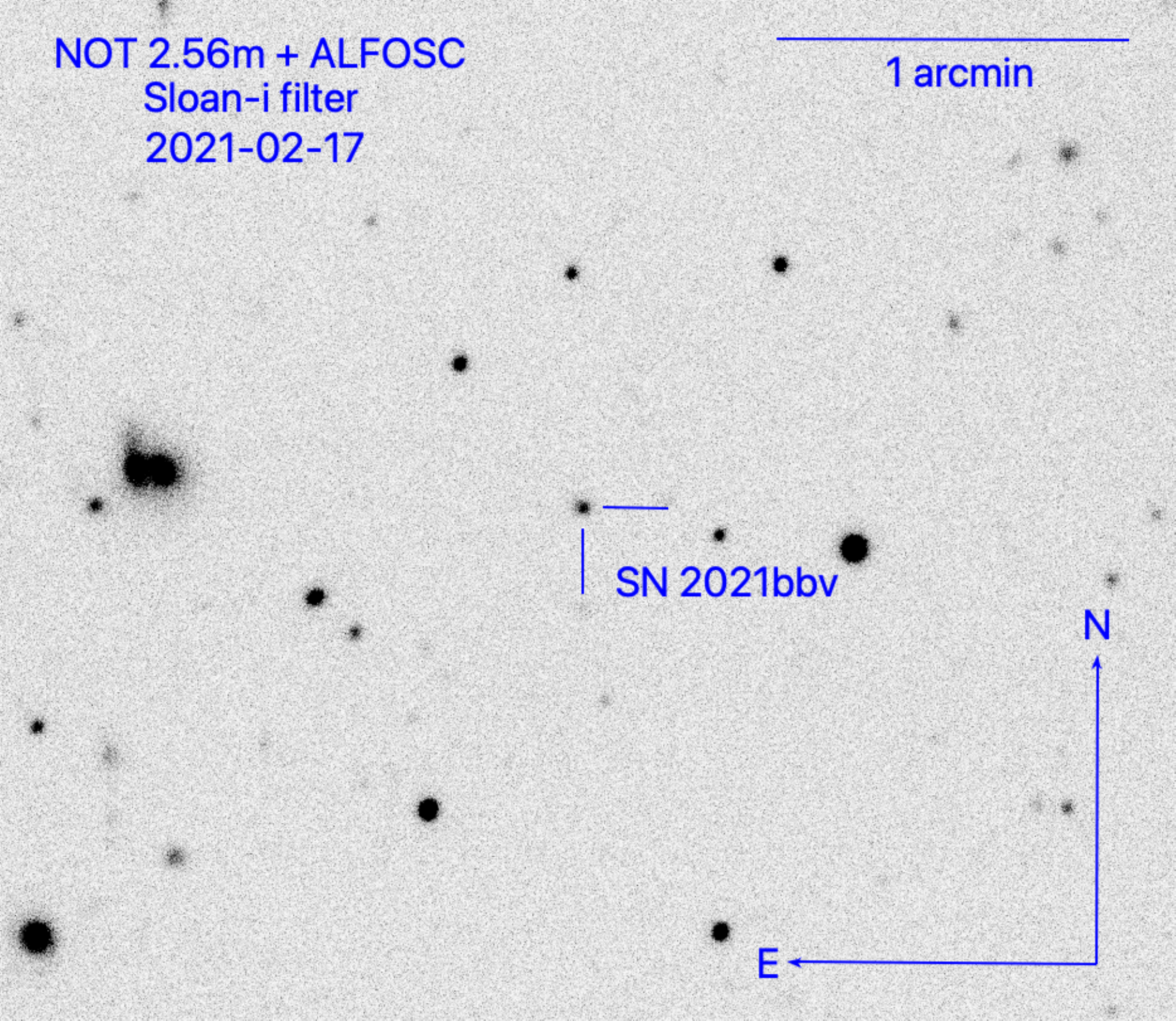}
\includegraphics[width=0.33\linewidth]{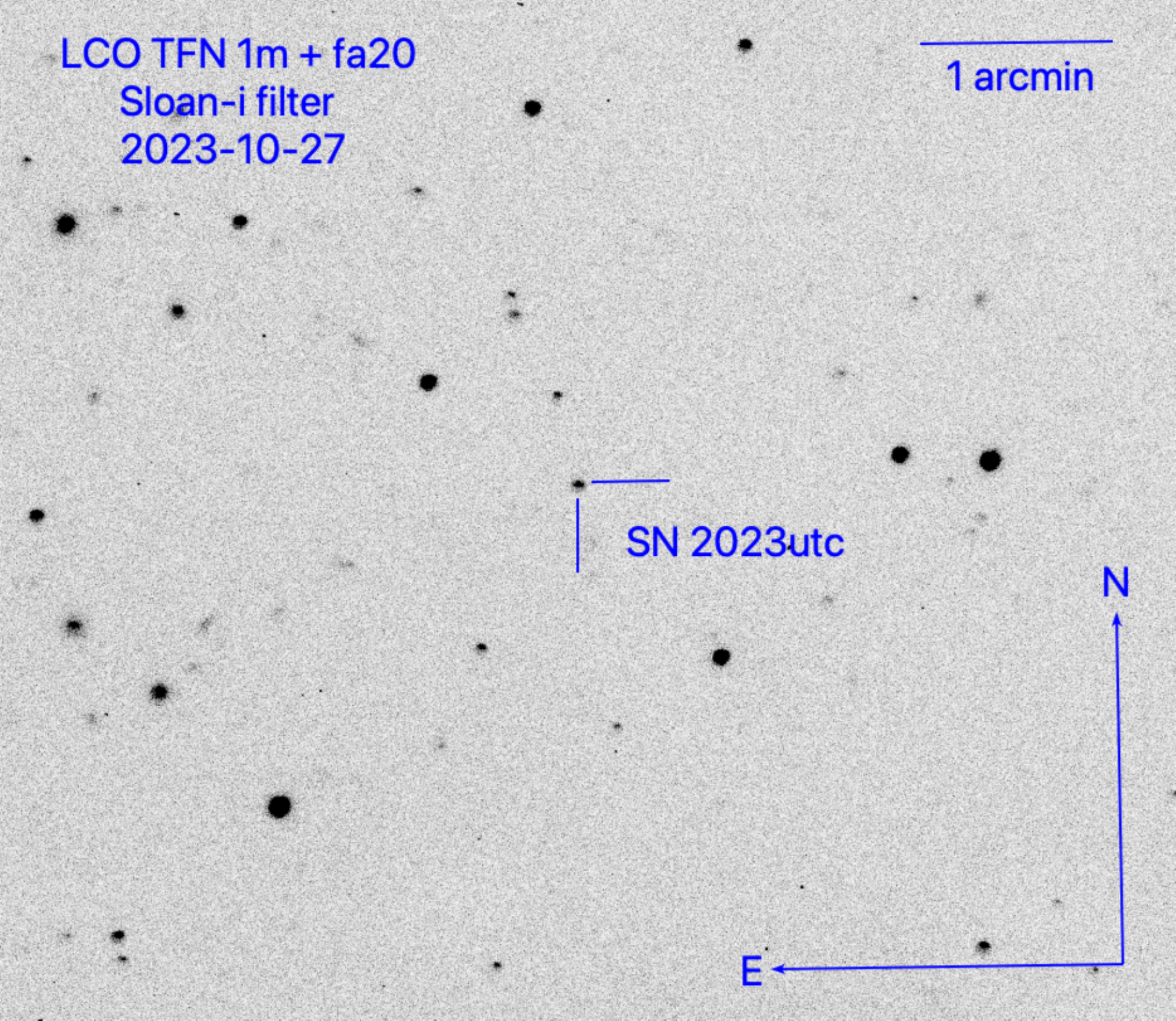}
\includegraphics[width=0.33\linewidth]{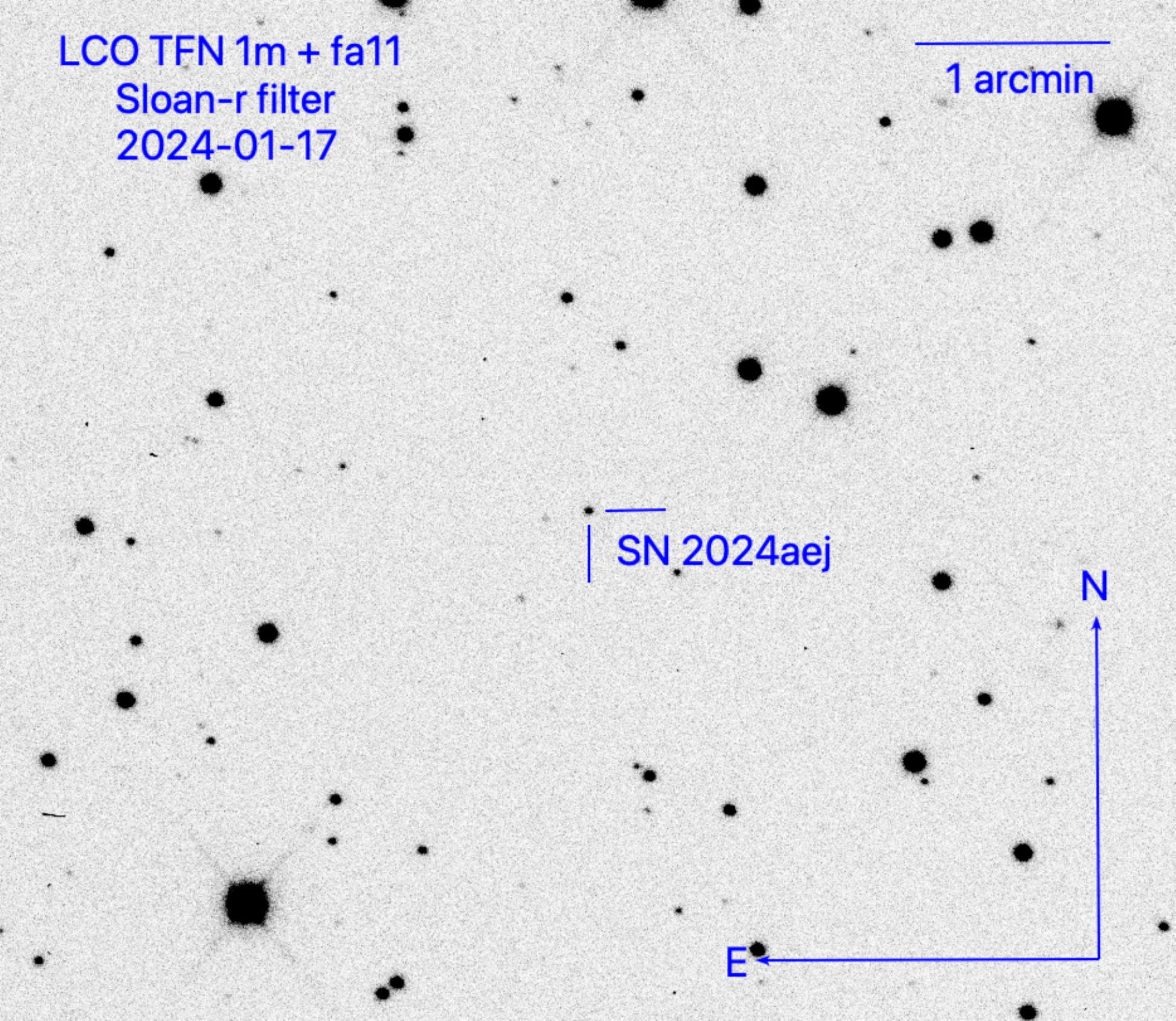}
\caption{
SN\,2020nxt in a NOT/ALFOSC image taken with a Sloan $i$-band filter on July 19, 2020; 
SN\,2020taz in a NOT/ALFOSC image taken with a Sloan $r$-band filter on October 1, 2020;
SN\,2021bbv in a NOT/ALFOSC image taken with a Sloan $i$-band filter on February 17, 2021;
SN\,2023utc in a LCO TFN/fa20 image taken with a Sloan $i$-band filter on October 27, 2023;
SN\,2024aej in a LCO TFN/fa11 image taken with a Sloan $r$-band filter on January 17, 2024.}
\label{fig:finderchart}
\end{center}
\end{figure*}

\begin{table*}
    \centering
    \caption{Basic information for the five Ibn SN host galaxies. }
    {\fontsize{9pt}{16pt}\selectfont
    \setlength{\tabcolsep}{8pt} 
    \begin{tabular}{ccccccccccc}
        \hline
        \hline
        Object      & Host Galaxy    &   Redshift   &  Distance        &  Distance Modulus & Radial Distance   & $E(B-V)_{\mathrm{MW}}$ \\
        SN          &                &              & ($\mathrm{Mpc}$) & ($\mathrm{mag}$)  & ($\mathrm{kpc}$)  &  ($\mathrm{mag}$)   \\
        \hline
        \hline
        2020nxt  & WISEA J223736.70+350006.5    &  0.0218(0.0003)  &  91.1(1.2)   & 34.80(0.03)  & 2.6(0.1) &  0.067     \\ 
        2020taz  & WISEA J222606.51+103337.3    &  0.0494(0.0005)  &  210.7(2.3)  & 36.62(0.02)  & 8.7(0.1) &  0.091    \\ 
        2021bbv  & SDSS J113020.86+085535.0     &  0.068(0.003)    &  294.1(13.6) & 37.34(0.10)  & 0.9(0.1) &  0.033    \\ 
        2023utc  & SDSS J091159.16+534304.2     &  0.014(0.003)    &  57.5(12.3)  & 33.80(0.46)  & 0.5(0.1) &  0.015    \\ 
        2024aej  & WISEA J014427.03+390545.2    &  0.063(0.003)    &  271.5(13.5) & 37.17(0.11)  & 6.0(0.3) &  0.055    \\ 
        \hline
        \hline
    \end{tabular}}
    \label{table:host_galaxies}
\end{table*}

\subsection{Individual Objects}
\subsubsection{SN 2020nxt}

SN\,2020nxt (also known with the survey designations ATLAS20rzv and PS20geh) was first detected by the Asteroid Terrestrial-impact Last Alert System \citep[ATLAS, equipped with the ACAM1 camera mounted on the ATLAS Haleakala Telescope;][]{Tonry2018PASP..130f4505T, Smith2020PASP..132h5002S, Shingles2021TNSAN...7....1S}, on 2020 July 03.54 UT (MJD= 59033.54; UT dates are used throughout the article) at the \textit{orange} filter magnitude of $o$ = 17.19 mag \citep[AB,][]{Tonry2020TNSTR2022....1T}. {The last ATLAS  non-detection was on 2020 June 30.56 (MJD = 59030.56) in the $o$ band, with an estimated upper limit of 20.76 mag.\footnote{\url{https://fallingstar-data.com/}}} The spectrum of this transient showed a match with the Type Ibn SN~2006jc, about 10 days past explosion. This supports the classification of SN 2020nxt as a Type Ibn SN \citep{Srivastav2020TNSCR2148....1S}. SN\,2020nxt, at RA~=~$22^{h}37^{m}36\fs235$, Dec~=~$+35\degr00\arcmin07\farcs68$~(all coordinates in this paper are given in J2000), is possibly associated with a galaxy named SDSS J223736.60+350007.4 ({2MASXJ22373660+3500074}), with the SN being 1\farcs15 north and 5\farcs76 west from the galaxy centre (see Fig.~\ref{fig:finderchart}). 

We adopt the redshift of the host galaxy, \( z = 0.0218 \pm 0.0003 \), from \citet{Wang2024MNRAS.530.3906W}. Assuming a standard cosmology with $H_{0}=73$ km s$^{-1}$ $\rm{Mpc}^{-1}$, $\Omega_{\mathrm{M}}=0.27$, $\Omega_{\mathrm{\Lambda}}=0.73$ \citep[used consistently throughout this paper;][]{Spergel2007ApJS..170..377S}, we estimated a luminosity distance $d_L$~=~91.1 $\pm$ 1.2\,Mpc ($\mu$ ~=~ 34.80 $\pm$ 0.03\,mag)\footnote{The uncertainty in the distance does not incorporate the contribution from the Hubble constant, and similar assumptions apply to all distance estimates presented in this paper.} for SN\,2020nxt. The Milky Way extinction in the SN direction is $E(B - V)_{MW} =$ 0.067 mag  \citep{Schlafly2011ApJ...737..103S}.

\subsubsection{SN 2020taz}
The discovery of SN\,2020taz (also known as ATLAS20baee, PS20hil and ZTF20abyznqs) was announced by the Panoramic Survey Telescope and Rapid Response System \citep[Pan-STARRS;][]{Wainscoat2016IAUS..318..293W} on 2020 September 11.32 (MJD = 59103.32) with a PanSTARRS-w filter brightness of 19.02 mag \citep[AB,][]{Chambers2020TNSTR2783....1C}. A prediscovery detection was obtained by the ATLAS survey on 2020 September 08.44 (MJD = 59100.44), with the object having an $o$-band magnitude of $o$ = 20.54$\pm$0.24 mag.\footnote{\url{https://fallingstar-data.com/}} The last non-detection, by the ZTF survey \citep[][]{Bellm2019PASP..131a8002B}, was on 2020 September 07.33 (MJD = 59099.33) in the $r$ band, with an estimated limit of 20.13 mag.\footnote{\url{https://alerce.online/object/ZTF20abyznqs}} Soon after the discovery, SN\,2020taz was classified as a Type Ibn SN \citep{Graham2020TNSCR2927....1G}. Its coordinates are RA~=~$22^\mathrm{h}26^\mathrm{m}06\fs270$, Dec~=~$+10^\circ33'29\farcs65$.

SN\,2020taz is located 3\farcs60 west and 7\farcs74 south of the centre of its predicted host galaxy, WISEA J222606.51+103337.3 (PGC 1381253). The location of SN\,2020taz is shown in Fig.~\ref{fig:finderchart}. Adopting the recessional velocity of $v$ = 14796~$\pm$~140~km s$^{-1}$ \citep[hence a redshift z = 0.0494 $\pm$ 0.0005;][]{Mould2000ApJ...529..786M}, and adopting the same standard cosmological model, we obtain a luminosity distance of $d_{L} = 210.7\pm2.3$~Mpc, hence a distance modulus $\mu_{L} = 36.62\pm0.02$~mag. The Milky Way extinction is $E(B - V)_{MW} =$ 0.091 mag in this direction \citep{Schlafly2011ApJ...737..103S}. 

Given the total apparent magnitude of PGC 1381253, as provided by HyperLeda\footnote{\url{http://atlas.obs-hp.fr/hyperleda/}} ($B = 17.04 \pm 0.50$), we derive a total absolute $B$-band magnitude of $-20.17$, after correction for Galactic extinction. Using the luminosity-metallicity relation from \citet{Tremonti2004ApJ...613..898T}, we estimate the overall oxygen abundance of PGC 1381253 to be $12 + \log(\text{O/H}) = 8.99$ dex. Based on the radial metallicity gradient model by \citet{Pilyugin2004A&A...425..849P}, we estimate the oxygen abundance at the supernova (SN) site to be $12 + \log(\text{O/H})_{\text{SN}} = 8.67$ dex. This value is slightly above the mean oxygen abundances reported at the SN location for Type Ibn SNe host galaxies, such as $8.63 \pm 0.42$ dex by \citet{Pastorello2015MNRAS.449.1954P} and $8.45 \pm 0.10$ dex by \citet{Taddia2015A&A...580A.131T}.

\subsubsection{SN 2021bbv}
The discovery of SN\,2021bbv (Gaia21akw, ZTF21aagyidr, PS21afl), attributed to the Gaia Photometric Science Alerts (GPSA)\footnote{\url{http://gsaweb.ast.cam.ac.uk/alerts}}, is dated 2021 January 24.00 (MJD = 59238.00). The object was observed in a Gaia-G filter image, with a magnitude of 18.7 mag \citep{Hodgkin2021TNSTR.241....1H}. However, an earlier detection was reported by the ZTF on 2021 January 20.45 (MJD = 59234.45), at a magnitude of $g~=~ 20.45\pm0.33$~mag.\footnote{\url{https://alerce.online/object/ZTF21aagyidr}} The last non-detection provided by ATLAS was on 2021 January 18.52 (MJD = 59232.52) in the cyan ($c$) band, to a limit of 20.53 mag.\footnote{\url{https://fallingstar-data.com/}} Soon after its discovery, the SN was classified as a Type Ibn event \citep{Gonzalez2021TNSCR.258....1G}. The SN coordinates are RA~=~$11^\mathrm{h}30^\mathrm{m}20\fs830$, Dec~=~$+08^\circ55'34\farcs68$ (see Fig.~\ref{fig:finderchart}). 

SN\,2021bbv is located 0.55\farcs west and 0.36\farcs south of the centre of its predicted host galaxy, SDSS J113020.86+085535.0. Due to the lack of distance information, we inferred the kinematic distance of the host galaxy from measuring the central wavelength of the most prominent narrow He I ($\lambda_{0}$~=~5875.6 \AA) line in the SN spectra, and we obtained a redshift of $z$~=~0.068$\pm$0.003. Then, we obtain a luminosity distance of $d_{L} = 294.1\pm13.6$~Mpc, hence a distance modulus $\mu_{L} = 37.34\pm0.10$~mag. The Milky Way extinction towards SN\,2021bbv is $E(B - V)_{MW} =$ 0.033 mag  \citep{Schlafly2011ApJ...737..103S}.

\subsubsection{SN 2023utc}
SN\,2023utc (KATS23T003, ATLAS23ukm, ZTF23abjjrwv, PS23lcq) was discovered by the Xingming Sky Survey (XOSS)\footnote{\url{http://xjltp.china-vo.org/about-xingming.html}} project on 2023 October 11.94 (corresponding to MJD = 60228.94). The object was observed in an unfiltered image with magnitude of 18.15~$\pm$~0.21 \citep{Zhang2023TNSTR2566....1Z}. The last ZTF non-detection in the $g$ band was on 2023 October 07.45 (MJD = 60224.45), to a limit of 19.84 mag. Soon after discovery, the SN was classified as a Type Ibn event by \citet{Taguchi2023TNSCR2735....1T}. The SN coordinates are RA~=~$09^\mathrm{h}11^\mathrm{m}59\fs155$, Dec~=~$+53^\circ43'02\farcs60$~(see Fig.~\ref{fig:finderchart}).

SN\,2023utc is possibly associated with the host galaxy SDSS J091159.16+534304.2. Missing alternative distance estimates, and due to the lack of host galaxy lines in the spectra (see Section~\ref{sec:spectroscopy}), we measured the central wavelength of the narrow He I SN lines and obtained a redshift of $z$ = 0.014$\pm$0.003. Then, we infer a Hubble flow distance of $d_{H} = 57.5\pm12.3$~Mpc, hence a distance modulus $\mu_{H} = 33.80\pm0.46$~mag. The Milky Way extinction towards this direction is $E(B - V)_{MW} =$ 0.015 mag \citep{Schlafly2011ApJ...737..103S}. 

Given the apparent magnitude of SDSS J091159.16+534304.2 ($g = 20.86 \pm 0.05$) as reported by SDSS\footnote{\url{https://skyserver.sdss.org/dr16/}}, we derive an absolute $g$-band magnitude of $-13.00$, after correcting for Galactic extinction. This places it among the faintest host galaxies for CC SNe \citep{Li2011MNRAS.412.1441L}. Using the luminosity-metallicity relation from \citet{Tremonti2004ApJ...613..898T}, we estimate the overall oxygen abundance of SDSS J091159.16+534304.2 to be $12 + \log(\text{O/H}) = 7.613$~dex, which is subsolar (adopting a solar metallicity of $12 + \log(\text{O/H}) = 8.69$~dex; see e.g., \citealt{vonSteiger2016ApJ...816...13V, Vagnozzi2019Atoms...7...41V, Asplund2021A&A...653A.141A}), and lower than most SNe Ibn host galaxies \citep{Taddia2015A&A...580A.131T, Pastorello2015MNRAS.449.1954P}. Due to the lack of detailed information on the host galaxy, we are unable to accurately estimate the oxygen abundance at the SN location.

\subsubsection{SN 2024aej}
The ATLAS discovery of SN\,2024aej (ATLAS24awb, ZTF24aabwvws) is dated 2024 January 14.32 (MJD = 60323.32), at an $o$-band brightness of 19.10~$\pm$~0.14 mag \citep{Tonry2024TNSTR.151....1T}. However, an earlier detection is reported by ZTF on 2024 January 14.22 (MJD = 60323.22), at a magnitude of $g$ = 18.85~$\pm$~0.13 mag.\footnote{\url{https://lasair-ztf.lsst.ac.uk/objects/ZTF24aabwvws/}} The last ATLAS non-detection in the $o$ band is dated 2024 January 11.32 (MJD = 60320.32), to a limiting magnitude of 20.81 mag. Soon after its discovery, the SN was classified as a Type Ibn event by the Global Supernova Project \citep{Terreran2024TNSCR.232....1T}. The SN coordinates are RA~=~$01^\mathrm{h}44^\mathrm{m}27\fs388$, Dec~=~$+39^\circ05'47\farcs20$ (see Fig.~\ref{fig:finderchart}).

SN\,2024aej is associated with a galaxy named WISEA J014427.03+390545.2. Due to the lack of alternative distance estimates, as for \object{SN\,2023utc}, we measured the central wavelength of the narrow He~I SN lines, and obtained a redshift of $z$ = 0.063~$\pm$~0.003. Then, we infer a luminosity distance of $d_{L} = 271.5~\pm~13.5$~Mpc, hence a distance modulus $\mu_{L} = 37.17~\pm~0.11$~mag. The Milky Way extinction in the SN direction is $E(B - V)_{MW} =$ 0.055 mag \citep{Schlafly2011ApJ...737..103S}.

\subsection{Interstellar reddening} \label{sec:reddening}
{The information for the host galaxies of the five SNe~Ibn in our sample is summarized in Table~\ref{table:host_galaxies}.

Usually, dust extinction within the SN host galaxies can be inferred through empirical relations between the equivalent width (EW) of the narrow Na~I doublet (Na~I~D) absorption line at the galaxy redshift, as observed in the early-time SN spectra, and the colour excess \citep[see, e.g.,][]{Turatto2003fthp.conf..200T, Poznanski2012MNRAS.426.1465P}. Unfortunately, for our SN sample, the host galaxy extinction cannot be firmly constrained as a consequence of the modest S/N and limited spectral resolution of our spectra (see Sect. \ref{sec:spectroscopy}). 

We have closely examined the Na\,\textsc{i}~D region in all available spectra (see Fig.~\ref{fig:NaID} in Appendix~\ref{sec:NaID}), but the S/N of the spectra is insufficient to identify or place meaningful limits on the presence of host-galaxy Na\,\textsc{i}~D absorption. In many cases, this spectral region is blended with He\,\textsc{i}~$\lambda$5876 emission, and any potential Na\,\textsc{i}~D absorption is indistinguishable from noise patterns. Consequently, we can only report conservative $3\sigma$ upper limits on EW, from which we estimate corresponding upper limits on the extinction values within the host galaxies (see Appendix~\ref{sec:NaID}).

Moreover, we caution that even a marginal detection of Na\,\textsc{i}~D would not necessarily provide a reliable estimate of the host-galaxy extinction in the context of interacting transients such as SNe~Ibn. As discussed by \citet{Byrne2023MNRAS.524.2978B} and \citet{Kochanek2012ApJ...759...20K}, Na\,\textsc{i}~D equivalent widths can be significantly affected by photoionisation in the circumstellar environment, as well as by geometric and radiative transfer effects, making them a poor proxy of extinction under such conditions. For all above reasons, we adopt a conservative approach and assume that the total line-of-sight reddening is solely due to the Galactic contribution. We acknowledge that this introduces a source of systematic uncertainty in our analysis.}

\section{Photometry}\label{sec:photometry}

\subsection{Observations and Data Reductions}\label{subsec:data}

We conducted comprehensive multiband follow-up campaigns for SNe 2020nxt, 2020taz, 2021bbv, 2023utc, and 2024aej in the Johnson-Cousins $UBV$ and Sloan $ugriz$ filters, with monitoring started shortly after the SN discoveries. Details on instrumental configurations are provided in Table~\ref{table_telescope} (Appendix~\ref{sec:facilities}). 

All raw images were pre-processed using standard reduction procedures in \textsc{iraf}\footnote{\url{http://iraf.noao.edu/}} \citep{Tody1986SPIE..627..733T, Tody1993ASPC...52..173T}, which included bias, overscan, and flat-field corrections \citep[see, e.g.,][]{Cai2018MNRAS.480.3424C}. For faint objects, multiple exposures were taken and subsequently combined to enhance the S/N ratio. Photometry was performed using the dedicated pipeline {\sl ecsnoopy},\footnote{{\sl ecsnoopy} is a package for SN photometry using PSF fitting and/or template subtraction developed by E. Cappellaro. A package description can be found at \url{http://sngroup.oapd.inaf.it/snoopy.html}.} which incorporates several photometric packages, including {\sc sextractor}\footnote{\url{www.astromatic.net/software/sextractor/}} \citep[][]{Bertin1996A&AS..117..393B} for source extraction, {\sc daophot}\footnote{\url{http://www.star.bris.ac.uk/~mbt/daophot/}} \citep[][]{Stetson1987PASP...99..191S} for magnitude measurements by fitting the point spread function (PSF), and {\sc hotpants}\footnote{\url{http://www.astro.washington.edu/users/becker/v2.0/hotpants.html}} \citep[][]{Becker2015ascl.soft04004B} for PSF-matched image subtraction. 
The SN instrumental magnitudes were measured using the PSF-fitting method, with the sky background first subtracted by fitting a low-order polynomial to the surrounding region. The PSF was modeled by fitting isolated and non-saturated field star profiles in the SN field. The PSF model was then subtracted from the original images to re-estimate the local background, and residuals were inspected to assess the fit quality. 
In the case of SN 2020taz, a straightforward PSF-fitting approach was adopted because of its location on the outskirts of its host galaxy. For SNe\,2020nxt, 2021bbv, 2023utc, and 2024aej, template subtraction was employed to mitigate the background contamination.\footnote{Sloan Digital Sky Survey (SDSS) templates were used for the Sloan images of SN\,2020nxt, while Johnson-Bessel templates were obtained with the Liverpool Telescope (LT) on 2020 October 08. In the case of SN 2021bbv, the $U$ band template images were acquired with LCO-fa20 on 2024 March 05, and the $BVugriz$ templates were taken with the Nordic Optical Telescope (NOT) on 2024 February 12, approximately three years after the discovery. In the case of SN\,2023utc, $UBVgri$ templates were obtained with LCO-fa11 on 2024 November 1, one year after discovery. Lastly, for SN 2024aej, $UBVgri$ templates were collected using LCO-fa16 on 2024 September 18, around seven months after the SN had faded.}

Once the SN instrumental magnitudes were obtained, we used zero points (ZPs) and colour terms (CTs) of each instrument for the photometric calibration to a standard system. ZPs and CTs were determined through observations of standard stars in the photometric nights among the SN observations. Johnson-Bessel magnitudes were calibrated using the \citet{Landolt1992AJ....104..340L} catalog, while Sloan ones were directly derived from the SDSS DR 18 catalog \citep{Almeida2023ApJS..267...44A}. As the field of SN\,2024aej was not sampled by SDSS, the Sloan-filter photometry of this SN was calibrated using reference stars taken from the Pan-STARRS catalog. A local sequence of standard stars in the vicinity of the SN was used to correct the zero points obtained in non-photometric nights and improve the SN calibration accuracy. 

Photometric uncertainties were estimated through an artificial star experiment. Multiple artificial stars of known magnitudes were evenly distributed near the position of the SN in the PSF-fit residual image and subsequently processed through PSF fitting. The total photometric uncertainties were then determined by combining (in quadrature) the errors from the artificial star experiment, the PSF fit, and the zero-point correction. When template subtraction was employed, no artificial star experiment was performed. Instead, the background uncertainty was determined from the root mean square (RMS) of the residuals in the background after subtracting the PSF-fitted source.

Additionally, we gathered archival data from public sources, including ATLAS, ZTF, Pan-STARRS, and the All-Sky Automated Survey for Supernovae (ASAS-SN). The ATLAS $o$- and $c$-band light curves were generated using the ATLAS Forced Photometry service\footnote{\url{https://fallingstar-data.com/forcedphot/}} \citep{Shingles2021TNSAN...7....1S}. 
A script provided by \citet{Young_plot_atlas_fp} was used to stack ATLAS photometric data, using a rolling-window technique to identify and exclude spurious data points and by binning the data in 1-day intervals. ZTF $g$- and $r$-band light curves were obtained through the ALeRCE \citep{Forster2021AJ....161..242F} and Lasair \citep{Smith2019RNAAS...3a..26S} brokers. Pan-STARRS1 (PS1) images were processed with the PS1 Image Processing Pipeline \citep[IPP;][]{Waters2020ApJS..251....4W, Magnier2020ApJS..251....3M, Magnier2020ApJS..251....5M, Magnier2020ApJS..251....6M} and calibrated using the Pan-STARRS DR1 catalogue \citep{Flewelling2020ApJS..251....7F}. 
Finally, some $g$-band data were collected from the ASAS-SN Sky Patrol\footnote{\url{https://asas-sn.osu.edu}} \citep{Shappee2014ApJ...788...48S, Kochanek2017PASP..129j4502K, Hart2023arXiv230403791H}. These data use aperture photometry on co-added image-subtracted frames for each epoch, excluding flux contributions from reference images in the final light curve. 

In the case of SN~2020nxt, in addition to ground-based observations, we obtained ultraviolet ($uvw2$, $uvm2$, $uvw1$) and $UBV$ images with the {\sl Neil Gehrels Swift Observatory} using the UVOT instrument \citep{Gehrels2004ApJ...611.1005G}. Ultraviolet (UV) and optical photometry from \textit{Swift}/UVOT were retrieved from the NASA \textit{Swift} Data Archive\footnote{\url{https://heasarc.gsfc.nasa.gov/cgi-bin/W3Browse/swift.pl}} and processed using the standard UVOT data analysis software, {\tt HEASoft}\footnote{\url{https://heasarc.gsfc.nasa.gov/lheasoft/download.html}} \citep[version 6.19;][]{Blackburn1999ascl.soft12002B}, alongside standard calibration data.
For photometry, a $5^{''}$ aperture was used to measure the source flux, with the sky background estimated within a manually selected uniform $25^{''}$ region without bright stars.
The host galaxy flux\footnote{The data were measured on 2021 January 22.} was subtracted from the source flux. The apparent magnitudes for the five SNe Ibn are given in Tables~\ref{table:mag_2020nxt}-\ref{table:mag_2024aej} (Appendix~\ref{tables:LC_mag}), and the light curves are shown in Fig.~\ref{fig:lc}.

{No $k$-correction was applied to our photometric data, as all supernovae in our sample are located in the local Universe ($z < 0.1$). To perform an accurate $k$-correction, comprehensive knowledge of the spectral evolution across all relevant phases is required, including the effects of reddening, to enable reliable interpolation between photometric bands. However, for our sample, the spectroscopic coverage is sparse and does not adequately span the full photometric wavelength range or temporal evolution. To avoid introducing additional systematic uncertainties due to these limitations, we refrain from applying $k$-corrections.}

\subsection{Light curves and the comparison with other SNe Ibn}\label{subsec:lightcurve}

\subsubsection{Apparent light curves}
\label{SubSubSec:Apparent light curves}

\begin{figure*}[ht]
    \centering
    \includegraphics[width=0.44\linewidth]{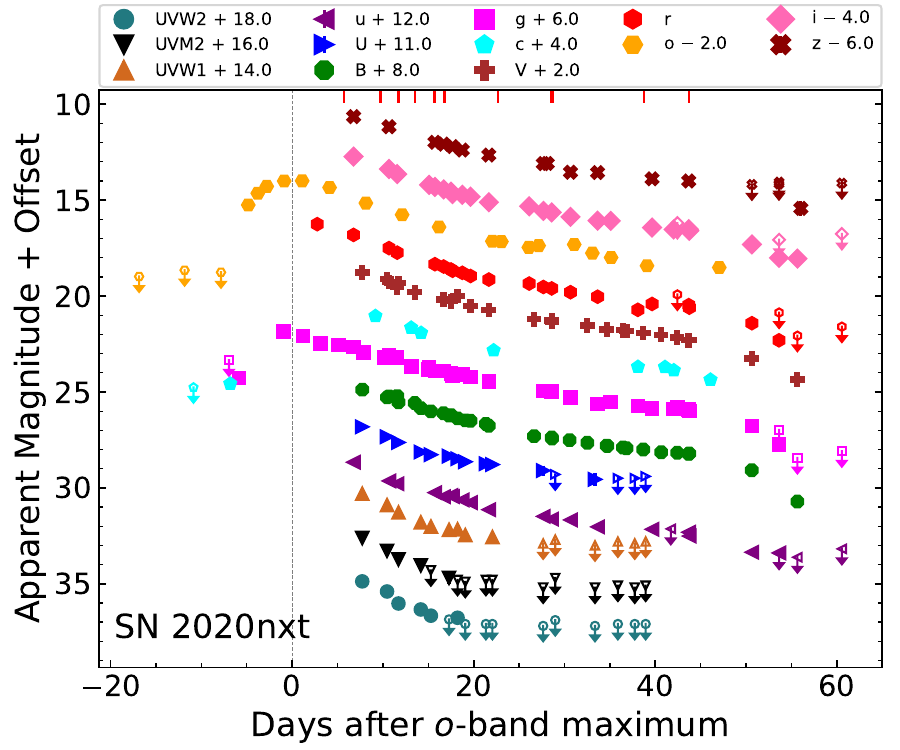}
    \includegraphics[width=0.44\linewidth]{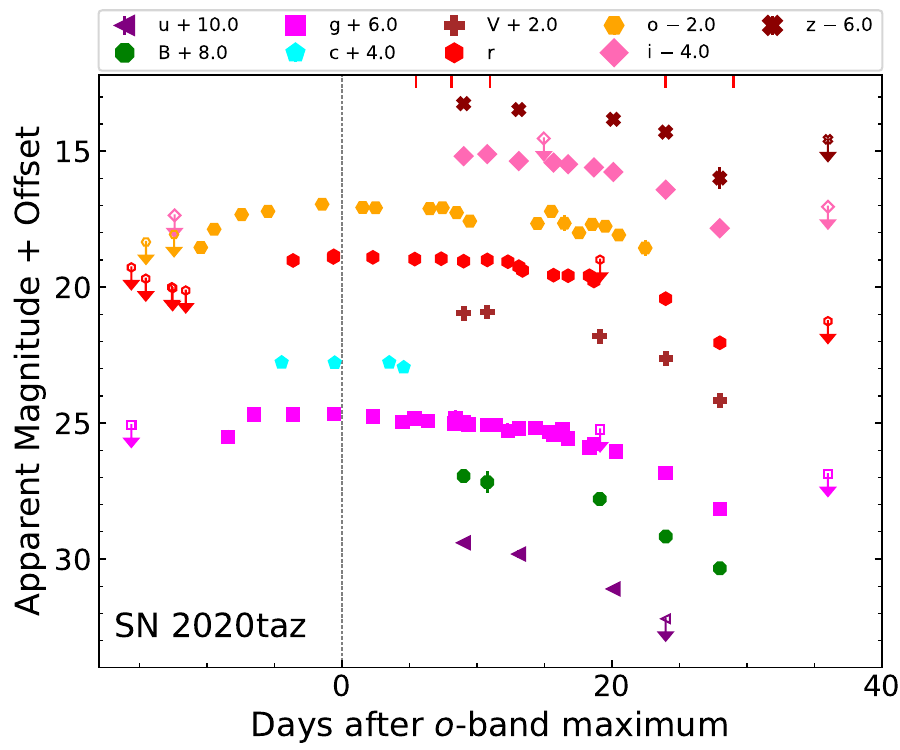}
    \includegraphics[width=0.44\linewidth]{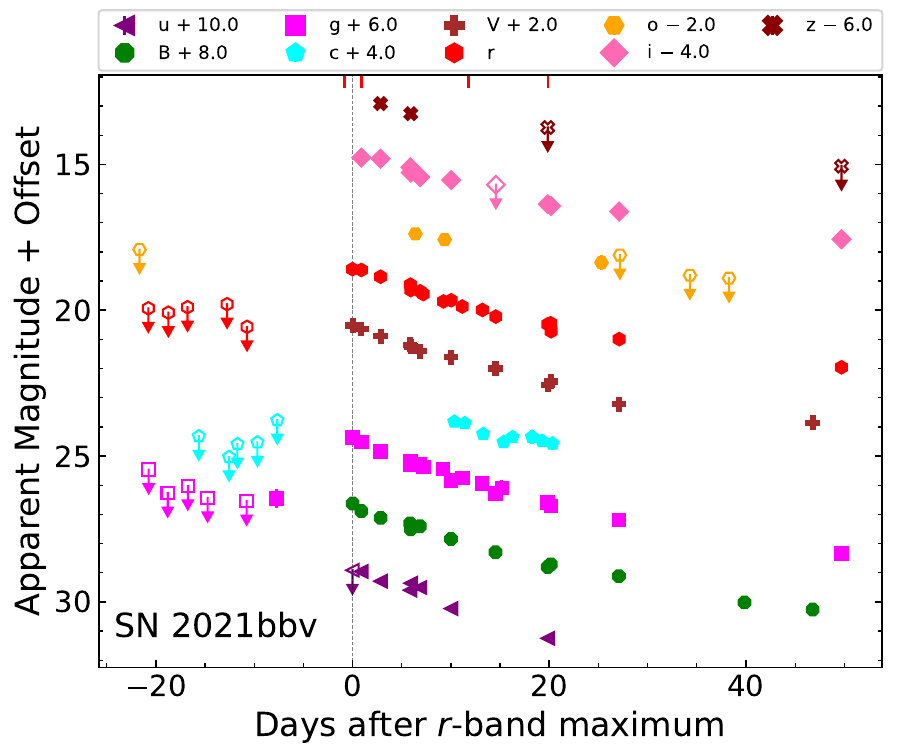}
    \includegraphics[width=0.44\linewidth]{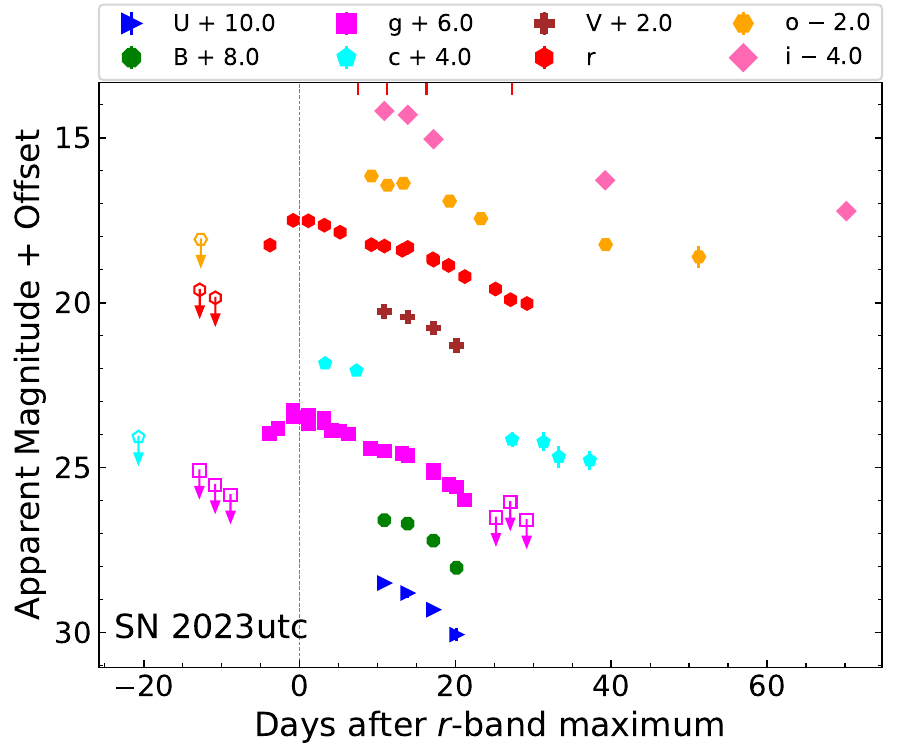}
    \includegraphics[width=0.44\linewidth]{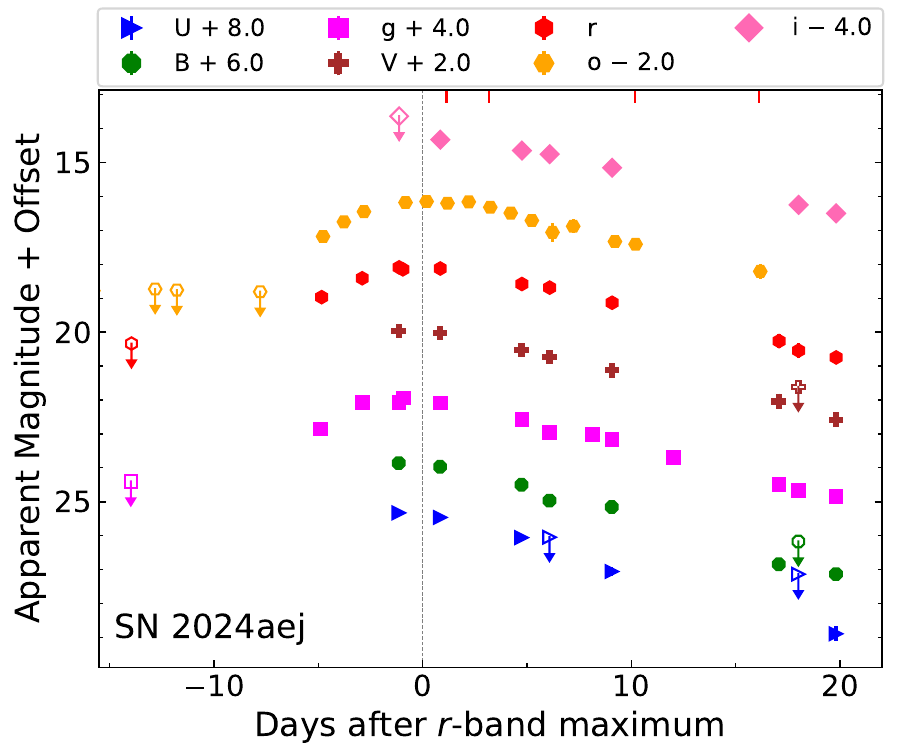}
    \caption{UV and optical light curves of SNe 2020nxt, 2020taz, 2021bbv, 2023utc, 2024aej. 
    The dashed vertical line marks the $o/r$-band maximum light as the reference epoch. 
    The epochs of our spectra are marked with vertical solid red lines on the top.
    The upper limits are marked by empty symbols with arrows. For clarity, the light curves for the different bands are shifted with arbitrary constants as reported in the legend. Usually, magnitude errors are smaller than the symbols.}
    \label{fig:lc}
\end{figure*}

Our follow-up observations for each of the five SNe commenced shortly after their discovery and continued for several months. Figure \ref{fig:lc} presents the apparent light curves of these events. 

\begin{figure*}
\begin{center}
\includegraphics[width=0.88\columnwidth]{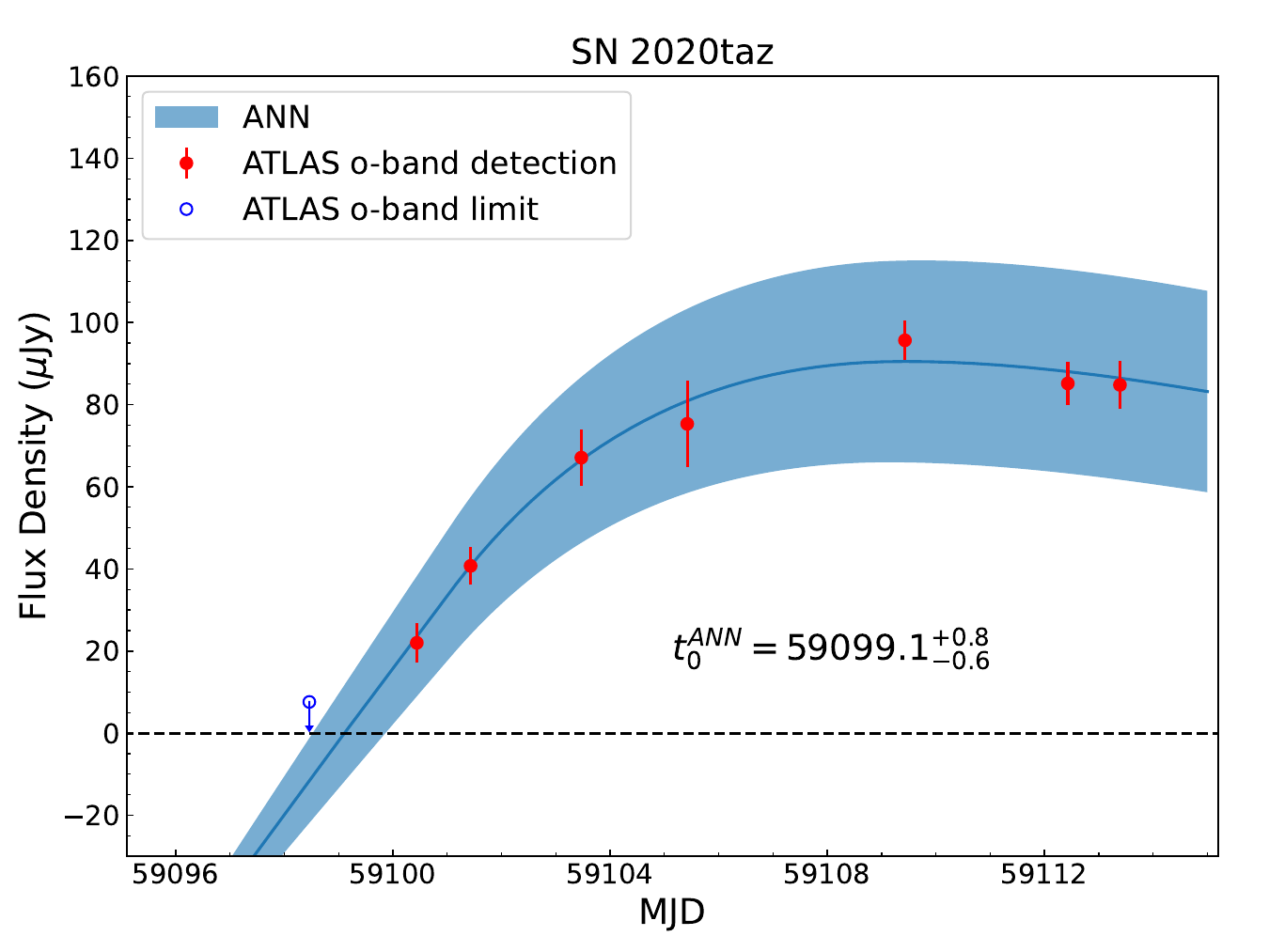}
\includegraphics[width=0.88\columnwidth]{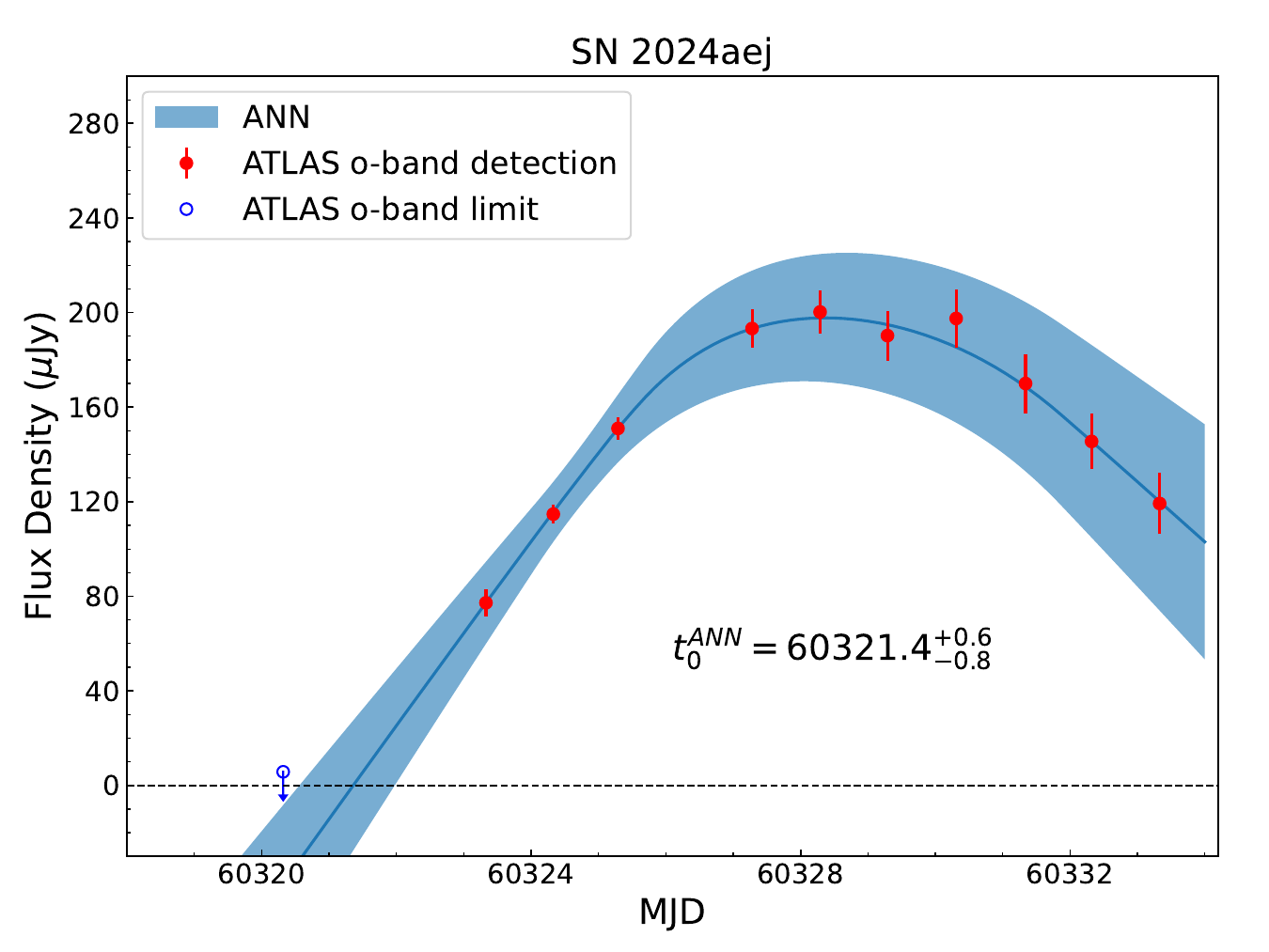}
\caption{Constraints on the explosion epochs of SNe\,2020taz (left) and 2024aej (right). Early time ATLAS $o$-filter data are shown in the flux space. The zero flux level is marked by a horizontal dashed line. Red circle dots indicate real detections, while blue circle is the latest detection limit. }
\label{fig:explosion_time}
\end{center}
\end{figure*}

\begin{table*}
    \centering
    \caption{Light-curve parameters for SNe Ibn.}
    {\fontsize{9pt}{16pt}\selectfont
    \setlength{\tabcolsep}{3pt} 
    \begin{tabular}{llllllll}
    \hline
        Object  & MJD$_{\mathrm{exp.}}$        &   MJD$_{\mathrm{peak}}$          & $t_{\mathrm{rise}}$ & $M_{\mathrm{peak}}(r/R)$ & $L_{\mathrm{peak}}$      & $E_{\mathrm{rad.}}$ & Sources\\        
             SN &                              &                                  &      (d)            &  (mag)                   & ($10^{42}$~erg~s$^{-1}$) & ($10^{48}$ ~erg)    &\\
    \hline
    2020nxt  & 59031.1~$^{+~0.5}_{-~0.5}$   &   59038.4~$^{+~0.1}_{-~0.1}$($o$)  & 7.3~$\pm$~0.5       & $-$19.1~$\pm$~0.1($o$)& 8.36~$\pm$~0.28          &  8.56~$\pm$~0.42 & 1\\
    2020taz  & 59099.1~$^{+~0.8}_{-~0.6}$   &   59110.9~$^{+~1.5}_{-~0.5}$($o$)  & 11.8~$\pm$~1.7      & $-$18.0~$\pm$~0.2     & 3.58~$\pm$~0.24          &  6.91~$\pm$~0.61 & 1\\
    2021bbv  & 59233.5~$^{+~1.0}_{-~1.0}$   &   < 59242.2($r$)                   & < 9.7               & < $-$18.8             & 6.68~$\pm$~0.82          & 8.94~$\pm$~1.16  & 1\\
    2023utc  & 60226.9~$^{+~2.5}_{-~2.5}$   &   60233.3~$^{+~0.1}_{-~0.1}$($r$)  & 6.4~$\pm$~2.5       & $-$16.4~$\pm$~0.5     & 0.71~$\pm$~0.28          & 0.89~$\pm$~0.38  & 1\\
    2024aej  & 60321.4~$^{+~0.6}_{-~0.8}$   &   60328.1~$^{+~0.5}_{-~0.2}$($r$)  & 6.7~$\pm$~0.9       & $-$19.2~$\pm$~0.1     & 9.86~$\pm$~0.95          &  9.89~$\pm$~1.13& 1\\
    \hline
    2006jc   & -                            & 54012.3~$\pm$~4                   &  < 10               & < $-$18.61            &  5.92~$\pm$~1.43        & 7.08~$\pm$~1.53  &2\\
    2010al   & 55267.5~$\pm$~1.5            & 55283.8~$\pm$~1.1($R$)             & 16.0~$\pm$~1.9      & $-$18.86~$\pm$~0.21   & 5.25~$\pm$~0.85         & 10.85~$\pm$~1.79 &3\\
    OGLE12-006  & 56203.3~$\pm$~4.0         & 56217.6~$\pm$~1.8($I$)             & 13.5~$\pm$~4.2      &$-$19.65~$\pm$~0.19($I$)& 7.93~$\pm$~0.99         & 32.46~$\pm$~5.78 &4\\
    ASASSN-14ms & -                         & 57025.2~$\pm$~0.5($V$)             &  -                  &$-$20.33~$\pm$~0.15($V$)& 22.89~$\pm$~3.24        & 47.80~$\pm$~6.73 &5\\
    2014av   & 56760.0~$\pm$~3.8            & 56770.6~$\pm$~1.2($R$)             & 10.3~$\pm$~3.9      & $-$19.76~$\pm$~0.16    &  11.23~$\pm$~1.61        & 12.28~$\pm$~1.92 &6\\
    2015U    & 57062.6~$\pm$~0.4            & 57071.5~$\pm$~0.8($r$)             & 8.8~$\pm$~0.9       & $-$19.95~$\pm$~1.13   &  24.29~$\pm$~4.94        & 30.56~$\pm$~6.68 &7\\
    ASASSN-15ed & -                         & 57086.9~$\pm$~0.6($r$)             &   > 4.3             & $-$20.04~$\pm$~0.20    &  16.61~$\pm$~3.08        & 21.00~$\pm$~3.80 &8\\
    2018jmt  & 58455.0~$\pm$~0.2            & 58465.7~$\pm$~1.2($g$)             & 10.7~$\pm$~1.2      &$-$19.03~$\pm$~0.37($g$)&   7.16~$\pm$~1.48        & 10.78~$\pm$~2.27 &9\\
    2019cj   & 58482.2~$\pm$~1.1            & 58492.4~$\pm$~0.2($V$)             & 10.2~$\pm$~1.1      & $-$18.94~$\pm$~0.19($V$)&   5.48~$\pm$~0.87       & 9.93~$\pm$~1.61 &9\\
    2019uo   & 58499.4~$\pm$~1.2            & 58508.1~$\pm$~0.5($r$)             & 8.7~$\pm$~1.3       & $-$18.30~$\pm$~0.24    &   3.84~$\pm$~0.12        & 4.68~$\pm$~0.17  &10\\
    2019wep  & 58824.5~$\pm$~2.0            & 58828.5~$\pm$~2($V$)               & 4~$\pm$~3           & $-$18.18~$\pm$~0.95    &   2.54~$\pm$~0.19        & 2.71~$\pm$~0.18  &11\\
    2019kbj  & 58664.49~$\pm$~1.0           & 58670.1~$\pm$~0.26($r$)            & 5.6~$\pm$~1.0       & $-$18.99~$\pm$~0.24    &   8.29~$\pm$~0.32        & 9.67~$\pm$~0.66  &12\\
    2020bqj  & 58880.0~$\pm$~1.5            & 58884.2($r$)                       &  < 5.7              & $-$19.23~$\pm$~0.07    &  -                       & -                &13\\
    \hline
    \end{tabular}}
    \label{tab:LC_parameters}

  \begin{flushleft} 
  \textbf{Notes:} Object name (column 1), explosion MJD (column 2), peak MJD (column 3), rise time (column 4), peak absolute magnitude (column 5), peak quasi-bolometric luminosity (column 6), radiated energy (column 7), sources of data (column 8):\\
  1 = this paper;
  2 = \citet{Foley2007ApJ...657L.105F,Pastorello2007Natur.447..829P,Pastorello2008MNRAS.389..113P};
  3 = \citet{Pastorello2015MNRAS.449.1921P};
  4 = \citet{Pastorello2015MNRAS.449.1941P};
  5 = \citet{Vallely2018MNRAS.475.2344V,Wang2021ApJ...917...97W}
  6 = \citet{Pastorello2016MNRAS.456..853P};
  7 = \citet{Tsvetkov2015IBVS.6140....1T, Pastorello2015MNRAS.454.4293P, Shivvers2016MNRAS.461.3057S, Hosseinzadeh2017ApJ...836..158H};
  8 = \citet{Pastorello2015MNRAS.453.3649P};
  9 = \citet{Wang2024A&A...691A.156W};
 10 = \citet{Gangopadhyay2020ApJ...889..170G};
 11 = \citet{Gangopadhyay2022ApJ...930..127G};
 12 = \citet{Ben-Ami2023ApJ...946...30B};
 13 = \citet{Kool2021A&A...652A.136K}.
  \end{flushleft}
\end{table*}

When the survey monitoring cadence is low, the explosion epoch of a new SN is usually assumed to be the midpoint between the last non-detection and the first detection. In the case of SN 2020nxt, the explosion time is estimated from the last non-detection in the $o$ band (MJD = 59030.56) and the first detection in the $c$ band (MJD = 59031.58), hence MJD = 59031.1~$\pm$~0.5 days. 
In the case of SN~2021bbv, the explosion time is derived as the midpoint between the last non-detection in the $c$ band (MJD = 59232.52) and the first detection in the $g$ band (MJD = 59234.45), yielding MJD = 59233.5~$\pm$~1.0 days. 
Similarly, for SN~2023utc, the explosion time is estimated to be at MJD = 60226.9~$\pm$~2.5 days, based on the last non-detection (MJD = 60224.4) and the first detection (MJD = 60229.4), both in the ZTF $g$ band.

Artificial Neural Networks (ANNs) are a widely adopted machine-learning techniques, particularly well-suited for tasks involving regression and estimation. {\sl ReFANN}\footnote{\url{https://github.com/Guo-Jian-Wang/refann}} is an ANN-based code that utilises a supervised learning procedure and consists of three primary layers: input, hidden, and output \citep[for further details, see][]{Wang2020ApJS..249...25W, Wang2020ApJS..246...13W, Wang2021MNRAS.501.5714W}. 
In Fig. \ref{fig:explosion_time}, we show the reconstruction of the early-time light curve (from approximately two weeks past explosion) in the flux space using the {\sl ReFANN} tool. This method provides an explosion epoch of $t^{ANN}_{0} = 59099.1^{+0.8}_{-0.6}$ for SN 2020taz, and $t^{ANN}_{0} = 60321.4^{+0.6}_{-0.8}$ for SN 2024aej.

To estimate the peak magnitude for SNe 2020nxt, 2020taz, 2023utc, and 2024aej, we use the {\sl ReFANN} tool on the $r$-band or $o$-band light curve data, focusing on a period of approximately one week around the peak (see Fig.~\ref{fig:peak_time} in Appendix~\ref{appendix:peak_time}).
In the case of SN 2020nxt, we obtain a peak magnitude of $o = 15.9^{+0.1}_{-0.1}$ on MJD = 59038.4~$^{+0.1}_{-0.1}$. Similarly, for SN 2020taz, we derive a peak magnitude of $o = 19.0^{+0.2}_{-0.2}$ on MJD = 59110.9~$^{+1.5}_{-0.5}$. In the case of SN 2023utc, the peak magnitude is found to be $r = 17.5^{+0.1}_{-0.1}$ on MJD = 60233.3~$^{+0.1}_{-0.1}$, and for SN 2024aej, the peak magnitude is $r = 18.1^{+0.1}_{-0.1}$ on MJD = 60328.1~$^{+0.5}_{-0.2}$. 
Due to the lack of pre-peak data for SN 2021bbv, the peak time can only be estimated to be earlier than MJD = 59242.2, with a peak brightness of $r \leq$ 18.6 mag, which is likely close to the actual peak magnitude given the flat light curve observed at discovery in the $r$ and $i$ bands. 

We additionally estimate the post-maximum decline rates of the five SNe across both UV and optical bands by performing linear regression fits on the post-peak data. These results, which offer a comparison of the fading behaviour in different wavelengths, are presented in Table \ref{decline_rate} (Appendix \ref{appendix:decline_rate}). Given the observed changes in the slope of the light curves of SN 2020nxt at approximately +25 days and +45 days, we calculated the decline rates over three distinct time intervals. A minor scatter in the decline rates among the different filters can be noticed. The UV light curves exhibit a faster decline than the optical ones. During the first phase (e.g., $\gamma_{0-25}(g) = 11.70 \pm 0.34$ mag per hundred days), the decline is steeper than in the second phase (e.g., $\gamma_{25-45}(g) = 5.92 \pm 0.59$ mag per hundred days). In the last phase (e.g., $\gamma_{45-60} (g)= 32.82 \pm 2.15$ mag per hundred days), the decline rate becomes significantly faster than in the first phase. An accelerated decline in optical luminosity during the late phases is a frequently observed in SNe Ibn \citep[e.g.,][]{Mattila2008MNRAS.389..141M, Pastorello2015MNRAS.453.3649P, Wang2024A&A...691A.156W}.

Following the rise to the $o$-band maximum, the light curves of SN 2020taz exhibit a plateau during the first 10 days, 
followed by a magnitude decline, with a rate of $\gamma_{10-20}(g) = 9.75 \pm 1.44$ mag per hundred days. 
At later stages, the decline rate increases to $\gamma_{20-30}(g) = 27.78 \pm 3.25$ mag per hundred days. 
We also note a difference in the decline rates across the different filters. Specifically, the redder light curves decline more slowly compared to the bluer ones. This trend is particularly pronounced during the early decline phase, as shown by the decline rate values reported in Table \ref{decline_rate} (Appendix \ref{appendix:decline_rate}). 

SN 2021bbv exhibits a linear decline until +25 days, with a rate of $\gamma_{0-25}(B) = 10.15 \pm 0.42$ mag per hundred days. This initial phase is followed by a slightly slower linear decline with a rate of $\gamma_{25-50}(B) = 5.97 \pm 0.91$ mag per hundred days.
SN 2023utc exhibits a similar two-phase decline pattern as SN 2021bbv.
In the case of SN 2024aej, observations are limited to the first 20 days past maximum, during which it shows a decline rate similar to the initial phase of SN 2021bbv and SN 2023utc, with $\gamma_{0-20}(B) = 17.23 \pm 0.80$ mag per hundred days.

\subsubsection{Colour curves}

\begin{figure*}
\includegraphics[width=1.8\columnwidth]{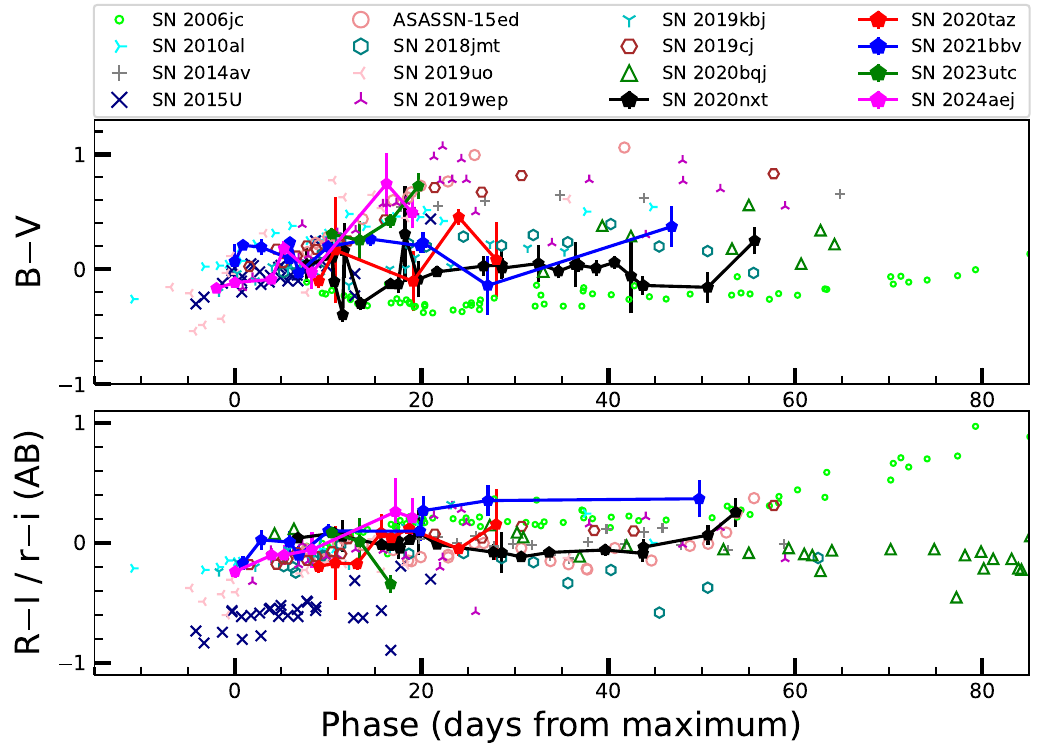}
\caption{Colour evolution of SNe\,2020nxt, 2020nxt, 2021bbv, 2023utc, 2024aej compared with a large sample of SNe Ibn from the literature. The colour curves have been corrected for Galactic extinction.}
\label{fig:colour_evolution}
\end{figure*}

The colour evolution of the five SNe of our sample is compared in Fig. \ref{fig:colour_evolution} with other Type Ibn events from the literature\footnote{The comparison sample includes SNe\,2006jc \citep{Pastorello2007Natur.447..829P}, 2010al \citep{Pastorello2015MNRAS.449.1921P}, 2014av \citep{Pastorello2016MNRAS.456..853P}, 2015U \citep{Pastorello2015MNRAS.454.4293P,Shivvers2016MNRAS.461.3057S}, ASASSN-15ed \citep{Pastorello2015MNRAS.453.3649P}, SNe\,2018jmt \citep{Wang2024A&A...691A.156W}, 2019uo \citep{Gangopadhyay2020ApJ...889..170G}, 2019wep \citep{Gangopadhyay2022ApJ...930..127G}, 2019kbj \citep{Ben-Ami2023ApJ...946...30B}, 2019cj \citep{Wang2024A&A...691A.156W}, and 2020bqj \citep{Kool2021A&A...652A.136K}.}. 

This comparison reveals a diversity in the colour evolution of Type Ibn SNe. 
In the case of SN 2020nxt, the $B~-~V$ colour remains nearly constant, fluctuating around 0 mag throughout the observed period. Similarly, for SNe 2020taz and 2021bbv, the $B~-~V$ colours remain close to 0.2 mag, resembling the early-stage behaviour of SN 2006jc.
In contrast, SN 2023utc shows a more pronounced evolution in its $B~-~V$ colour, increasing from approximately 0.3 mag at +10 days to around 0.7 mag by +20 days. At maximum light, SN 2024aej exhibits an initial $B~-~V$ colour close to $-$0.1 mag, but it transitions towards redder colours, reaching $B~-~V$ $\sim$ 0.7 mag by +20 days. 
The colour evolution of both SN 2023utc and SN 2024aej closely follows the early-stage evolution observed in ASASSN-15ed, SN 2010al, and SN 2014av. 

In the case of SN 2020nxt, the $r~-~i$ colour remains nearly constant around $\simeq0\,\rm{mag}$ between +5 and +40 days, resembling the behaviour of SN 2014av. However, at $t\gtrsim+55\,\rm{days}$, it shifts towards redder colours, reaching $\sim$ 0.3 mag. 
The $r~-~i$ colour for SNe 2020taz, 2021bbv, and 2024aej slowly increases, suggesting a gradual temperature decrease over time. Specifically, for SN 2020taz, the $r~-~i$ colour rises from $-$0.2 mag to 0.2 mag, while for SN 2021bbv it increases from $-$0.2 mag to 0.4 mag, and for SN 2024aej, it changes from $-$0.2 mag to 0.3 mag. 
The evolution of the $R~-~I$ / $r~-~i$ colour in these SNe is consistent with the trends observed in SNe 2010al, 2019kbj, and 2019uo, although the timescales can vary significantly among individual objects. In contrast with the other SNe Ibn and with its own $B~-~V$ trend, the SN 2023utc $r~-~i$ colour becomes bluer with time, from 0.1 mag to $-$0.3 mag. 

\begin{figure*}[htp]
\begin{center}
\includegraphics[width=1.9\columnwidth]{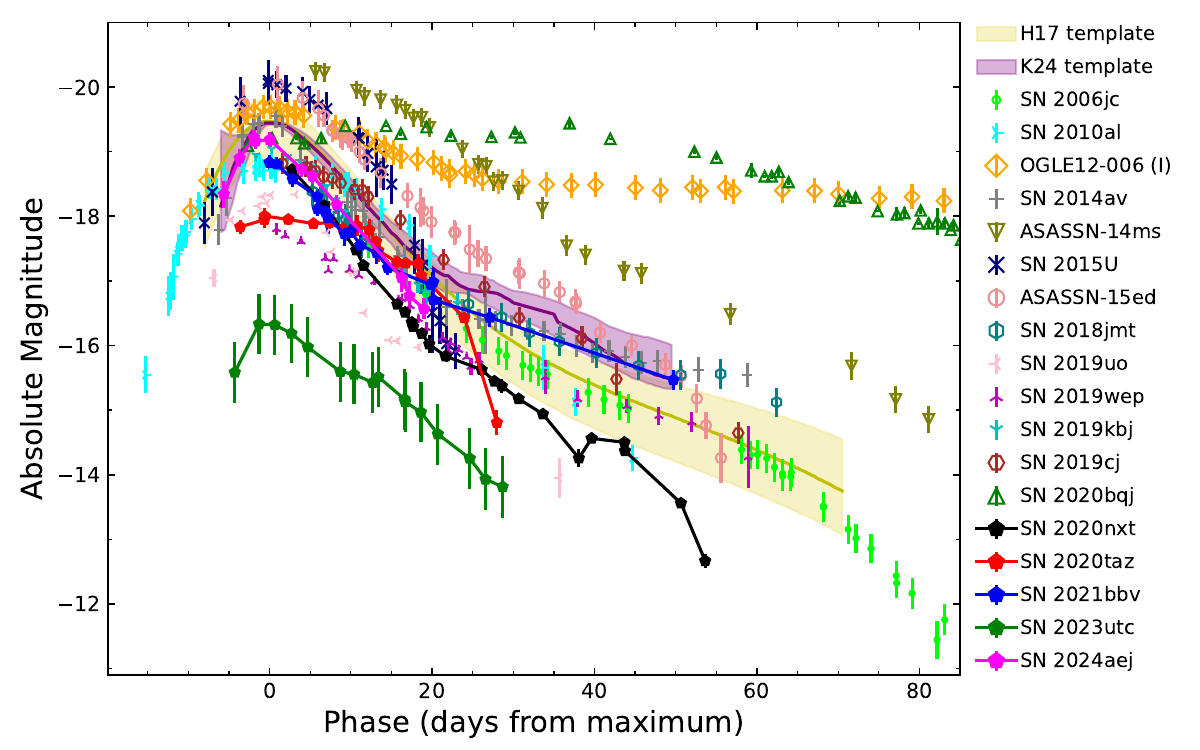}
\caption{$R/r$-band light curves of SNe\,2020nxt, 2020taz, 2021bbv, 2023utc, 2024aej, including the comparison SNe Ibn. Template $r$-band light curves for Type~Ibn SNe from \citet[][yellow]{Hosseinzadeh2017ApJ...836..158H} and \citet[][purple]{Khakpash2024ApJS..275...37K}.  }
\label{fig:Ibns_AbsoMag_ALL}
\end{center}
\end{figure*}

\subsubsection{Absolute magnitude light curves}

After applying the corrections for distance and extinction as described in Section \ref{sec:information}, we calculate the maximum absolute magnitude for SN 2020nxt to be $M_{o} = -19.1 \pm 0.1$ mag. The peak absolute magnitudes of the other objects of the sample are: $M_{o} = -17.8 \pm 0.2$ mag for SN 2020taz, $M_{r} = -16.4 \pm 0.5$ mag for SN 2023utc, and $M_{r} = -19.2 \pm 0.1$ mag for SN 2024aej. In the case of SN 2021bbv, only an upper limit of $M_{r} < -18.8$ mag can be inferred for the peak absolute magnitude in the $r$ band (see Table \ref{tab:LC_parameters}).

Figure \ref{fig:Ibns_AbsoMag_ALL} shows a comparison of the absolute $r$-band light curves for a subset of Type Ibn SNe. When $r$-band observations are not available, light curves in adjacent bands are used for comparison. For instance, for OGLE-2012-SN-006 \citep[hereafter OGLE12-006,][]{Pastorello2015MNRAS.449.1941P}, we used observations in the well-sampled $I$-band. 
Type Ibn SNe are usually quite luminous, with absolute $r$-band magnitudes between $-18$ and $-20$ mag. As shown in Fig. \ref{fig:Ibns_AbsoMag_ALL}, the light-curve shapes of SNe Ibn are quite diverse. 
Comparing the $r$-band light curves of SNe 2020nxt, 2021bbv, and 2024aej with those of other SNe Ibn, we find that in most cases they follow the behaviour of the template presented by \citet{Hosseinzadeh2017ApJ...836..158H} and \citet{Khakpash2024ApJS..275...37K} at around the peak brightness. 
Following the initial rise, the light curve of SN\,2020taz transitions into a plateau phase, similar to that observed in SN\,2020bqj, which lasts approximately 10 days at an $r$-band absolute magnitude of $-17.8$ to $-18.0$ mag, placing it at the fainter end of the SN Ibn sample. 
SN~2023utc is an evident outlier, as it is the faintest Type~Ibn supernova discovered to date, with an extremely faint $r$-band absolute magnitude of only $-16.4$ mag. This is largely below the average absolute magnitude of Type~Ibn supernovae ($M_r \sim -19~\mathrm{mag}$; \citealt{Pastorello2016MNRAS.456..853P}), and also significantly fainter than the transitional Type~IIn/Ibn SN~2005la ($M_R \sim -17.2~\mathrm{mag}$; \citealt{Pastorello2008MNRAS.389..131P}). The very faint absolute magnitude of SN\,2023utc poses new questions regarding the possible progenitors stars and the explosion mechanisms of SNe Ibn.

\subsubsection{Pseudo-bolometric light curves}

\begin{figure*}[htp]
\includegraphics[width=1.9\columnwidth]{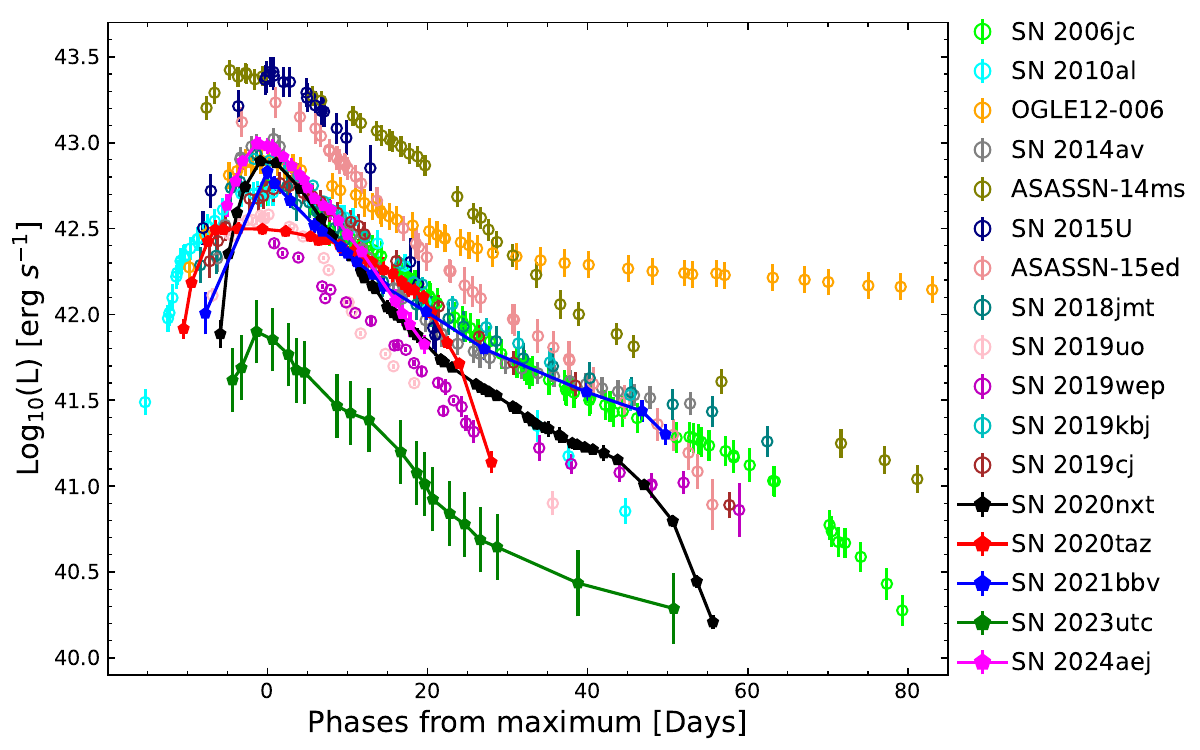}
\caption{Pseudo-bolometric light curves of SNe\,2020nxt, 2020taz, 2021bbv, 2023utc, 2024aej, compared with those of a sample of SNe Ibn.}
\label{fig:quasibolom}
\end{figure*}

A bolometric light curve can be derived by integrating the spectral energy distribution (SED) across the entire electromagnetic spectrum. However, in most cases, observations are not available for filters redder than the $I/i$ band or bluer than the $u$ band. Consequently, to facilitate reliable comparisons among SNe Ibn, we computed pseudo-bolometric light curves limited to the observed $B$ through $I/i$ bands. 
To obtain these, we first converted extinction-corrected magnitudes to flux densities and then integrated the SEDs at their effective wavelengths, under the assumption of negligible flux contributions outside this integration range. The resulting pseudo-bolometric light curves are shown in Fig.~\ref{fig:quasibolom}, and the peak luminosities are reported in Table~\ref{tab:LC_parameters}.
In most cases, the peak luminosities of our SN sample fall within the range of $3 \times 10^{42}$ erg s$^{-1}$ to $2 \times 10^{43}$ erg s$^{-1}$, with SN 2023utc being notably fainter at approximately $7 \times 10^{41}$ erg s$^{-1}$. The pseudo-bolometric light curve profiles of SNe 2020nxt, 2021bbv, 2023utc, and 2024aej show broad similarities to those of typical Type Ibn events (e.g., SN 2006jc and SN 2018jmt), as illustrated in Fig.~\ref{fig:quasibolom}. In contrast, the pseudo-bolometric light curve of SN\,2020taz displays a distinctive plateau near the luminosity peak. A sudden drop at +40 days observed in the optical light curve of SN 2020nxt, accompanied by a redward colour shift, suggests a potential early dust formation in a cool dense shell, as observed in SN~2006jc \citep{Mattila2008MNRAS.389..141M, Smith2008ApJ...680..568S, DiCarlo2008ApJ...684..471D}. 

To further analyse these light curves, we applied a non-parametric fit using the \texttt{ReFANN} code to reconstruct the pseudo-bolometric light curves and integrated them over the entire photometric evolution, with integration limits defined by the time range of the available photometric data. The resulting radiated energies range from $(1$–$32) \times 10^{48}$ erg, as listed in Table~\ref{tab:LC_parameters}. These values should be interpreted as lower limits due to incomplete wavelength and temporal coverages.

\begin{figure}[htp]
\includegraphics[width=1\columnwidth]{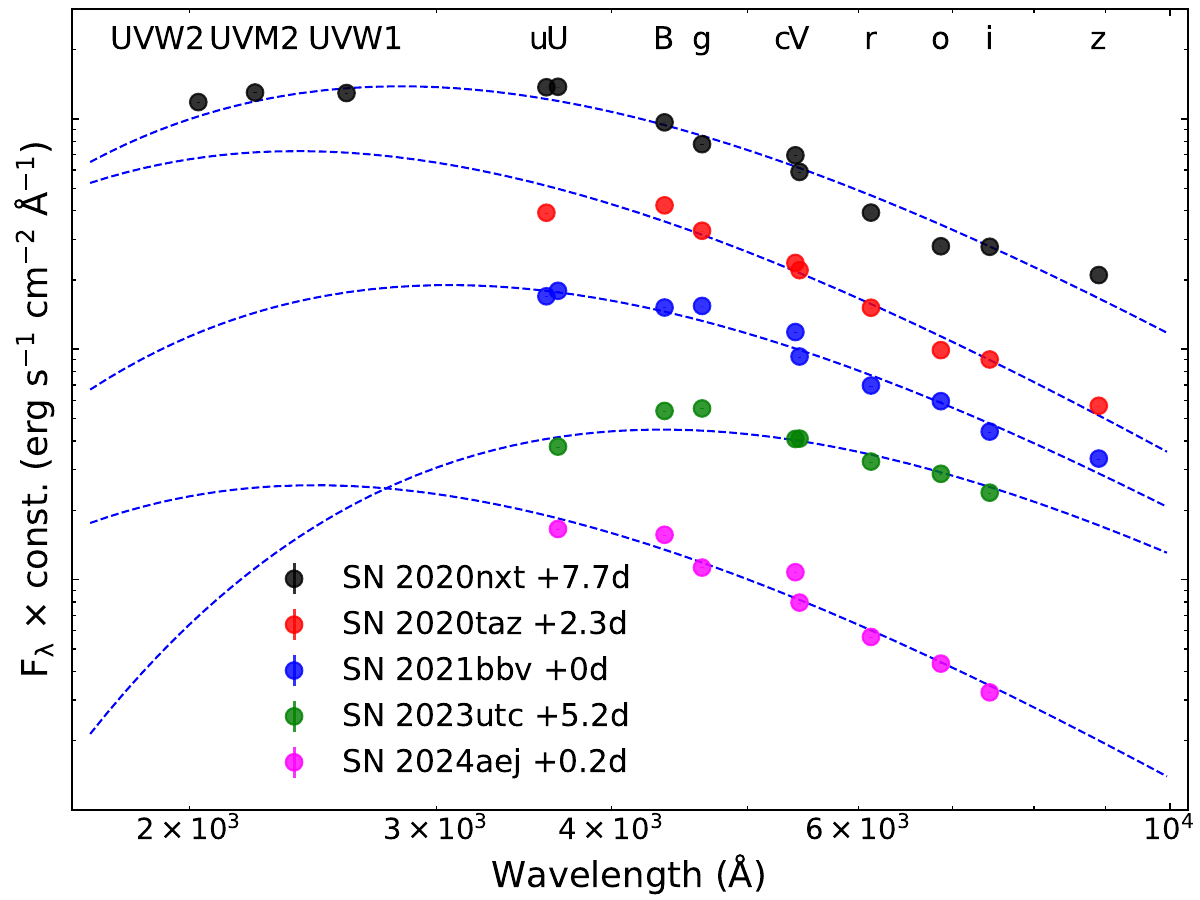}
\caption{Representative blackbody fits to the SEDs at epochs near the peak luminosity for each SN in our sample. The blue lines show the best-fitting blackbody functions. For visual clarity, the SEDs have been vertically offset by arbitrary constants.}
\label{fig:Five_SNe_SED}
\end{figure}

\begin{figure}[htp]
\includegraphics[width=1\columnwidth]{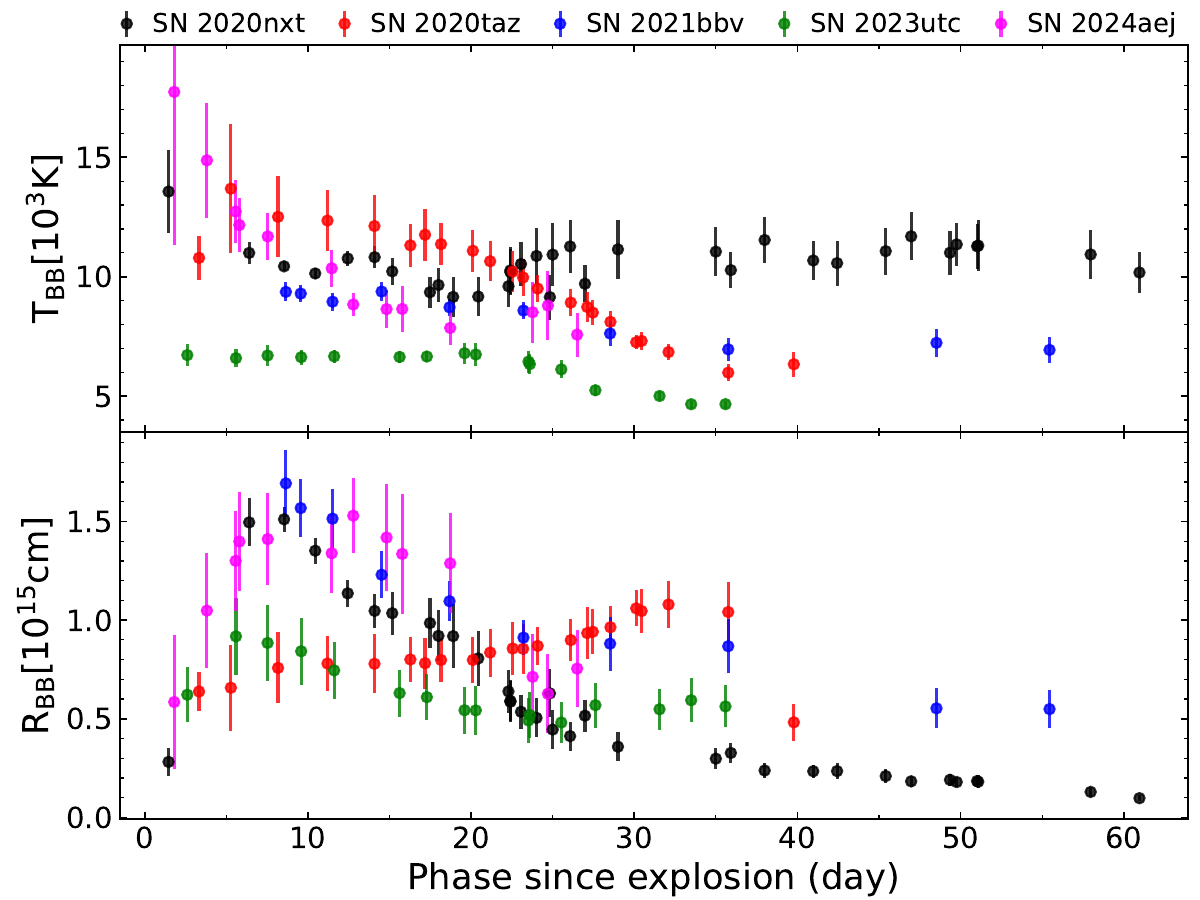}
\caption{Evolution of the blackbody temperature and radius for our SN sample.}
\label{fig:Five_Ibn_T_R}
\end{figure}

\begin{figure*}
\begin{center}
\includegraphics[width=2\columnwidth]{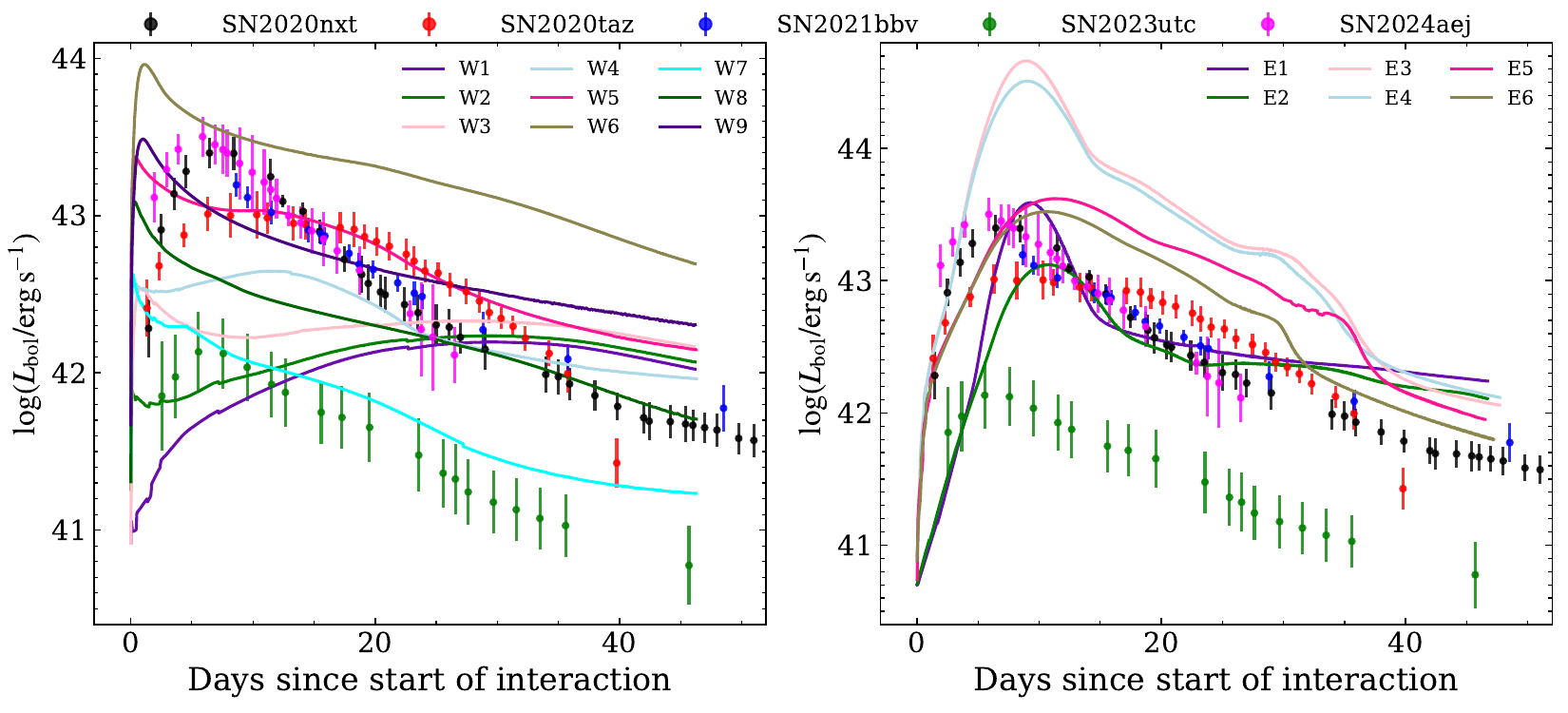}
\caption{Comparison between the pseudo-bolometric light curves of the five SNe~Ibn in our sample and representative interaction models from \citet{Dessart2022A&A...658A.130D}. The left panel displays simulations involving ejecta–wind interactions, while the right panel shows ejecta–ejecta interaction scenarios.}
\label{fig:Luc_model_comparison}
\end{center}
\end{figure*}

To facilitate a meaningful comparison between SNe 2020nxt, 2020taz, 2021bbv, 2023utc, 2024aej, and the comparison star, we constructed their pseudo-bolometric light curves based on the observed wavelength range. To more accurately estimate the full bolometric light curves of these five SNe, we fitted the broad-band photometry with a blackbody model, extrapolating the luminosity contributions from the blackbody tails outside the observed range. 
Figure~\ref{fig:Five_SNe_SED} shows, for each SN, representative blackbody fits to the SED at times near the peak luminosity. All five SEDs are well described by a single blackbody function. The evolution of the best-fitting blackbody temperatures and radii, along with the resulting bolometric luminosities, is presented in Figures~\ref{fig:Five_Ibn_T_R} and \ref{fig:Luc_model_comparison} (for the latter, see discussion in Sect. \ref{Sec:Dessart}). Note that we only focus primarily on the long-term evolution of the bolometric luminosity; short-duration fluctuations in the light curves are not captured in the final bolometric light curves. 

Except for SN~2020nxt, our sample lacks UV observations, and this may significantly affect the robustness of our blackbody fits, particularly at early times. \citet{Arcavi2022ApJ...937...75A} explored the systematic uncertainties in constraining hot blackbody parameters using optical photometry alone and found that for blackbody temperatures exceeding $\sim35000$~K, the inferred values may be overestimated by $\sim10000$~K. 
Bolometric luminosities can be overestimated or underestimated by a factor $\times~3-5$ at very high temperatures (above $\sim60000$~K).
However, as indicated by the relatively constant colours and the blackbody temperatures derived from the spectra (see Section~\ref{sec:spectroscopy}), the temperatures near and after the peak luminosity in our sample are typically below $20000$~K. These values are sufficiently low that the lack of UV data is unlikely to introduce significant systematic biases in the bolometric luminosity \citep{Arcavi2022ApJ...937...75A}.
In practice, UV coverage is currently available for only a subset of SNe~Ibn, mostly through \textit{Swift} observations. For other SNe~Ibn, including our sample, blackbody fitting to optical photometry remains a standard and widely accepted approach to estimate bolometric light curves \citep[e.g.][]{Shivvers2016MNRAS.461.3057S, Karamehmetoglu2021A&A...649A.163K, Gangopadhyay2022ApJ...930..127G}.

\subsection{Modeling the Multi-band Light Curves with the MOSFiT Framework} \label{Sec:MOSFiT}

\begin{figure*}
\begin{center}
\includegraphics[width=0.44\linewidth]{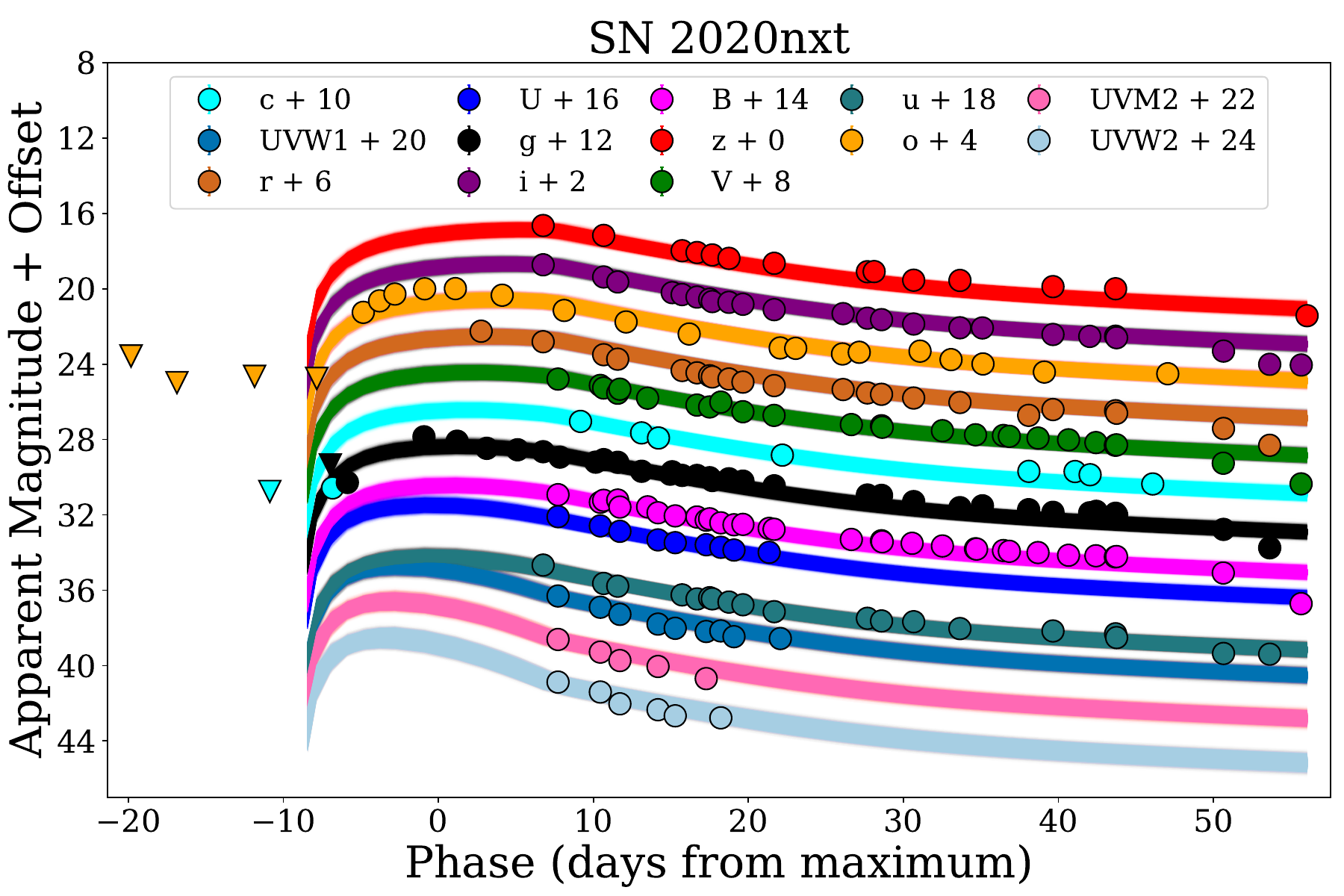}
\includegraphics[width=0.44\linewidth]{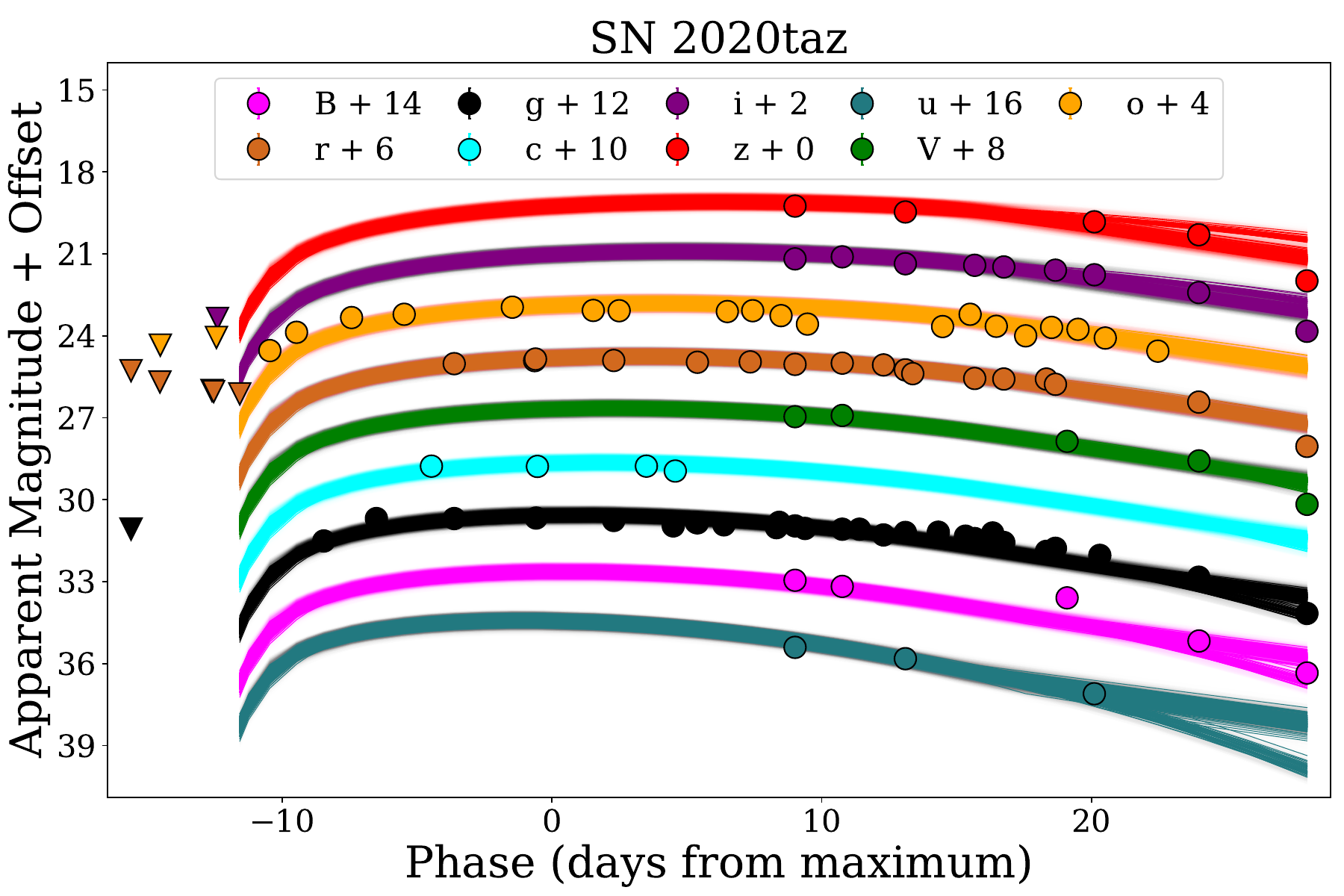}
\includegraphics[width=0.44\linewidth]{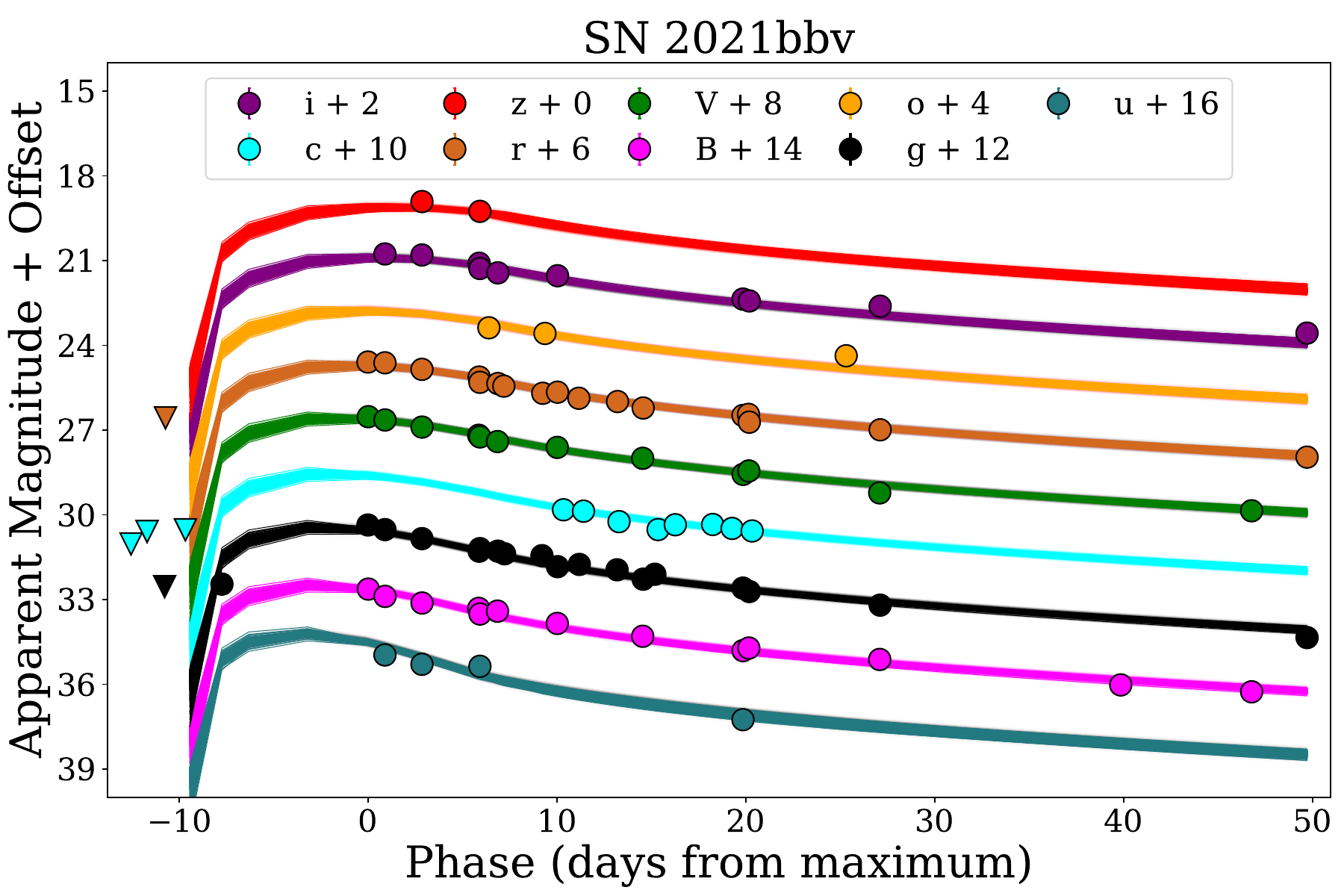}
\includegraphics[width=0.44\linewidth]{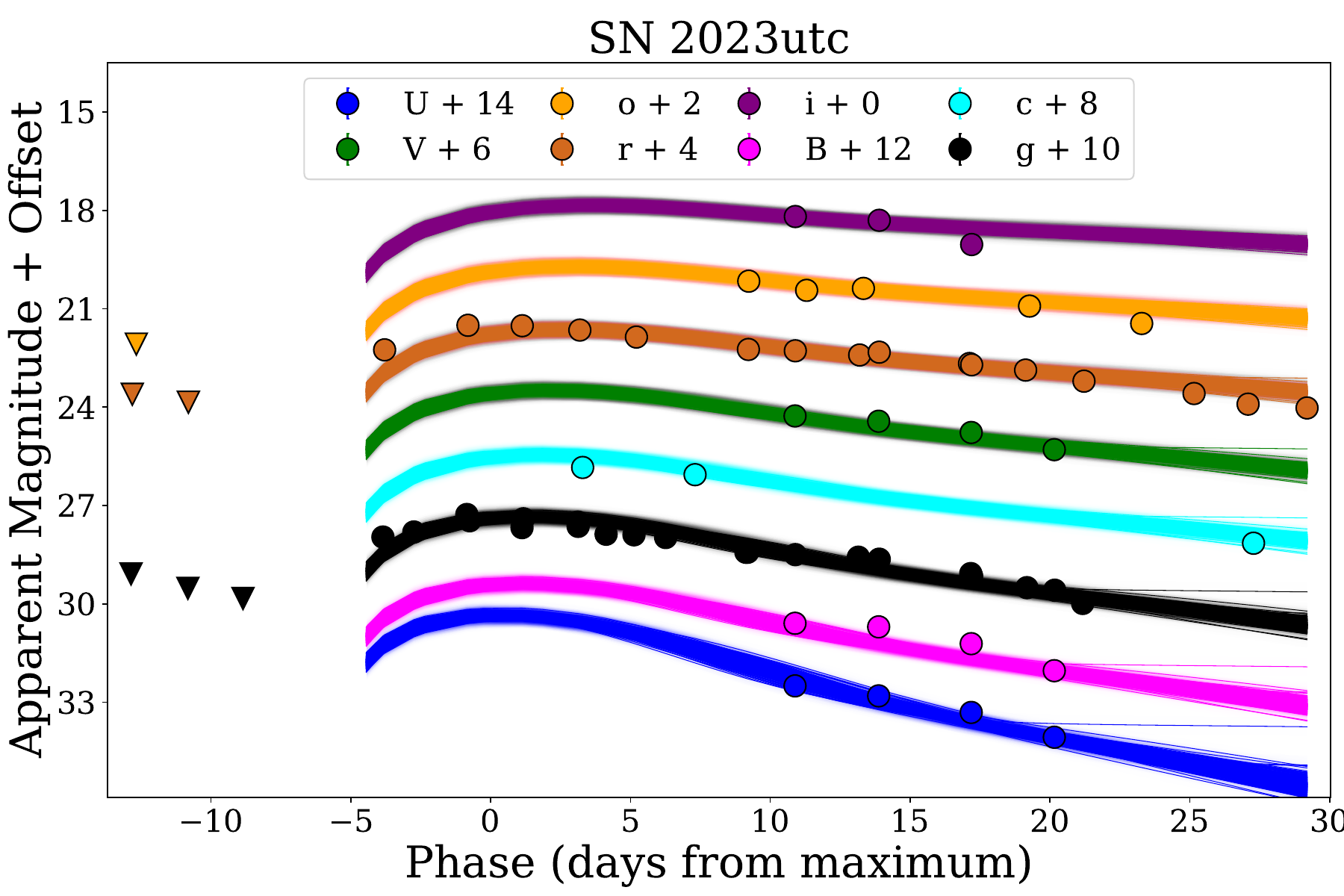}
\includegraphics[width=0.44\linewidth]{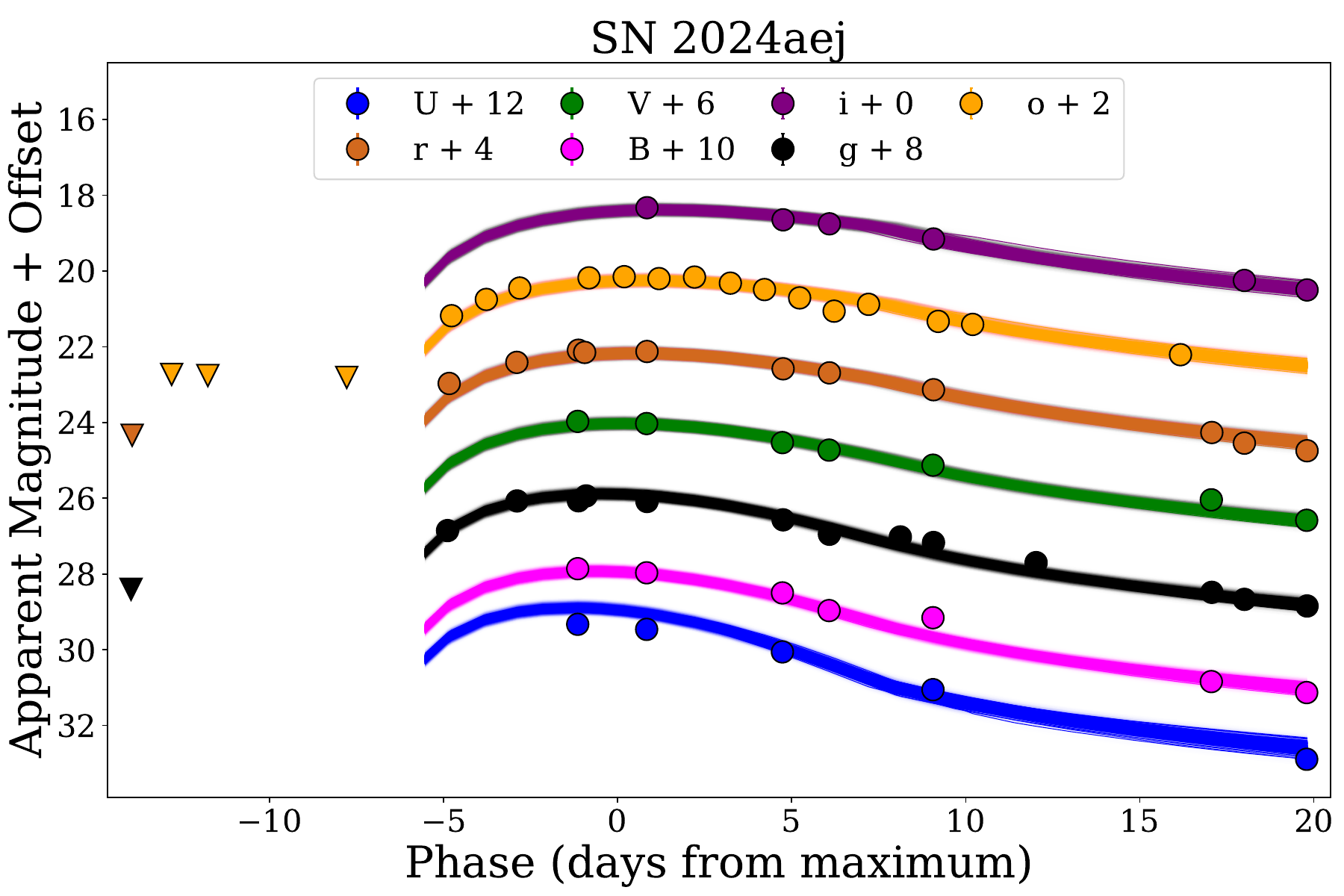}
\caption{Light curves from the \texttt{RD+CSI} model fitted to the multi-band photometry of five Type~Ibn SNe using the Monte Carlo code \texttt{MOSFiT}. For each filter, a representative subset of model light curves randomly drawn from the posterior distributions is shown to illustrate the range of the model fits. The latest pre-discovery upper limits are indicated by triangles.}
\label{fig:mosfit_lc}
\end{center}
\end{figure*}

\begin{table*}
    \centering
    \caption{Best-fit values of the six free parameters used in the hybrid \texttt{RD+CSI} model fitting to the multi-band light curves of the five SNe~Ibn in our sample, along with two derived physical parameters. All results are obtained using the \texttt{MOSFiT} code.
    }
    {
    \fontsize{9pt}{16pt}
    \selectfont
    \setlength{\tabcolsep}{12pt}
    \begin{tabular}{lcccccc}
\hline \hline
                                                         & SN 2020nxt                   & SN 2020taz                   & SN 2021bbv                  & SN 2023utc                   & SN 2024aej \\\hline
Nickel fraction ($f_{\rm Ni}$; \%)                        & $0.037_{-0.005}^{+0.004}$    & $0.001_{-0.000}^{+0.001}$    & $0.16_{-0.02}^{+0.02}$   & $0.018_{-0.002}^{+0.002}$    & $0.14_{-0.04}^{+0.03}$ \\

Kinetic energy ($E_{\mathrm{kin}}$; $10^{51}$ erg)       & $0.40_{-0.05}^{+0.06}$    & $0.36_{-0.05}^{+0.08}$    & $0.61_{-0.09}^{+0.08}$   & $0.06_{-0.01}^{+0.01}$    & $0.91_{-0.22}^{+0.24}$  \\

CSM mass ($M_{\rm CSM}$; M$_{\odot}$)                     & $0.18_{-0.04}^{+0.04}$    & $0.95_{-0.18}^{+0.19}$    & $0.21_{-0.04}^{+0.05}$   & $0.17_{-0.04}^{+0.06}$    & $0.25_{-0.06}^{+0.13}$ \\

Ejecta mass ($M_{\rm ej}$; M$_{\odot}$)                   & $1.06_{-0.04}^{+0.08}$    & $3.03_{-0.57}^{+0.71}$   & $0.90_{-0.09}^{+0.11}$   & $0.85_{-0.03}^{+0.07}$    & $0.85_{-0.17}^{+0.23}$ \\

CSM inner radius ($R_0$; AU)                             & $9.3_{-2.1}^{+2.6}$    & $49.0_{-9.85}^{+10.7}$    & $24.2_{-5.4}^{+6.0}$   & $17.7_{-3.9}^{+6.4}$    & $34.6_{-8.1}^{+17.6}$ \\

CSM density (log$_{10}~\rho_{\rm CSM}$; g cm$^{-3}$)    & $-10.41_{-0.39}^{+0.78}$    & $-9.05_{-0.35}^{+0.31}$    & $-7.68_{-1.08}^{+1.07}$   & $-7.04_{-0.94}^{+0.72}$    & $-8.06_{-0.82}^{+1.05}$ \\

\hline
Ejecta velocity ($v_{\mathrm{ej}} = \sqrt{2\times E_{\mathrm{kin}} / M_{\mathrm{ej}}}$; km s$^{-1}$)   
                                                        & $6153_{-404}^{+516}$       & $3481_{-417}^{+554}$         & $8237_{-732}^{+737}$       & $2728_{-150}^{+198}$         & $13480_{-764}^{+1023}$ \\

Nickel mass ($M_{\rm Ni}$ = $f_{\rm Ni}\times M_{\rm ej}$; M$_{\odot}$)                                
                                                         & $0.04_{-0.01}^{+0.01}$    & $0.002_{-0.001}^{+0.003}$    & $0.15_{-0.03}^{+0.04}$    & $0.015_{-0.002}^{+0.003}$    & $0.12_{-0.06}^{+0.06}$ \\

\hline\hline
\end{tabular}}
\label{tab:mosfit_params}
\vspace{0.5em} 
\begin{minipage}{\textwidth} 
\footnotesize
\noindent \textbf{Note.} The hybrid \texttt{RD+CSI} model with a constant CSM density profile ($s = 0$) is adopted for all fits. Quoted uncertainties represent the 68\% confidence intervals derived from the posterior distributions (see Appendix~\ref{appendix:corner} for the corresponding corner plots), and do not include systematic uncertainties arising from the simplified nature of the model or potential limitations in the data.
\end{minipage}
\end{table*}

The radioactive decay (RD) model of $^{56}$Ni, combined with circumstellar interaction (CSI), has been widely employed to interpret the light curves of Type Ibn SNe (e.g. \citealt{Karamehmetoglu2017AA...602A..93K, Kool2021A&A...652A.136K, Pellegrino2022ApJ...926..125P, Ben-Ami2023ApJ...946...30B}). A robust and widely used tool to model light curve is the \texttt{MOSFiT} code \citep{Guillochon2018ApJS..236....6G}, which has been successfully applied in recent studies of SNe~Ibn \citep[e.g.][]{Kool2021A&A...652A.136K, Farias2024ApJ...977..152F}.
In this work, we utilise \texttt{MOSFiT} to fit the full multi-band light curves, making use of a Monte Carlo approach which yields statistically robust parameter uncertainties and self-consistent fits all relevant variables. Crucially, \texttt{MOSFiT} accounts for the colour information by fitting each band independently.

The RD \texttt{MOSFiT} model describes any radioactive powering from the $^{56}$Ni decay through three key parameters: the nickel fraction ($f_{\rm Ni}$), the $\gamma$-ray opacity of the ejecta ($\kappa_\gamma$), and the optical opacity ($\kappa$). In addition, the code allows us to incorporate an ejecta–CSM interaction model, with free parameters characterising both the ejecta and the surrounding medium \citep[see][]{Villar2017ApJ...849...70V}. These include the ejecta mass ($M_{\rm ej}$), the kinetic energy ($E_{\rm kin}$), and the inner and outer density profiles of the ejecta ($\rho_{\rm ej,in} \propto r^{-\delta}$ and $\rho_{\rm ej,out} \propto r^{-n}$). The CSM is described by its inner radius ($R_0$), the density ($\rho_{\rm CSM}$), and the radial density profile ($\rho_{\rm CSM} \propto r^{-s}$), where $s=0$ corresponds to a constant-density shell and $s=2$ represents a steady wind.
\texttt{MOSFiT} also accounts for the explosion time ($t_{\rm exp}$), defined as relative to the first photometric detection, and includes nuisance parameters such as the minimum allowed temperature of the photosphere ($T_{\rm min}$) and a white-noise term ($\sigma$) that is added in quadrature to the photometric uncertainties to assess the fit quality.

Although the model involves a high-dimensional parameter space, the key physical parameters of interest in this study are $f_{\rm Ni}$, $M_{\rm ej}$, $M_{\rm CSM}$, $E_{\rm kin}$, and $R_0$. We fix five model parameters to standard values: $\delta = 0$, $n = 12$, $s = 0$, $\kappa=0.1$ cm$^2$ g$^{-1}$ and $\kappa_{\gamma}=0.027$ cm$^2$ g$^{-1}$. This results into a set of eight free parameters: $f_{\rm Ni}$, $M_{\rm ej}$, $M_{\rm CSM}$, $\rho_{\rm CSM}$, $R_0$, $E_{\rm kin}$, $T_{\rm min}$, and $\sigma$.
To explore the parameter space efficiently, we employ the dynamic nested sampling algorithm implemented via the \texttt{dynesty} package \citep{Speagle2020MNRAS.493.3132S}, as recommended for complex models. 
We initialise the sampler with 120 live points (also referred to as ``walkers'') and run the algorithm until the default convergence criterion in \texttt{dynesty} is satisfied. This criterion ensures that the uncertainties in both the model evidence and the posterior distributions fall below predefined thresholds. The number of iterations required to reach convergence varies depending on the complexity of the light-curve data: approximately 760 000 iterations for SN 2020nxt, 210 000 for SN 2020taz, 180 000 for SN 2021bbv, 150 000 for SN 2023utc, and 110 000 for SN 2024aej.

Figure~\ref{fig:mosfit_lc} presents the \texttt{MOSFiT} model light curves across the observed photometric bands, overplotted with the corresponding observational data. The best-fitting parameter values, which are most relevant to our analysis, are summarised in Table~\ref{tab:mosfit_params}. All parameters are well constrained by the data, with uncertainties representing the 68\% confidence intervals derived from the posterior distributions. 
Two parameters are omitted from the table: the noise term $\sigma$, which was consistently well constrained across the five samples (typically $\sigma \sim 0.2$), and the final plateau temperature $T_{\rm f}$, to which the key physical parameters are relatively insensitive \citep{Nicholl2017ApJ...850...55N, Kool2021A&A...652A.136K}. 
The corner plots illustrating the posterior probability distributions and the correlations among parameters for the model fits are shown in Appendix~\ref{appendix:corner}, Figs~\ref{fig:modfit_corner1} to \ref{fig:modfit_corner5}.

The inferred ejecta masses in our sample span a relatively narrow range, from a minimum of $\sim$0.85\,$M_{\odot}$ (SN~2023utc and SN~2024aej) to a maximum of $\sim$3.03\,$M_{\odot}$ (SN~2020taz). This upper edge is substantially lower than the average ejected mass of $\sim$16\,$M_{\odot}$ reported for some other SNe~Ibn, including SN~2019uo \citep{Gangopadhyay2020ApJ...889..170G}, PS15dpn \citep{Wang2020ApJ...900...83W}, and SN~2020bqj \citep{Kool2021A&A...652A.136K}. In contrast, the ejected masses derived for our events are broadly consistent with the theoretical maximum of $\sim$1.2\,$M_{\odot}$ predicted for typical SNe~Ibn progenitor systems (single or binary) from radiative-transfer models \citep{Dessart2022A&A...658A.130D}, and also fall within the $\sim$1–3\,$M_{\odot}$ grid explored in the analytical light-curve models by \citet{Pellegrino2022ApJ...926..125P}.
The kinetic energies inferred for our sample span a wide range ($E_{\mathrm{K}} \sim (0.06$–$0.91) \times 10^{51}$\,erg). These values are consistent with those estimated for other SNe~Ibn, such as $\sim$0.35$\times10^{51}$\,erg for SN~2023tsz \citep{Warwick2025MNRAS.536.3588W} and $\sim$$10^{51}$\,erg reported by \citet{Maeda2022ApJ...927...25M}, although SN~2023utc appears to represent a sort of lower limit for the kinetic energies observed in SNe Ibn.
The estimated CSM masses lie in the range $M_{\mathrm{CSM}} \sim 0.17$–$0.95$\,$M_{\odot}$, in agreement with prior estimates for similar events, including 0.2–1\,$M_{\odot}$ from \citet{Pellegrino2022ApJ...926..125P}, $\sim$0.73\,$M_{\odot}$ for SN~2019uo \citep{Gangopadhyay2020ApJ...889..170G}, and $\sim$1.6\,$M_{\odot}$ for SN~2020bqj \citep{Kool2021A&A...652A.136K}. The inner radii of the CSM shells are found to be $R_{\mathrm{CSM}} \sim 10$–$50$\,AU, with corresponding densities of $\rho_{\mathrm{CSM}} \sim 10^{-10}$–$10^{-7}$\,g\,cm$^{-3}$.
The synthesised $^{56}$Ni masses, computed using $M_{\mathrm{Ni}} = M_{\mathrm{ej}} \times f_{\mathrm{Ni}}$, are in the range 0.002–0.15\,$M_{\odot}$. These are consistent with $\leq$0.19\,$M_{\odot}$ reported by \citet{Pellegrino2022ApJ...926..125P}, as well as the $\leq$0.1\,$M_{\odot}$ reported by \citet{Maeda2022ApJ...927...25M}.
Using the relation $v_{\mathrm{ej}} = \sqrt{2 E_{\mathrm{K}} / M_{\mathrm{ej}}}$, we estimate the ejecta velocities to lie in the range $\sim$2700–13,500\,km\,s$^{-1}$, consistent with the $\sim$3300\,km\,s$^{-1}$ measured for SN~2020bqj \citep{Kool2021A&A...652A.136K} and $\sim$11300\,km\,s$^{-1}$ for SN~2019kbj \citep{Ben-Ami2023ApJ...946...30B}.

\subsection{Comparison with Radiation-Hydrodynamic Interaction Models} \label{Sec:Dessart}

To investigate the interaction-powered scenarios relevant to Type Ibn SNe, \citet{Dessart2022A&A...658A.130D} conducted one-dimensional multi-group radiation-hydrodynamics simulations using the \textsc{HERACLES} code. Their models comprise an inner shell representing the SN ejecta and an outer shell mimicking the CSM, formed either through explosive mass ejection or via a super-Eddington wind. By systematically varying the ejecta and CSM properties, they explored the resulting bolometric light curves focusing on the rise time, the peak luminosity, and the radiated energy during the high-luminosity phase. The fiducial inner ejecta model adopted $E_{\rm kin} = 7 \times 10^{50}$\,erg, $M_{\rm ej} = 1.49\,M_\odot$, and $M(^{56}\mathrm{Ni}) = 0.08\,M_\odot$, with scaled-down variants to test lower-energy explosions. The outer shell configurations included both $1\,M_\odot$ ejecta-like CSM with kinetic energies of $10^{47}$–$10^{49}$\,erg and wind-like CSM with a terminal velocity of $v_\infty = 1000$\,km\,s$^{-1}$ and mass-loss rates spanning $10^{-3}$–$10^{-1}\,M_\odot$\,yr$^{-1}$. These simulations yielded a broad range of light-curve morphologies and properties of the cold dense shell \citep[CDS; see Table~1 of][]{Dessart2022A&A...658A.130D}.

In Figure~\ref{fig:Luc_model_comparison}, we compare the pseudo-bolometric light curves of the five SNe~Ibn in our sample with representative interaction models from \citet{Dessart2022A&A...658A.130D}. The left panel shows simulations with ejecta–wind interaction, while the right panel presents ejecta–ejecta configurations, similar to those discussed in \citet{Woosley2021ApJ...913..145W}.
It is important to note that the models from \citet{Dessart2022A&A...658A.130D} are not tailored to reproduce individual events, but rather to explore the general behaviour of interacting transients across a wide parameter space. They highlight key observational features such as the rise time, the peak luminosity, and the time-integrated bolometric output during the high-luminosity phase, demonstrating how variations in ejecta and circumstellar medium properties produce diverse light-curve morphologies.

Their simulations show that standard-energy explosions of helium stars embedded in dense, wind-like CSM can reach peak luminosities of a few $10^{44}$\,erg\,s$^{-1}$ within a few days, similar to luminous events like AT\,2018cow. Weaker winds, on the other hand, lead to Type~Ibc-like transients with double-peaked light curves and more modest luminosities of $\sim10^{42.2}$–$10^{43}$\,erg\,s$^{-1}$.
The observed rise time of $\sim$10 days in our SN Ibn sample suggests the need for relatively high mass-loss rates ($\dot{M} \gtrsim 0.1\,M_\odot$\,yr$^{-1}$) to reproduce the light-curve shapes. For SN\,2023utc, that shows a lower luminosity, models with reduced ejecta density or velocity may provide a more suitable match, as the shock power becomes less relevant and the luminosity accordingly diminishes.
A viable configuration that accounts for both the radiative and kinematic properties of several objects in our sample (SNe~2020nxt, 2020taz, 2021bbv, and 2024aej) requires low-mass and low-energy ejecta colliding with a massive outer shell, as exemplified by models E1, E2, E5, or E6 in \citet{Dessart2022A&A...658A.130D}. This interaction scenario naturally explains the persistence of narrow spectral features and moderate luminosities. Such a setup may occur when a low-mass helium star in a binary system undergoes substantial envelope stripping via mass transfer, followed by a nuclear flash or enhanced wind phase shortly before the core collapse, and ultimately explodes with a relatively low kinetic energy. For SN\,2023utc, an even lower-velocity or lower-density inner shell than those currently modelled are likely required to match the observed photometric evolution.

\subsection{Correlations of the physical parameters}\label{Sec:correlations}

\begin{figure*}
\begin{center}
\includegraphics[width=1.94\columnwidth]{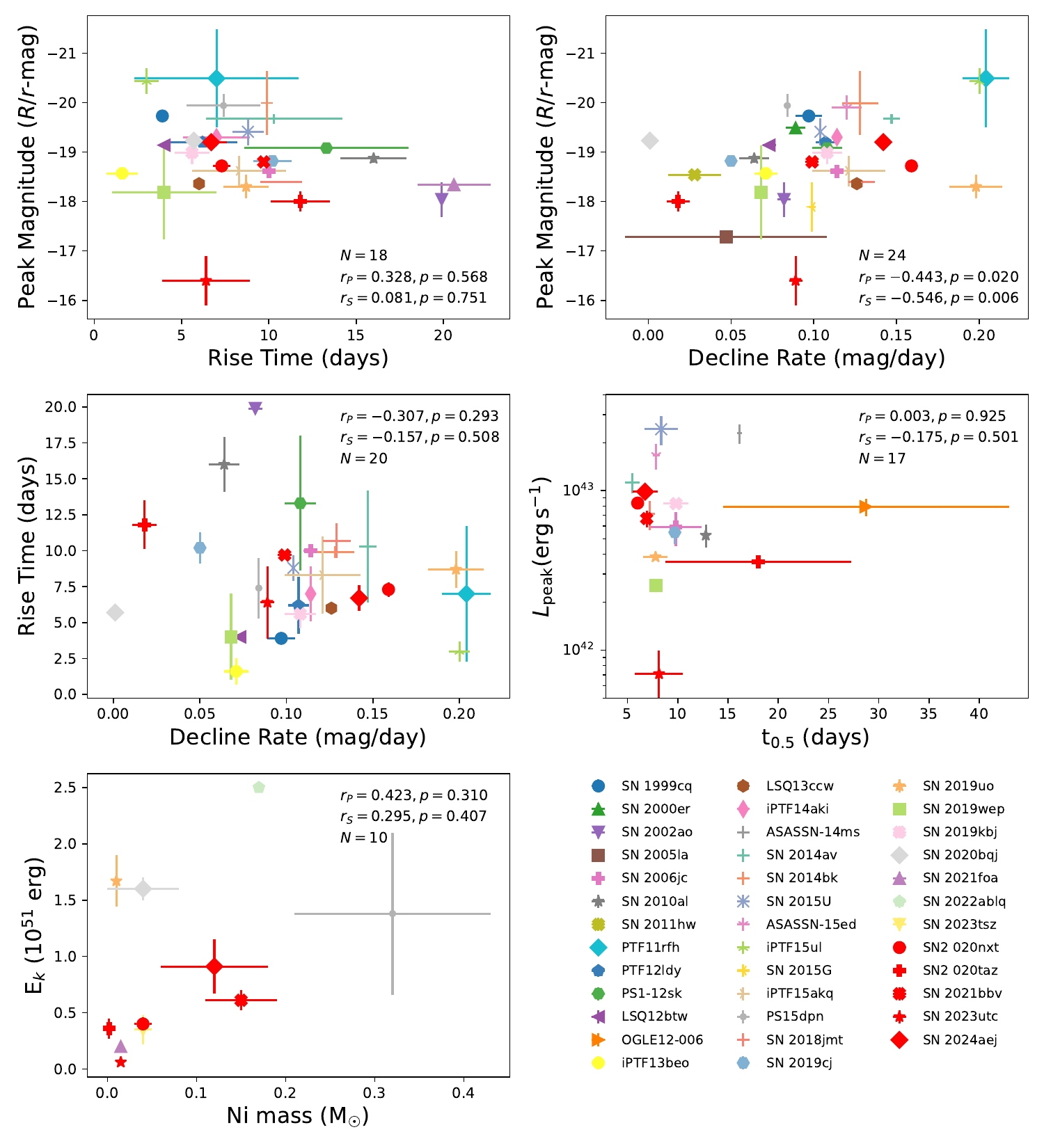}
\caption{
Relationships between parameters inferred from the light curves of SNe~Ibn. 
\textit{top left} — $R/r$-band peak magnitude versus rise time; 
\textit{top right} — $R/r$-band peak magnitude versus $R/r$-band decline rate; 
\textit{centre left} — rise time versus $R/r$-band decline rate; 
\textit{centre right} — peak luminosity ($L_{\mathrm{peak}}$) versus $t_{0.5}$ (the time required for the luminosity to decline by half from its peak); 
\textit{bottom left} — kinetic energy ($E_{\mathrm{K}}$) versus synthesised $^{56}$Ni mass. 
The weighted Pearson correlation coefficient ($r_P$), the Spearman rank correlation coefficient ($r_S$), and the associated $p$-values (probability of chance correlation) are provided for each parameters pair. Statistics exclude data points with rise-time and peak magnitude limits.}
\label{fig:phase_space}
\end{center}
\end{figure*}

Comparing the properties of our five SNe with those of other SNe Ibn provides valuable insights for a more precise characterisation of this SN type. \citet{Wang2024A&A...691A.156W} presented a $V$-band phase-space diagram for Type Ibn SNe, suggesting the existence of correlations between peak magnitude, rise time, and decline rate. 
Here, we extend this approach by examining both photometric observables and derived physical parameters for SNe~Ibn. Specifically, we present in Fig. \ref{fig:phase_space} the following phase-space diagrams: $R/r$-band peak magnitude versus rise time; $R/r$-band peak magnitude versus $R/r$-band decline rate; rise time versus $R/r$-band decline rate; peak bolometric luminosity ($L_{\mathrm{peak}}$) versus $t_{0.5}$ (the time required for the luminosity to decline by half from peak); and ejecta kinetic energy ($E_{\mathrm{K}}$) versus synthesised $^{56}$Ni mass.
Our new sample comprises SNe~2020nxt, 2020taz, 2021bbv, 2023utc, and 2024aej.\footnote{Data for the comparison objects are taken from \citet{Matheson2000AJ....119.2303M,Pastorello2007Natur.447..829P,Pastorello2008MNRAS.389..113P,Pastorello2008MNRAS.389..131P,Mattila2008MNRAS.389..141M,Sanders2013ApJ...769...39S,Gorbikov2014MNRAS.443..671G,Morokuma2014CBET.3894....1M,Pastorello2015MNRAS.449.1921P,Pastorello2015MNRAS.449.1954P,Pastorello2015MNRAS.453.3649P,Pastorello2015MNRAS.454.4293P,Pastorello2015MNRAS.449.1941P,Pastorello2016MNRAS.456..853P,Hosseinzadeh2017ApJ...836..158H,Vallely2018MNRAS.475.2344V,Wang2020ApJ...900...83W,Gangopadhyay2020ApJ...889..170G,Kool2021A&A...652A.136K,Wang2021ApJ...917...97W,Gangopadhyay2022ApJ...930..127G,Farias2024ApJ...977..152F,Pellegrino2024ApJ...977....2P,Wang2024A&A...691A.156W,Warwick2025MNRAS.536.3588W}.}
These comparisons allow us to place our events within the broader context of the SN~Ibn population and to assess the extent to which the observed diversity is reflected in the explosion and physical parameters.

To quantify potential correlations among these parameters, we calculate the weighted Pearson correlation coefficient ($r_P$), which accounts for the uncertainties under the assumption of Gaussian errors; the Spearman rank correlation coefficient ($r_S$), which is non-parametric and equally weights all points; and the associated $p$-values (see Fig.~\ref{fig:phase_space}). No statistically significant correlation is found between the peak magnitude and the rise time, or between the rise time and the decline rate. However, we observe a moderate negative correlation between the peak magnitude and the decline rate, indicating that more luminous events tend to evolve more rapidly after maximum light.
No clear correlation is evident between $L_{\mathrm{peak}}$ and $t_{0.5}$, although most SNe~Ibn exhibit $t_{0.5}$ values in the range of 5–15\,d and peak luminosities around $10^{43}$\,erg\,s$^{-1}$. A weak positive trend is apparent between the kinetic energy and the synthesised $^{56}$Ni mass, consistent with the finding of \citet{Pellegrino2024ApJ...977....2P} that fainter, slower-evolving, interaction-powered transients generally exhibit lower explosion energies and produce less $^{56}$Ni.

As shown in Fig.~\ref{fig:phase_space}, SN 2023utc displays a peak magnitude lower than those of most Type Ibn SNe, while the decline rate of SN 2020taz is comparable to those of SNe 2011hw and 2020bqj, both exhibiting slower rates. In contrast to the extreme phase-space locations of SNe 2023utc and 2020taz, SNe 2020nxt, 2021bbv, and 2024aej occupy central positions in the diagrams, suggesting that they represent more typical events within the Ibn sample.

\section{Spectroscopy}\label{sec:spectroscopy}

Our spectral sequences for SNe 2020nxt, 2020taz, 2021bbv, 2023utc and 2024aej were obtained using multiple instrumental configurations, which are listed in Table~\ref{table_telescope} (Appendix~\ref{sec:facilities}). Basic parameters for the spectra are reported in Tables~\ref{table:speclog_2020nxt}-\ref{table:speclog_2024aej} (Appendix~\ref{Spec_log}).

The spectral data were processed using standard procedures, combining \texttt{IRAF} tool and optimised pipelines like \texttt{FLOYDS},\footnote{\url{https://lco.global/documentation/data/floyds-pipeline/}} depending on the instrumental configuration and observatory setup. The preliminary reduction steps included corrections for bias, overscan, and flat-fielding of the two-dimensional frames to account for specific instrumental effects. Subsequently, the one-dimensional (1D) spectra were optimally extracted, ensuring a robust S/N ratio for the target spectra. Wavelength calibration was performed using arc lamp spectra obtained during the same observing run and instrumental configuration as the science data. Night-sky emission lines were cross-checked to refine the wavelength solution. Flux calibration was achieved using spectrophotometric standard stars observed under similar conditions. A sensitivity function derived from the standard star spectra was applied to the transient's spectra. To ensure consistency, the flux-calibrated spectra were compared against coeval broadband photometric data, and correction factors were applied in cases of significant discrepancies. Finally, the strongest telluric absorption bands, primarily from $\mathrm{O}_2$ and $\mathrm{H}_2\mathrm{O}$, were removed using spectra of early-type standard stars, which exhibit a nearly featureless continuum in the telluric absorption regions. The fully processed spectra,  presented in Fig.~\ref{fig:Five_SN_spectra}, provide a robust basis for analysing the properties of the five SN Ibn.

\subsection{Spectroscopic evolution and line identification}
\begin{figure*}
\begin{center}
\includegraphics[width=1.99\columnwidth]{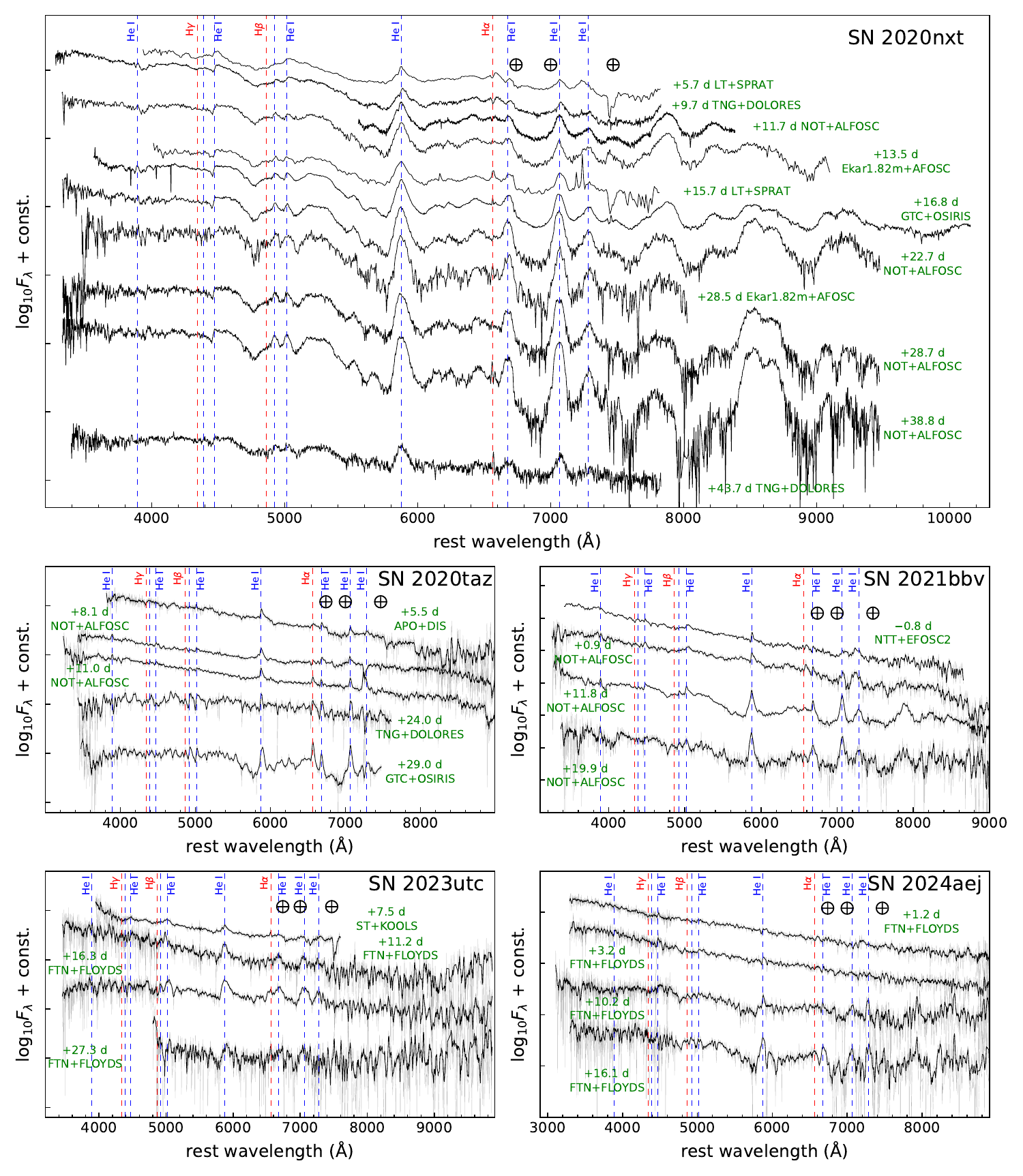}
\caption{Spectral sequences of the five SNe Ibn. The dashed vertical lines indicate the main H and He I transitions, while the $\oplus$ symbol marks the strongest telluric absorption bands. All spectra are corrected for redshift and extinction. Gray lines represent smoothed spectra (originally with a lower S/N) processed with a Savitzky-Golay filter.}
\label{fig:Five_SN_spectra}
\end{center}
\end{figure*}

\begin{figure*}
\begin{center}
\includegraphics[width=2\columnwidth]{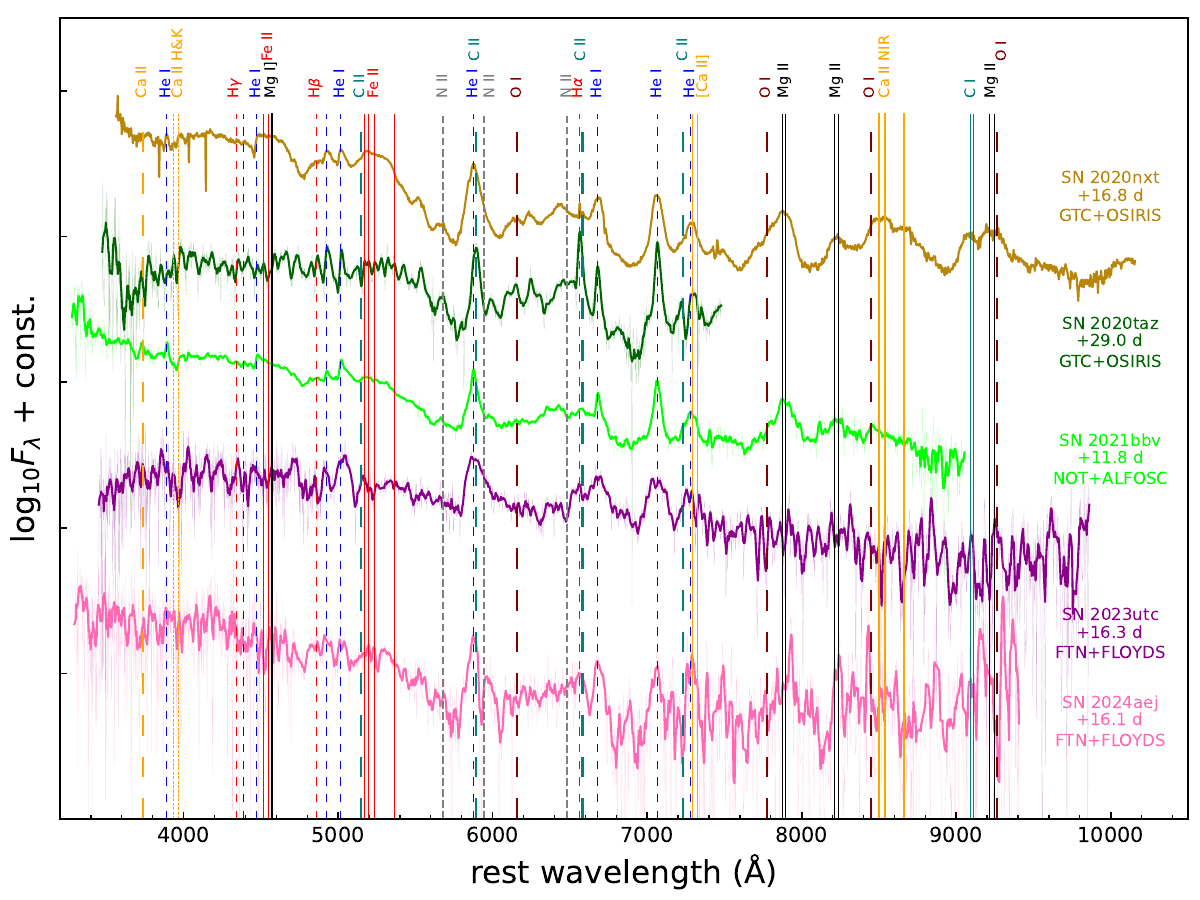}
\caption{Line identification in the highest-resolution late-time spectra of the five SNe presented in this paper. Spectra are corrected for redshift and reddening, with indicated phases from the maximum light.}
\label{fig:Line_identification}
\end{center}
\end{figure*}

{\bf SN 2020nxt} -- \citet{Wang2024MNRAS.530.3906W} reported 19 spectra for this object; we supplement the available dataset with 10 additional spectra. Our spectral sequence is shown in Fig.~\ref{fig:Five_SN_spectra}. Over the entire observational period, the spectra exhibit a moderate evolution.

The first spectrum (phase +5.7~d), previously reported by \citet{Srivastav2020TNSCR2148....1S} and \citet{Wang2024MNRAS.530.3906W}, is the classification spectrum. It shows a blue pseudo-continuum, and a blackbody fit to it yields a photospheric temperature of $T_{\rm BB} = 11800 \pm 3300~{\rm K}$. In the second spectrum (phase +9.7~d), the blackbody temperature increases to $T_{\rm BB} = 17500 \pm 6200~{\rm K}$. The third spectrum (phase +11.7~d) lacks sufficient blue-wavelength coverage, making the temperature fit unreliable. The fourth to tenth spectra (taken at phases +13.5, +15.7, +16.8, +22.7, +28.5, +28.7, and +38.8~d) show minimal evolution, with temperatures fluctuating between $11000$ and $14000~{\rm K}$. The  last spectrum (phase +43.7~d)  still displays a blue continuum, with the temperature decreasing to $T_{\rm BB} = 9700 \pm 3500~{\rm K}$.

In Fig.~\ref{fig:Five_SN_spectra}, we mark the strongest He~\textsc{i} lines and the lines of the Balmer series. The He~\textsc{i} lines, particularly He~\textsc{i}~$\lambda5876$ and $\lambda7065$, are the most prominent features in the spectra of SN~2020nxt. Additionally, a weak and narrow H$\alpha$ line with a P~Cygni profile is observed in the first three spectra, with an absorption minimum blue-shifted by approximately $100{-}300~{\rm km~s}^{-1}$. However, other prominent Balmer lines typical of type IIn SNe are not securely detected in SN~2020nxt.

We conducted a detailed line identification using the spectrum with the highest S/N ratio, obtained at phase +16.8~d (Fig.~\ref{fig:Line_identification}). This spectrum exhibits several broad bump features, including the following:
\begin{itemize}
    \item $7600{-}8000~\text{\AA}$:  a blend of O~\textsc{i}~$\lambda7774$ and Mg~\textsc{ii}~$\lambda\lambda7877{-}7896$;
    \item $8100{-}8300~\text{\AA}$:  Mg~\textsc{ii}~$\lambda\lambda8214{-}8235$;
    \item $8300{-}8800~\text{\AA}$:  a blend including O~\textsc{i}~$\lambda8446$ and the near-infrared Ca~\textsc{ii} triplet;
    \item $9000{-}9400~\text{\AA}$:  a blend including C~\textsc{i}~$\lambda\lambda9095{-}9112$, Mg~\textsc{ii}~$\lambda\lambda9218{-}9244$, and O~\textsc{i}~$\lambda\lambda9261{-}9266$;
    \item $4450{-}4650~\text{\AA}$: likely a blend including He~\textsc{i}~$\lambda4471$, Mg~\textsc{i}]~$\lambda4571$, and Fe~\textsc{ii}~$\lambda\lambda4303{-}4352$;
    \item $5100{-}5400~\text{\AA}$: likely a blend including C~\textsc{ii}~$\lambda5145$ and Fe~\textsc{ii}~$\lambda\lambda5018,5169,5198,5235$.
\end{itemize}

We also tentatively identify features at $\lambda5680$ and $\lambda6482$ as N~\textsc{ii}, and [Ca~\textsc{ii}] $\lambda\lambda7292,7324$. Whilst these features are usually expected in core-collapse SNe, we do not securely detect lines typical of thermonuclear SNe, such as S~\textsc{ii} and Si~\textsc{ii}. \\

{\bf SN 2020taz} -- The very rapid evolution and the modest apparent magnitude of SN~2020taz limited its spectroscopic monitoring to only five epochs. The first spectrum (phase +5.5~d) is the blue continuum-dominated classification spectrum, with a blackbody temperature of $T_{\rm BB} = 15000 \pm 3500~{\rm K}$. Its main features include He~\textsc{i} lines ($\lambda5876, \lambda6678, \lambda7065$) with P~Cygni profiles and velocities of approximately $800~{\rm km~s^{-1}}$, as well as a weak P~Cygni H$\alpha$ line.

The second and third spectra (phases +8.1~d and +11.0~d, respectively) show a redder continuum with a decreasing blackbody temperature of approximately $T_{\rm BB} \sim 11400~{\rm K}$, while the H$\alpha$ feature becomes more prominent with time. The fourth spectrum (phase +24.0~d) has a low S/N ratio and shows a significantly cooler temperature of $T_{\rm BB} = 6700 \pm 1800~{\rm K}$. The final spectrum (phase +29.0~d) reveals an increase in the prominence of the H$\alpha$ line, with a P~Cygni absorption minimum blue-shifted by $\sim 800~{\rm km~s^{-1}}$. Weak H$\beta$ and H$\gamma$ lines are also detected, while the blackbody temperature increases slightly to $T_{\rm BB} = 7400 \pm 2200~{\rm K}$.

Similar to the type~Ibn prototype SN~2006jc \citep{Foley2007ApJ...657L.105F, Pastorello2007Natur.447..829P, Smith2008ApJ...680..568S, Mattila2008MNRAS.389..141M}, SN~2020taz exhibits a very weak (almost undetectable) H$\alpha$ line during the early phases, which becomes more pronounced only at later epochs. In contrast, type~IIn SNe typically display a narrow H$\alpha$ line (with full width at half maximum, ${\rm FWHM} \leq 1000~{\rm km~s^{-1}}$) that dominates in strength the He~\textsc{i} lines at all phases.\\

{\bf SN 2021bbv} -- The four spectra of SN~2021bbv exhibit predominantly a blue continuum, with the photospheric temperature gradually decreasing from $T_{\rm BB} \sim 14000~{\rm K}$ to $8600~{\rm K}$. The prominent features in the spectra include the He~\textsc{i} lines ($\lambda\lambda5876, 6678, 7065$), which become more pronounced over time. The evolution of the He~\textsc{i} line velocities is modest; this will be discussed in detail in Section~\ref{sec:He_velocity}.
For a more detailed identification of the spectral lines, we used the third spectrum (phase +11.8~d; see Fig.~\ref{fig:Line_identification}). Broad features, similar to those observed in SN~2020nxt, are detected at $4450{-}4650~\text{\AA}$, $5100{-}5400~\text{\AA}$, and $7600{-}8000~\text{\AA}$. However, the NIR Ca~\textsc{ii} triplet was not securely identified. 
Furthermore, we were unable to confidently identify the Balmer series lines or spectral features typical of thermonuclear SNe, such as S~\textsc{ii} and Si~\textsc{ii}.\\

{\bf SN 2023utc} -- The four noisy spectra show a blue continuum, with the strongest feature being He~\textsc{i} $\lambda5876$. The photospheric temperature initially increases, and then decreases over time: $T_{\rm BB} = 11900 \pm 5900~{\rm K}$ in the first spectrum (phase +7.5~d), it peaks at $T_{\rm BB} = 15600 \pm 2900~{\rm K}$ in the second spectrum (phase +11.2~d), and declines to $T_{\rm BB} = 4800 \pm 1300~{\rm K}$ in the last spectrum (phase +27.3~d).
For the line identification, the third higher S/N spectrum (phase +16.3~d) was considered (see Fig.~\ref{fig:Line_identification}). We identify H$\alpha$, H$\beta$, H$\gamma$, and He~\textsc{i} $\lambda\lambda4471, 4921, 5016, 5876, 6678, 7065, 7281$. However, alternative identifications for some features, such as C~\textsc{ii} $\lambda6578$, N~\textsc{ii} $\lambda4803$ and [Ca~\textsc{ii}] $\lambda\lambda7291, 7324$, cannot be ruled out.\\

{\bf SN 2024aej} -- Figure~\ref{fig:Five_SN_spectra} shows that SN~2024aej exhibits spectral properties similar to those of SN~2023utc. Four spectra were obtained, all quite noisy ratios. They show a blue continuum dominated by prominent He~\textsc{i} lines typical of Type~Ibn SN. Due to the low S/N, additional spectral lines cannot be securely identified (see Fig.~\ref{fig:Line_identification}).
The photospheric temperature evolves from $T_{\rm BB} = 15600 \pm 3000~\rm{K}$ in the first spectrum (phase +1.2~d), to a peak of $T_{\rm BB} = 17800 \pm 3600~\rm{K}$ in the second spectrum (phase +3.2~d), before decreasing to $T_{\rm BB} = 10500 \pm 2100~\rm{K}$ in the final spectrum (phase +16.1~d). 

\subsection{Comparison of Type Ibn SN spectra}
\begin{figure*}
\begin{center}
\includegraphics[width=2\columnwidth]{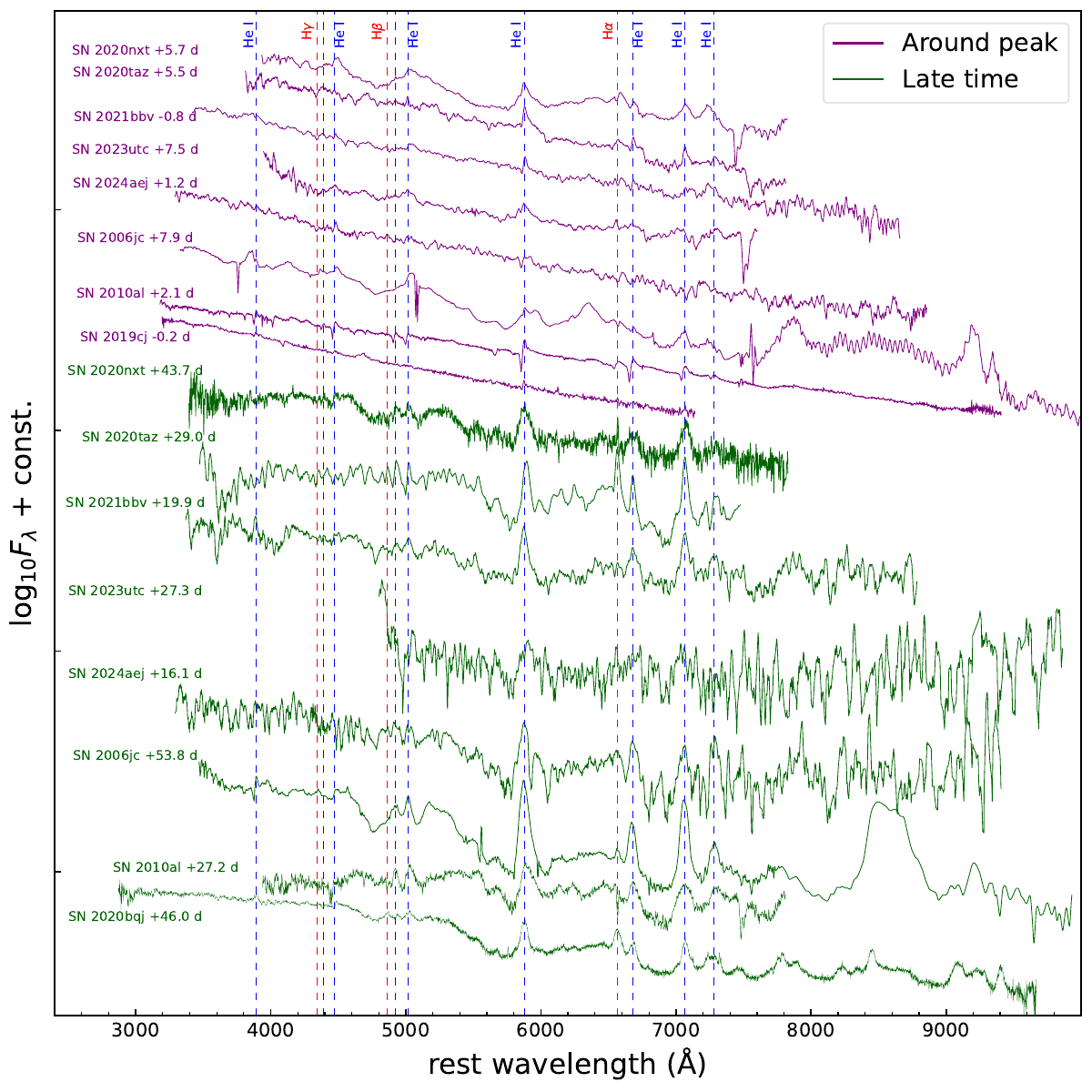}
\caption{Comparison of around-peak (purple) and late-time (green) spectra of SNe 2020nxt, 2020taz, 2021bbv, 2023utc, and 2024aej with other SNe Ibn at similar phases. All spectra are corrected for redshift and extinction. Significant He I features are marked by blue dashed lines, while Balmer features by red dashed lines.}
\label{fig:sepc_IBN_comparison}
\end{center}
\end{figure*}

In Fig.~\ref{fig:sepc_IBN_comparison}, we compare the spectra of SNe~2020nxt, 2020taz, 2021bbv, 2023utc, and 2024aej obtained near the maximum light and at late phases with those of other SNe~Ibn at similar phases. 
In the top part of Fig.~\ref{fig:sepc_IBN_comparison}, the spectra of these five supernovae near peak brightness are compared to those of SNe~2006jc, 2010al, and 2019cj. The spectra of type~Ibn SNe exhibit remarkably similar blue continua with prominent He~\textsc{i} lines in emission. However, subtle differences are also present. Specifically, H$\alpha$ emission is weak but detectable in the spectra of SNe~2006jc, 2010al, 2020nxt, 2020taz, and 2023utc, while its presence is uncertain in SNe~2021bbv, 2024aej, and 2019cj. 
None of the objects exhibit clear flash-ionisation features. However, it should be emphasised that the earliest spectrum of our five events was obtained only after maximum light, later than the timing when flash-ionisation features were observed in other Ibn events. For instance, in SN~2010al \citep{Pastorello2015MNRAS.449.1921P}, flash-ionisation features appeared 8~days before the maximum light and disappeared 4~days later. Similarly, in SN~2019cj \citep{Wang2024A&A...691A.156W}, prominent flash features were observed 2.1~days before maximum. Therefore, given the timing of the earliest spectra for these five supernovae, we cannot rule out the presence of flash-ionisation features at earlier stages.

\begin{figure*}
\begin{center}
\includegraphics[width=2\columnwidth]{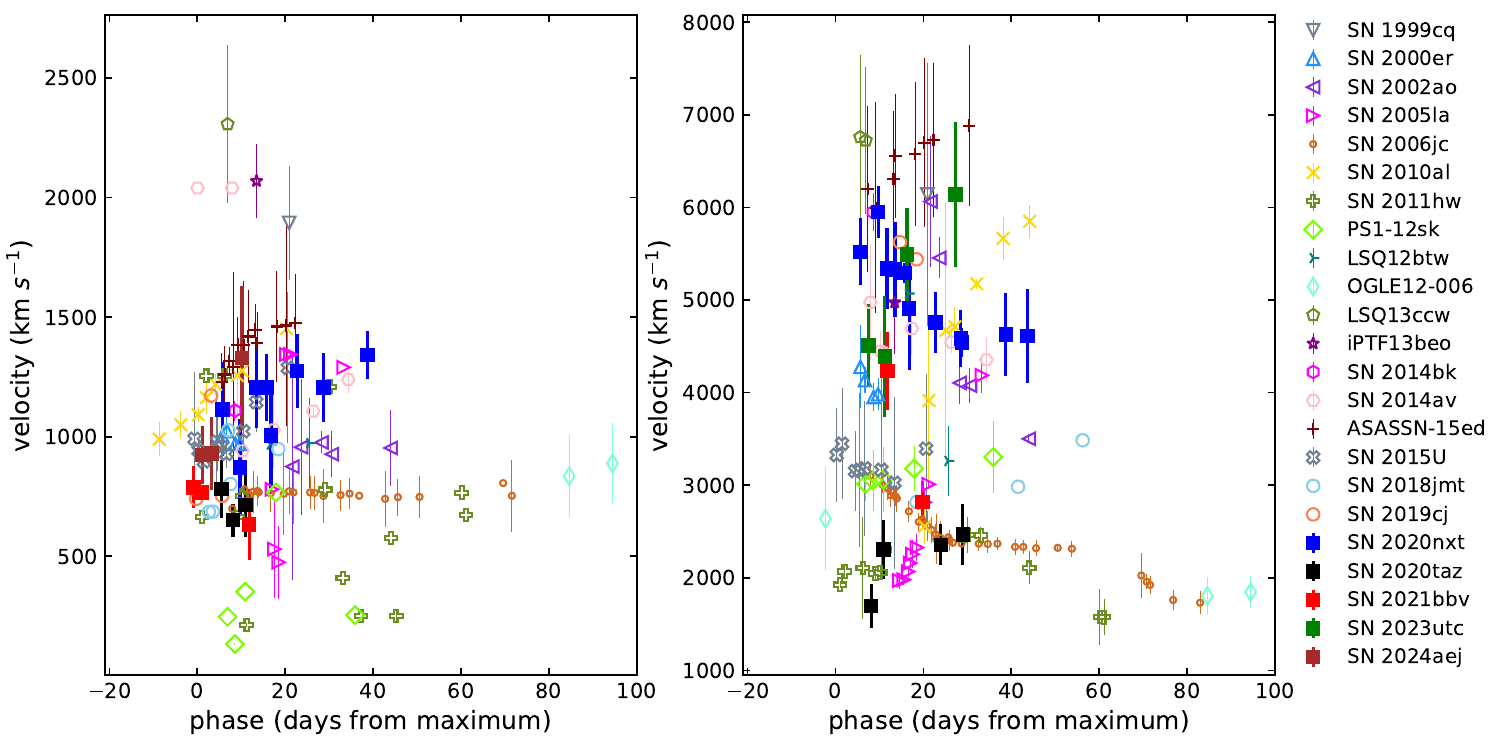}
\caption{Evolution of He~\textsc{i} line velocities. \textit{Left panel:} Temporal evolution of the velocities associated with the narrow He~\textsc{i} line components, which trace the unshocked CSM. \textit{Right panel:} Velocity evolution of the broader He~\textsc{i} emission components, reflecting the dynamics of the shocked gas region. Data for comparison SNe~Ibn are adopted from \citet{Pastorello2016MNRAS.456..853P} and \citet{Wang2024A&A...691A.156W}.}
\label{fig:HeVelocity}
\end{center}
\end{figure*}

In the lower part of Fig.~\ref{fig:sepc_IBN_comparison}, we compare the last spectra of SNe~2020nxt, 2020taz, 2021bbv, 2023utc, and 2024aej with those of SNe~2006jc, 2010al, and 2020bqj at similar phases. 
At late times, the emission lines become more prominent, allowing for a more reliable spectral identification. 
SNe~2020nxt, 2020taz, 2023utc, 2006jc, and 2020bqj show strikingly similar features, including prominent He~\textsc{i} lines. 
A notable characteristic shared by these events is the absorption feature in the 4600--5200~\AA\ region, which is likely due to Fe~\textsc{ii} line absorptions. A similar feature is also observed in stripped-envelope SNe. The blue pseudo-continuum and the decrease in flux beyond 5400~\AA\ appear to be common features among most interaction-powered CC~SNe, and is due to a blend of prominent emission lines of Fe \citep{Smith2012MNRAS.426.1905S, Stritzinger2012ApJ...756..173S, Pastorello2015MNRAS.449.1921P}. 
Although some heterogeneity exists within the Type Ibn SN population, their spectra generally exhibit consistent pseudo-continuum shapes, as well as line identification and profiles. This suggests that the physical conditions in the line-forming regions of the CSM are not significantly different among these SNe.

\subsection{Velocity evolution of the He I lines}
\label{sec:He_velocity}

To constrain the properties of the stellar wind and the nature of the line-emitting regions, we examine the velocity evolution of spectral lines. SNe powered by interaction typically exhibits lines with multiple-width components, originating from gas located in different regions \citep{Chevalier1994ApJ...420..268C, Chugai1997Ap&SS.252..225C, Pastorello2016MNRAS.456..853P}. The narrow lines (with velocities ranging from a few hundred to $\leq 2000~\mathrm{km~s^{-1}}$) are likely produced in the unshocked CSM, constituted by material lost by the progenitor star prior to the SN explosion. 
Broader components, with typical velocities ranging from several thousand to $\sim 10^4~\mathrm{km~s^{-1}}$, are likely produced in shocked gas regions, distinct from the electron-scattering wings often observed in interacting SNe.

Following \citet{Pastorello2016MNRAS.456..853P}, the velocity of the He-rich ejecta can be determined by measuring the wavelength of the blue-shifted absorption core of the P~Cygni profile, when such a profile is clearly identified. If a P~Cygni profile is not detected, the velocity is roughly estimated from the FWHM of the He~\textsc{i} emission line. The FWHM is derived after the line profile has been deblended using a combination of Gaussian and Lorentzian fitting functions. 
The evolution of the narrow and the broader components of the He~\textsc{i} lines for the SN sample is presented in Fig.~\ref{fig:HeVelocity}.
For SNe 2023utc and 2024aej, the noisy spectra prevent the measured velocities from accurately constrain the true physical properties of the gas.

The velocity of the narrow He~\textsc{i} components attributed to the unshocked CSM is a proxy for the stellar wind velocity prior to the SN explosion. Fig.~\ref{fig:HeVelocity} shows that the velocities of the narrow He~\textsc{i} lines in the SN Ibn sample span a wide range of values, from a few hundred to approximately $2500~\mathrm{km~s^{-1}}$. However, while SNe 2005al, 2011hw, and PS1-12sk exhibit very low wind velocities (below $500~\mathrm{km~s^{-1}}$), more frequently the narrow He~\textsc{i} components have velocities ranging from $500$ to $1500~\mathrm{km~s^{-1}}$. 
In most cases, the narrow-line velocities exhibit a temporal grow: in SN~2020nxt the $v_{\text{FWHM}}$ increases from $900~\mathrm{km~s^{-1}}$ to $1300~\mathrm{km~s^{-1}}$, while in SN~2010al the P~Cygni velocity rises from $1000~\mathrm{km~s^{-1}}$ to $1500~\mathrm{km~s^{-1}}$. Occasionally, the line velocities show a negligible evolution with time, like in the cases of SNe 2020taz and 2021bbv, where the line velocities 
remain relatively stable at around $700~\mathrm{km~s^{-1}}$. These differences may suggest distinct progenitor, mass-loss and interaction scenarios.

The velocity evolution of the broader components of He~\textsc{i} lines is also shown in Fig.~\ref{fig:HeVelocity}. These broader components exhibit more pronounced temporal changes, reflecting heterogeneity in the ejecta velocities and in the density profiles of the interacting material. In some cases, the velocities of broader He~\textsc{i} components increase over time. For instance, in SN 2005la the gas velocity rises from approximately $2000~\mathrm{km~s^{-1}}$ shortly after discovery to around $4200~\mathrm{km~s^{-1}}$ three weeks later. Similarly, in SN 2020taz $v_{\text{FWHM}}$ increases from $1700~\mathrm{km~s^{-1}}$ to $2500~\mathrm{km~s^{-1}}$. 
In contrast, some cases exhibit a narrowing trend over time. For example, in SN 2006jc the intermediate components of He~\textsc{i} lines decreases from $3100~\mathrm{km~s^{-1}}$ to $1700~\mathrm{km~s^{-1}}$ in four months. This trend is also observed in other obejcts \citep[SNe 2000er, 2002ao, 2014av;][]{Pastorello2016MNRAS.456..853P}. This is also observed in some SNe of our sample (SNe~2020nxt and 2021bbv), suggesting a decline in the velocity of the shocked gas region, likely caused by increasing density in the CSM gas distribution. 
A non-monotonic trend is seen in SN 2011hw, where the broader components velocity initially increases from $1900~\mathrm{km~s^{-1}}$ to $2500~\mathrm{km~s^{-1}}$, before declining to approximately $1600~\mathrm{km~s^{-1}}$ by $\sim60$ days post-maximum. Such peculiar evolution could be attributed to a complex density distribution in the shocked gas region.

\subsection{Modeling the Spectra}

\begin{figure*}
\begin{center}
\includegraphics[width=1\columnwidth]{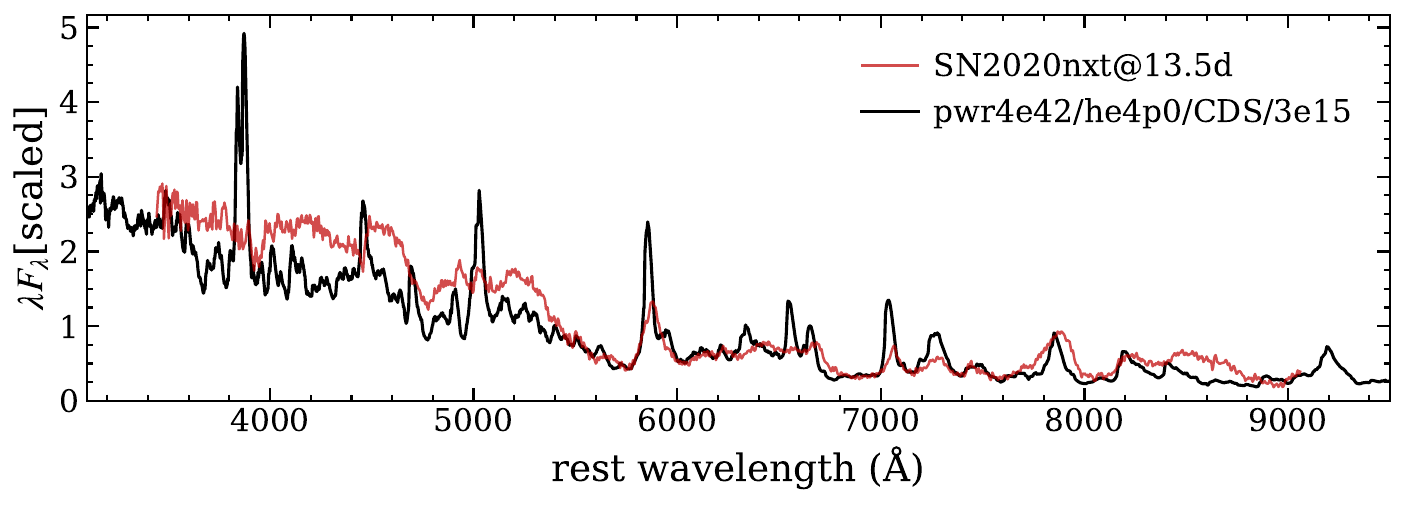}
\includegraphics[width=1\columnwidth]{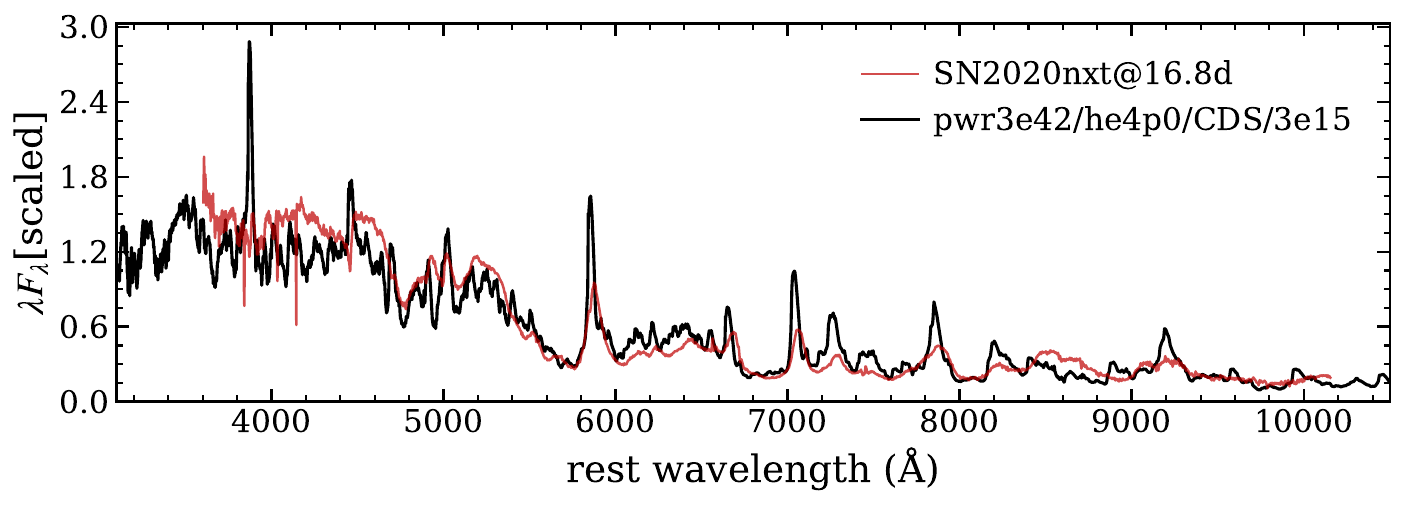}
\includegraphics[width=1\columnwidth]{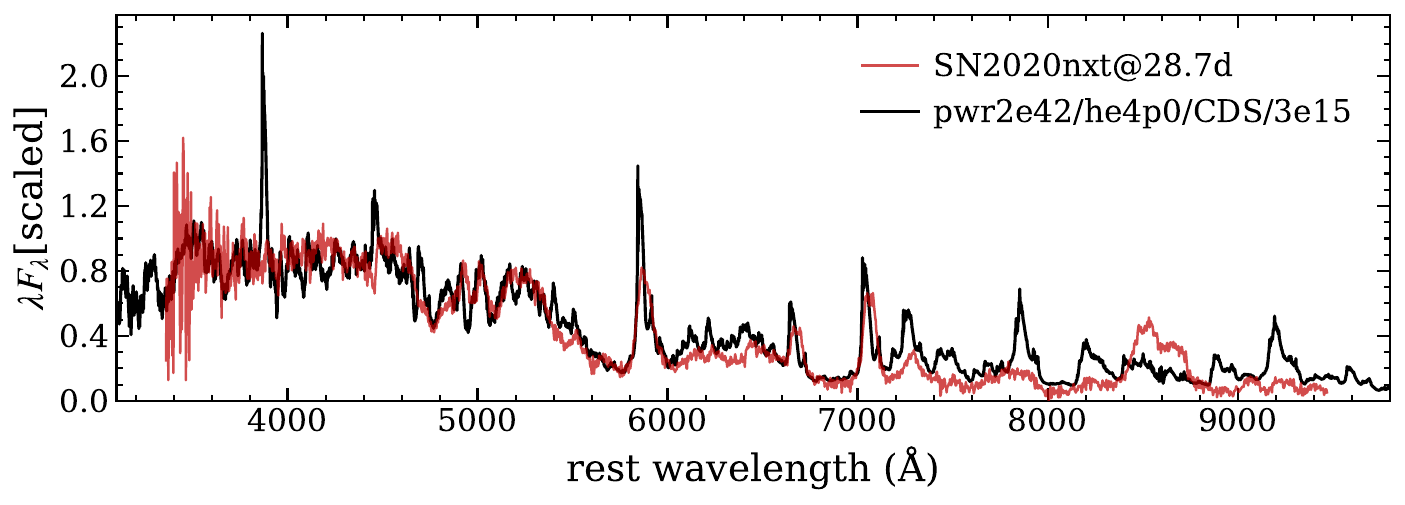}
\includegraphics[width=1\columnwidth]{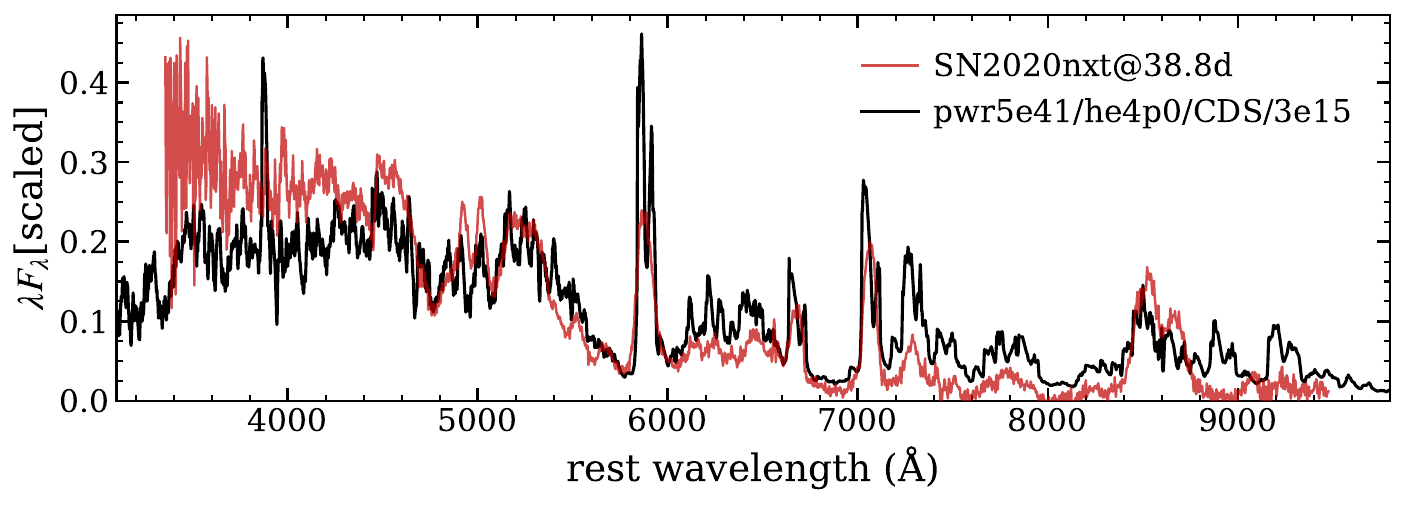}
\includegraphics[width=1\columnwidth]{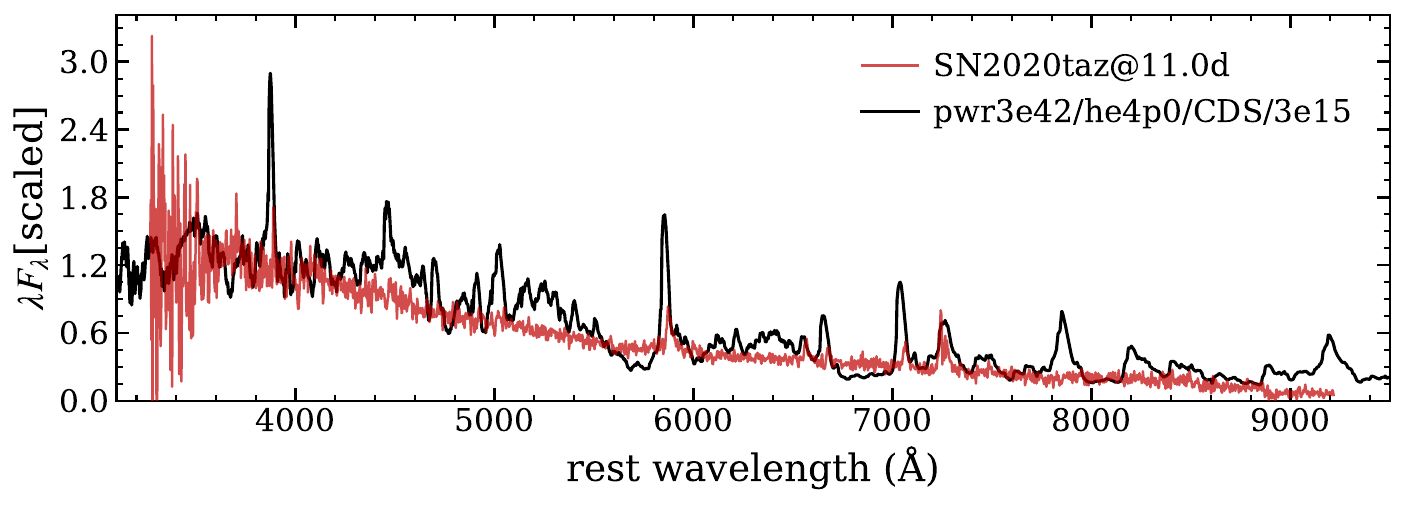}
\includegraphics[width=1\columnwidth]{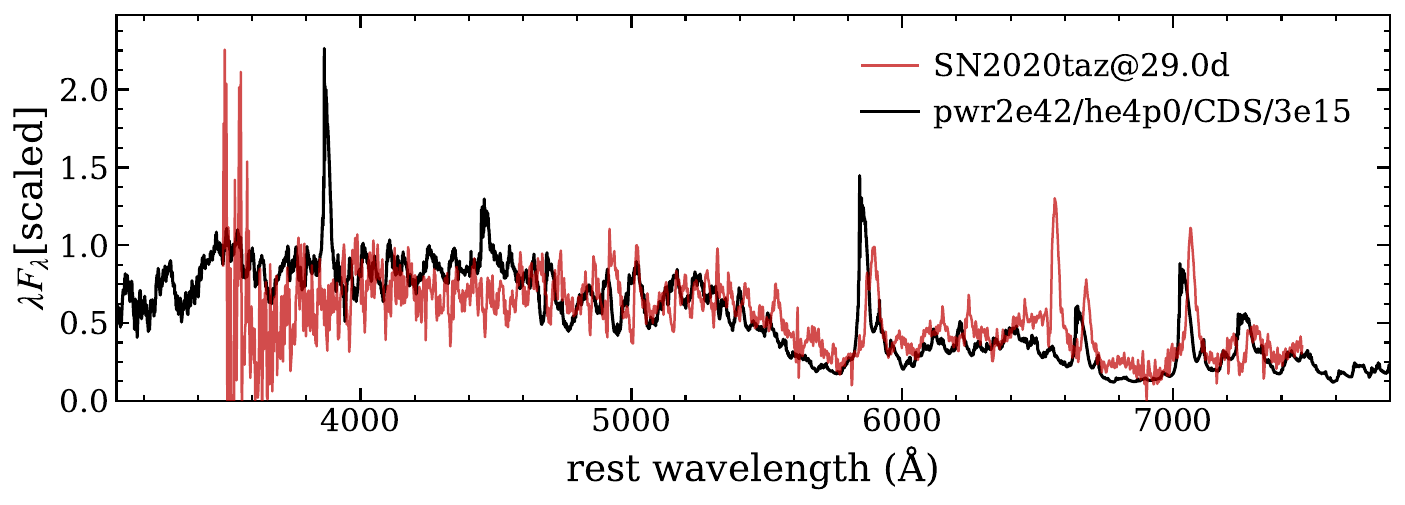}
\includegraphics[width=1\columnwidth]{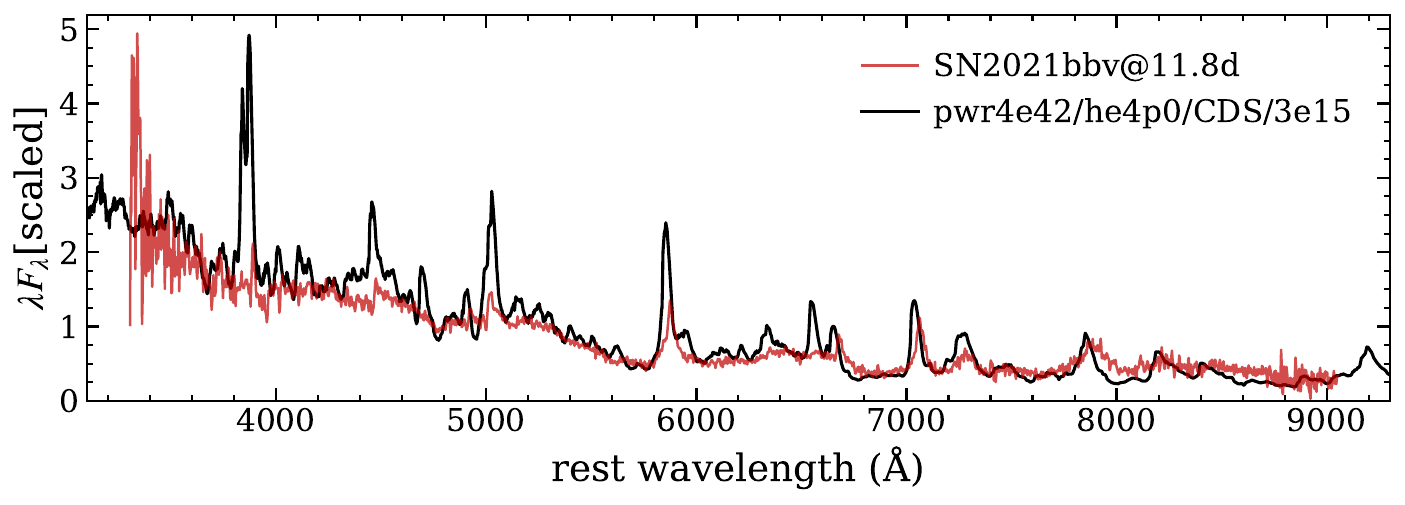}
\includegraphics[width=1\columnwidth]{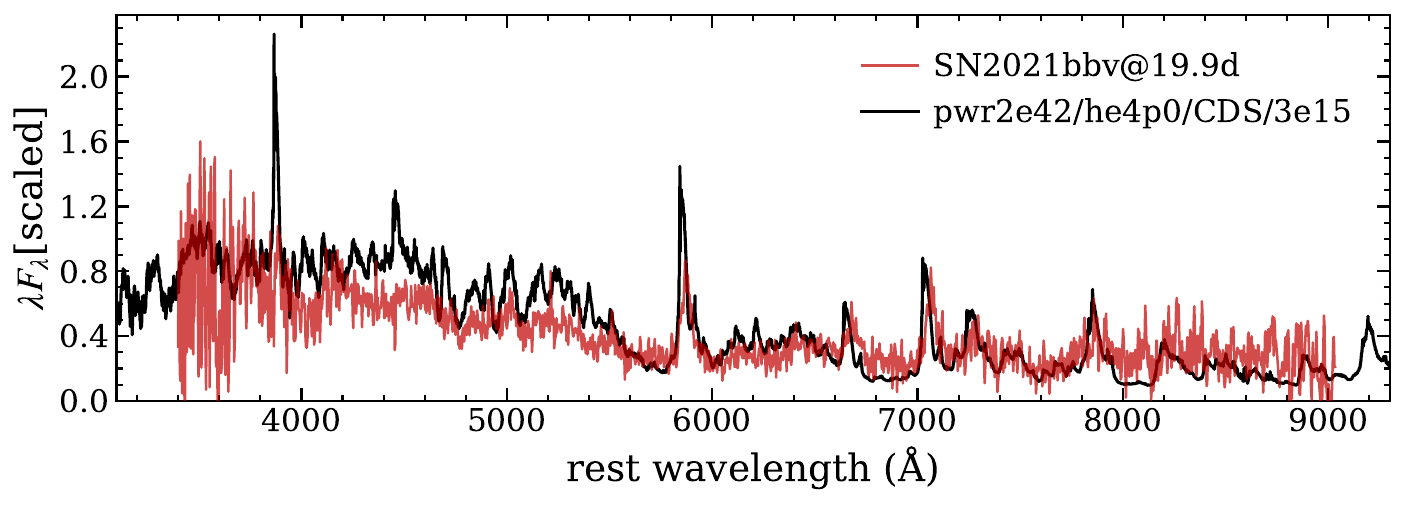}
\includegraphics[width=1\columnwidth]{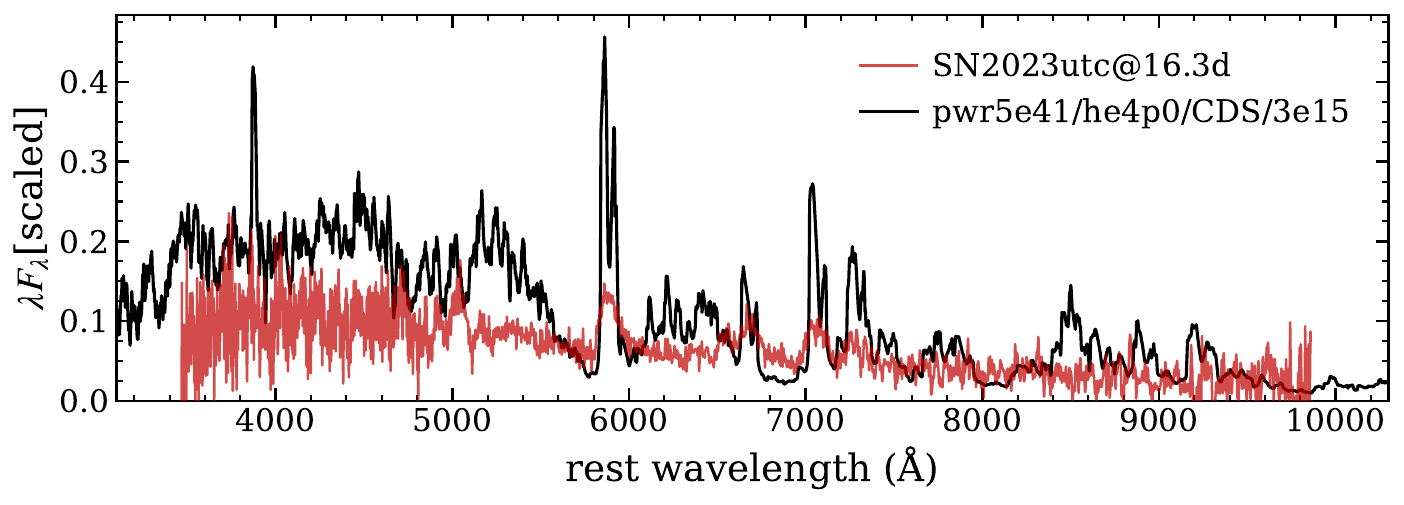}
\includegraphics[width=1\columnwidth]{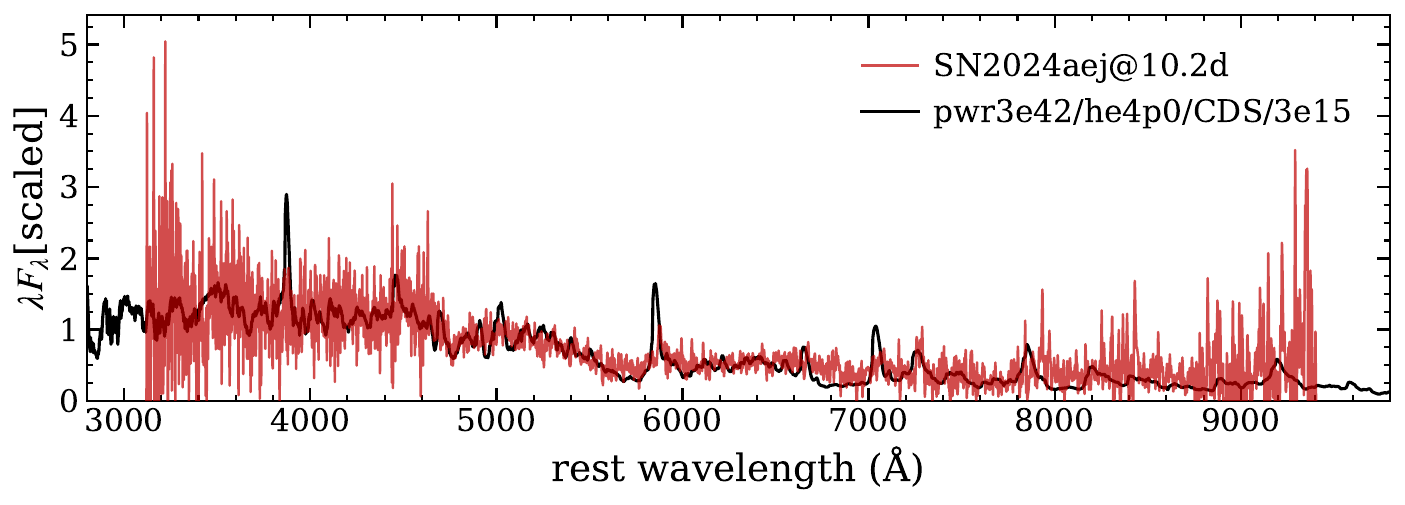}
\caption{Comparison between synthetic spectra from model \texttt{he4p0} and observed spectra of five SNe~Ibn at multiple epochs after bolometric maximum light. No smoothing has been applied to either the observed or model spectra. The synthetic spectra are based on simulations from \citet{Dessart2022A&A...658A.130D} and \citet{Wang2024MNRAS.530.3906W}, as well as newly computed models incorporating updated parameters.}
\label{fig:spectral_comparison_model}
\end{center}
\end{figure*}

To constrain the progenitor and ejecta properties for our SNe~~Ibn sample, we compare observed spectral sequences with a set of non-local thermodynamic equilibrium (NLTE) radiative-transfer simulations computed using \textsc{CMFGEN}. These include models from \citet{Dessart2022A&A...658A.130D}, along with a few additional models with adjusted parameters (Dessart, priv. comm.). The simulations assume interaction between low-mass ($\lesssim 1~M_\odot$), moderate-energy ($\sim10^{50}$~erg) ejecta and a slowly expanding, He-rich circumstellar shell of comparable mass. A dense, thin CDS forms as a result of this interaction and dominates the emission at late times.

In this quasi-steady configuration, hydrodynamical evolution is neglected, and the CDS is treated as a chemically mixed zone with a Gaussian density profile centred at 2000~km~s$^{-1}$ and rescaled to match the total ejected mass. The energy from the radioactive decay and the residual interaction is non-thermally deposited into the CDS, allowing an accurate treatment of the ionisation and excitation conditions. Although these simulations are not tailored to specific SNe, they allow us the exploration of spectral diversity and parameter degeneracies. Notably, \citet{Dessart2022A&A...658A.130D} demonstrated that similar spectral features can result from different combinations of CDS mass, radius, and input power.

The persistent presence of \ion{He}{i} lines throughout the spectral evolution suggests a helium-rich progenitor. Following the approach of \citet{Wang2024MNRAS.530.3906W}, we adopt helium-rich progenitor models for our analysis. Based on the helium-star evolutionary models of \citet{Woosley2019ApJ...878...49W}, progenitors with the zero-age helium core masses between 3--4~$M_\odot$ (e.g., models \texttt{he3} and \texttt{he4}) are appropriate for reproducing the observational characteristics of SNe~Ibn. For this study, we employ model \texttt{he4}, which is characterised by a total mass of 1.62~$M_\odot$, comprising 0.92~$M_\odot$ of helium, 0.31~$M_\odot$ of oxygen, 0.03~$M_\odot$ of magnesium, 0.0014~$M_\odot$ of calcium, and assumes solar metallicity.
Figure~\ref{fig:spectral_comparison_model} presents a comparison between synthetic spectra from model \texttt{he4p0} and observed spectra of five SNe~Ibn at multiple epochs after the maximum light. 
To isolate the effect of the luminosity evolution, only the power varies with time, while the CDS radius ($3\times10^{15}$~cm) and velocity (2000~km~s$^{-1}$) are fixed. This simplification enables a broad comparison with observed spectral evolution, although we note that adopting a time-dependent composition would better capture specific line changes. These power values are broadly consistent with the bolometric light curve inferred from observations (see Fig.~\ref{fig:Luc_model_comparison}).

The radiative-transfer models presented by \citet{Dessart2022A&A...658A.130D} are not designed to reproduce individual SNe~Ibn, but instead aim to explore the general spectral diversity arising from ejecta-CSM interaction across a broad parameter space. As such, their applicability to specific events remains limited. A notable example of this limitation lies in the prediction of strong Fe\,\textsc{ii} emission features blueward of 5500\,\AA\ in most models that do not assume nearly pure helium composition. However, such features are absent in several observed SNe~Ibn, including the 11.0\,d spectrum of SN~2020taz and the 16.3\,d spectrum of SN~2023utc. The presence of a well-defined continuum and the absence of Fe\,\textsc{ii} lines in these early spectra imply a dense and hot environment, inconsistent with the relatively cool, optically thin conditions required for Fe\,\textsc{ii} emission. These discrepancies suggest that the underlying assumptions of a quasi-steady, dense shell configuration may not hold at early phases, when the ejecta and circumstellar material have not yet formed a coherent, shocked shell. 
In such cases, models with a smaller shell radius and a higher power, implying higher density and temperature, are likely to provide a more accurate representation of the ionisation conditions (as shown in Appendix~\ref{appendix:model_spectra}, Fig.~\ref{fig:model_spectra_low}, where a model with a reduced radius better reproduces the observed spectra of SN\,2020taz).
However, the modelling framework adopted by \citet{Dessart2022A&A...658A.130D} does not accommodate this early-time hydrodynamical evolution. Therefore, caution must be adopted when interpreting early-time spectra using these models, and alternative approaches incorporating time-dependent hydrodynamics may be required to fully capture the physical conditions in the pre-shock or early interaction stages.

The models of \citet{Dessart2022A&A...658A.130D} do not include hydrogen, and therefore fail to reproduce Balmer lines such as the prominent H$\alpha$ feature observed in the 29.0\,d spectrum of SN\,2020taz. The presence of hydrogen can significantly modify the ionisation balance and increase the optical depth, thereby affecting the emergent spectrum. Additionally, synthetic spectral lines in these models are sometimes broader than those observed. This mismatch may be attenuated by adopting lower velocities for the CDS in tailored models, which would yield narrower features while preserving the overall spectral morphology. Another recurring discrepancy is the presence of central absorption dips in several He\,\textsc{i} lines (e.g., He\,\textsc{i}\,$\lambda5876$) in the synthetic spectra. These features arise due to a marginal optical thickness in the shell and are not frequently present in the observed spectra of SNe~Ibn. Their absence may reflect small-scale asymmetries or clumping in the real ejecta-CSM interaction region, which are not captured by the current spherically symmetric, one-dimensional modelling approach. Incorporating multidimensional effects or introducing shell inhomogeneities may help resolve this tension.

A detailed analysis of SN\,2020nxt has already been presented by \citet{Wang2024MNRAS.530.3906W}, and our results are broadly consistent with their findings. Here, we briefly summarise a few key aspects. Synthetic spectra computed at 13.5, 16.8, 28.7, and 38.8\,d post-maximum, using power inputs from $4 \times 10^{42}$ to $5 \times 10^{41}$\,erg\,s$^{-1}$, reproduce the temporal strengthening of the Ca\,\textsc{ii} near-infrared triplet due to decreasing ionisation and optical depth. This feature, absent in the models of \citet{Dessart2022A&A...658A.130D}, is explained here as an ionisation effect rather than a calcium overabundance. While Mg\,\textsc{ii} lines are consistently overestimated, their declining strength with time is qualitatively captured. Additionally, models underestimate the strength of metal lines blueward of 5300\,\AA\ in SN\,2020nxt at 13.5\,d, while overestimating them in the 19.9\,d spectrum of SN\,2021bbv. Although some discrepancies remain in reproducing individual features, the overall spectral evolution is broadly consistent with observations, supporting the scenario of ejecta--CSM interaction in SNe~Ibn.

\section{Discussion and concluding remarks} \label{sec:Discussion}
\subsection{Observables and physical parameters}

In this paper, a detailed analysis of the photometric and spectroscopic properties of five type~Ibn SNe was presented. The main outcomes can be summarised here:

\begin{itemize}
\item The rise times of their light curves to maximum brightness range from 6 to 12~days, which is typical within the SN~Ibn population (see Fig.~\ref{fig:phase_space} and Table~\ref{tab:LC_parameters}). 
\item At maximum light, our sample exhibits a wide range of luminosities, with peak magnitudes spanning from $M_r \sim -16$~mag for SN~2023utc to $M_r \sim -19$~mag for SN~2024aej. These provide pseudo-bolometric luminosities in the range of $(1$--$10) \times 10^{42}~\mathrm{erg~s^{-1}}$ and radiated energies $(1$--$10) \times 10^{48}~\mathrm{erg}$, with light curve durations spanning from 25 to $170~\mathrm{days}$.
\item The post-peak light curve decline shows two distinct trends. In some cases, the decline is initially steep, with rates of $\gamma(r) \approx 9$--$16~\mathrm{mag~(100~days)^{-1}}$, followed by a slower decline with $\gamma(r) \approx 4$--$7~\mathrm{mag~(100~days)^{-1}}$ (e.g., SN~2020nxt, SN~2021bbv, and SN~2023utc; see Section~\ref{SubSubSec:Apparent light curves}). In other cases, a shallow plateau phase is observed at early phases \citep[resembling the one observed in SN~2020bqj;][]{Kool2021A&A...652A.136K}, followed by a faster decline (e.g., SN~2020bqj with $\gamma(r) \approx 5.5~\mathrm{mag~(100~days)^{-1}}$ and SN~2020taz with $\gamma(r) \approx 24~\mathrm{mag~(100~days)^{-1}}$). 
\item The colour evolution of our sample is consistent with that observed in other type~Ibn SNe. The nearly constant colour observed in most SNe Ibn is indicative of little temperature evolution across this SN type, although with some exceptions (see Fig.~\ref{fig:colour_evolution}). The similar colour evolution may reflect a shared power source in the Type Ibn SN sample, driven by a sustained interaction between the ejecta and the CSM.
\item  
We performed multi-band light-curve modelling using \texttt{MOSFiT}, adopting the \texttt{RD+CSI} framework to constrain the properties of both the ejecta and the CSM. The inferred ejecta masses ($M_{\rm ej}$) span a relatively broad range of $\sim$1–3\,$M_{\odot}$, with corresponding kinetic energies ($E_{\rm Kin}$) on the order of $(0.1$–$1) \times 10^{51}$\,erg. For the CSM, the shock-swept mass ($M_{\rm CSM}$) is estimated to lie between $\sim$0.2 and 1\,$M_{\odot}$, with inner radii in the range of $\sim$10–50\,AU. The upper limits on the synthesised $^{56}$Ni mass for all five SNe~Ibn in our sample are $\lesssim$0.2\,$M_{\odot}$, consistent with values inferred for other SNe~Ibn in the literature (e.g. \citealt{Pellegrino2022ApJ...926..125P}; \citealt{Kool2021A&A...652A.136K}; \citealt{Farias2024ApJ...977..152F}). We note a general trend whereby fainter and more slowly evolving SNe Ibn tend to exhibit lower explosion energies—regulated primarily by the ejecta mass ($M_{\rm ej}$) and velocity ($v_{\rm ej}$)—as well as reduced amounts of synthesised $^{56}$Ni.
\item Spectroscopically, the sample exhibits relatively slow spectral evolution, characterised by a hot blue continuum with superimposed  prominent He~\textsc{i} emission lines. Most spectra exhibit blackbody temperatures exceeding $10000~\mathrm{K}$ during the early phases, with a decline at later stages. Narrow He~\textsc{i} lines, indicative of unshocked CSM, exhibit velocities of $\sim~1000~\mathrm{km~s^{-1}}$. In some cases, such as SN~2020nxt, a weak and narrow H$\alpha$ line with a P~Cygni profile ($v \sim 100$--$300~\mathrm{km~s^{-1}}$) is detected. However, other Balmer lines as well as spectral features typical of thermonuclear supernovae are absent or very weak. A common characteristic of the sample is the dominance of a blue pseudo-continuum in the mid to late-time  spectra, accompanied by a significant flux decline beyond $5400~\text{\AA}$.
\end{itemize}

\subsection{Progenitor and explosion scenarios}

Despite several plausible scenarios proposed to explain the origins of type~Ibn SNe, their exact nature remains debated \citep{Pastorello2007Natur.447..829P,Sanders2013ApJ...769...39S,Maund2016ApJ...833..128M}. Hypotheses include thermonuclear explosions of helium white dwarfs, core-collapse events from moderate-mass helium stars in binary systems \citep{Maund2016ApJ...833..128M, Sun2020MNRAS.491.6000S}, and the explosions of massive Wolf-Rayet stars \citep{Pastorello2007Natur.447..829P,Maeda2022ApJ...927...25M}. 

\subsubsection{Thermonuclear explosions of helium white dwarfs} \label{Sec:thermonuclear}

The He-shell detonation on a white dwarf (sometimes labelled as Type~.Ia SN explosions) would potentially explain some of the properties observed in our sample. Our events belong to the well-populated group of fast-evolving SNe~Ibn, with most exhibiting rise times of less than 10 days. Candidate SNe~.Ia, such as SN~2002bj \citep{Poznanski2010Sci...327...58P}, are also fast-evolving, He-rich transients with rapid light-curve evolution, characterised by rise times of $1-6$ days. Notably, dimmer type~.Ia SNe tend to exhibit faster rises. Their peak magnitudes, ranging from $M_V \sim -15$ to $-18$ mag \citep{Perets2010Natur.465..322P, Kasliwal2010ApJ...723L..98K, Perets2011ApJ...730...89P, Fesen2017ApJ...848..130F} are comparable to that of the faintest member of our sample, SN~2023utc.

On the other hand, other parameters determined for our sample, including the ejecta mass and the CSM properties, are inconsistent with the expectations for explosions originating from very low progenitor masses, such as thermonuclear supernovae from helium white dwarfs. Another argument against the white dwarf detonation scenario is the absence of spectroscopic features typically associated with type~Ia SNe, such as prominent \ion{S}{ii} and \ion{Si}{ii} lines.
For these reasons, we conclude that our sample is inconsistent with being thermonuclear explosions of He white dwarfs.

\subsubsection{Core-collapse SNe from moderate-mass He stars in binary systems}

While a thermonuclear origin cannot be definitely ruled out for a fraction of SNe Ibn \citep[see Sect. \ref{Sec:thermonuclear}, and discussion in][]{Sanders2013ApJ...769...39S}, observational evidences suggest that most SNe~Ibn are interacting CC SNe. In particular, the analysis of the spectra can provide key evidence for interpreting the explosion scenario. The spectral features identified in Fig.~\ref{fig:Line_identification} exhibit some similarity with CC SNe transitioning to the nebular phase. While the [O~\textsc{i}] $\lambda\lambda6300, 6364$ doublet is not definitively detected, the enhanced He~\textsc{i} $\lambda7281$ line may be attributed to the emergence of the [Ca~\textsc{ii}] $\lambda\lambda7291, 7324$ doublet, a hallmark feature of CC SNe in their nebular phase. While on-going CSM interaction may inhibit the detection of the [O~\textsc{i}] $\lambda\lambda6300, 6364$ doublet \citep[as suggested for Type IIn SNe similar to SN~2009ip; see, e.g.,][]{Brennan2022MNRAS.513.5666B}, its weakness may also be due to the lower mass of the progenitor's core. A plausible scenario for our sample is that they originate from the explosions of moderate-mass ($M_{\rm ZAMS}$~$<$~20~$M_\odot$ $-$ 25~$M_\odot$) He stars in binary systems, producing partially stripped CC~SNe. This interpretation is supported by the studies of recent SNe~Ibn.

\citet{Sun2020MNRAS.491.6000S} detected a point source at the location of the prototypical SN~Ibn 2006jc with the HST at late times, identifying it as the surviving binary companion with an estimated initial mass of $M_2 \leq 12.3^{+2.3}_{-1.5} \, M_\odot$, with the primary being only slightly more massive.
This finding suggests that SNe~Ibn may arise from interacting binary systems with moderate-mass progenitors. Additionally, we observe that most SNe~Ibn occupy a confined region in the relationships between light curve parameters, as shown in Fig.~\ref{fig:phase_space}, indicating some degree of homogeneity in the progenitor properties and hence the explosion scenario.

Furthermore, recent studies of SN~2023fyq have provided compelling evidence for a binary-origin scenario involving a low-mass helium star and a compact companion. \citet{Dong2024ApJ...977..254D} reported a three-year-long precursor activity prior to the explosion, followed by a double-peaked post-explosion light curve with a peak luminosity of $\sim10^{43}$\,erg\,s$^{-1}$. They interpret the precursor emission as a consequence of unstable mass transfer in a close binary system, which leads to the formation of a massive ($\sim$0.6\,M$_\odot$) equatorial disc. The interaction between the SN ejecta and this disc is proposed to power the second peak in the light curve. The early-time light curve suggests the presence of dense, extended material ($\sim$0.3\,M$_\odot$ at $\sim$3000\,R$_\odot$), likely ejected weeks before core collapse due to either silicon-burning instabilities or binary-driven mass loss.
\citet{Tsuna2024OJAp....7E..82T} provide further support this scenario by modelling the binary interaction between a $\sim$2.5--3\,M$_\odot$ helium star and a neutron star. Their simulations show that super-Eddington accretion onto the compact object can drive a long-lasting wind, producing optical/UV transients with luminosities of $\sim10^{40}$--$10^{41}$\,erg\,s$^{-1}$ over several years, consistent with the observed precursor of SN~2023fyq. The final explosion—either due to core collapse or a merger—produces an interaction-powered transient with characteristics resembling those of SNe~Ibn.
Although none of the SNe~Ibn in our sample shows evidence of pre-explosion outbursts or double-peaked light curves, their spectral features and light-curve properties—such as peak magnitudes and post-maximum decline rates—are broadly consistent with the expectations from binary-interaction scenarios. This suggests that they may represent a different manifestation of binary evolution leading to the Type Ibn phenomenon.

Additional insights supporting moderate-mass He progenitor systems for most SNe~Ibn are given by \cite{Wang2024MNRAS.530.3906W}, who performed radiative-transfer simulations on the spectral evolution of SN 2020nxt. Their results suggest that its progenitor was a 4~$M_{\odot}$ He-star, which lost $\sim$ 1~$M_{\odot}$ of its He-rich envelope prior to the explosion, implying an interacting binary system, which is in line with the outcomes of our modelling in Sect. \ref{Sec:MOSFiT}.

Finally, \citet{Dessart2022A&A...658A.130D} investigated the light-curve and spectral properties of Type Ibn SNe by performing both radiation-hydrodynamics simulations and NLTE radiative-transfer calculations, using the results of the former primarily to inform the initial conditions of the latter. The narrow spectral lines and moderate peak luminosities observed in most SNe~Ibn suggest that they originate from low-energy explosions of relatively low-mass ($\lesssim 5$~$M_\odot$) helium stars, likely in interacting binary systems. These progenitors are thought to collide with a dense, helium-rich CSM, possibly of ejecta origin, located at a radius of $\sim 10^{15}$\,cm. NLTE radiative-transfer models for the resulting slow-moving, dense shell formed and powered by ejecta-CSM interaction show good agreement with the late-time spectra of several SNe~Ibn, such as SN~2006jc, SN~2011hw, and SN~2018bcc. These models favour a shell composition of approximately 50\% helium, solar metallicity, and a total ejecta plus CSM mass of 1--2~$M_\odot$. A lower helium fraction in the shell suppresses He\,\textsc{i} lines, thus disfavouring higher-mass configurations for typical SNe~Ibn. Our bolometric light curves and spectral sequences are broadly consistent with the predictions from the \citet{Dessart2022A&A...658A.130D} models, supporting the applicability of their framework to our SN~Ibn sample.

\subsubsection{The explosions of massive Wolf-Rayet stars}

Another possible  scenario for SNe~Ibn is the explosion of higher-mass ($M_{\rm ZAMS}$~$>$~25~$M_\odot$) WR stars. This scenario provides a straightforward interpretation of the typical CSM composition and velocities observed in type~Ibn SNe, as well as the mass-loss events occurring prior to the explosion.
Key evidence supporting this interpretation is the wind velocity of $\sim$1000~km~s$^{-1}$ inferred for our SN sample, consistent with the speeds expected in WR stellar winds. Several studies have suggested that WR stars may experience fallback during their explosions, leading to the production of little or no $^{56}$Ni \citep{Woosley1995ApJS..101..181W, Zampieri1998ApJ...502L.149Z, Maeda2007ApJ...666.1069M, Moriya2010ApJ...719.1445M}. Furthermore, the absence of [O~\textsc{i}] $\lambda\lambda$6300, 6364 lines in late-time spectra, as noted by \citet{Valenti2009Natur.459..674V}, is an argument used for supporting fallback in the explosion of very massive stars.

Massive stars with initial masses exceeding $\sim$18~$M_{\odot}$ can lose significant amounts of mass via strong stellar winds without requiring a binary companion, leaving behind a C+O core surrounded by a He-rich CSM \citep{Heger2003ApJ...591..288H, Langer2012ARA&A..50..107L}. 
Stars with even larger masses (70 to 140~$M_{\odot}$) may undergo pulsational pair-instability (PPI) events.
In particular, some PPI models have been shown to reasonably reproduce the light curves of SNe Ibn  \citep{Woosley2017ApJ...836..244W, Karamehmetoglu2021A&A...649A.163K}.
However, a major challenge to this interpretation is that our sample appears to consist of single supernova-like events without evidence of prior eruptions. Pre-supernova outbursts have been observed in only a handful of Type~Ibn SNe, such as SN~2006jc \citep{Pastorello2007Natur.447..829P},
SN~2019uo \citep{Strotjohann2021ApJ...907...99S}, SN~2022pda (Cai et al., in preparation), SN~2023fyq \citep{Brennan2024A&A...684L..18B}, and the transitional Type Ibn/IIn SN~2021foa \citep{Reguitti2022AA...662L..10R, Farias2024ApJ...977..152F}. 
Unfortunately, as our sample is located in distant galaxies, we lack direct evidence of the progenitor stars or observations deep enough to capture their pre-explosion activity.

\subsection{Concluding remarks}

The available dataset for our sample does not provide sufficient constraints to definitively determine the progenitor masses or the explosion mechanisms of type~Ibn SNe. However, several lines of evidence suggest that, in most cases, these events are terminal CC SNe originating from massive stars. Whether the progenitors are high-mass WR stars in binary systems or lower-mass helium stars remains an open question. 
Future observational facilities, such as the Chinese Space Station Telescope (CSST)\footnote{\url{http://nao.cas.cn/csst/}} and the Vera C. Rubin Observatory,\footnote{\url{https://www.lsst.org/}} will be instrumental in improving the sampling frequency of type~Ibn SNe. These advancements will not only enhance our ability to detect and characterise such events, but also will play a crucial role in refining existing theoretical models and furthering our understanding of this enigmatic subclass of SNe.

\bibliographystyle{aa} 
\bibliography{Ibnref.bib} 
\begin{appendix}

\onecolumn 


\section{Host galaxy reddening}
\label{sec:NaID}
\begin{figure*}[htp]
\includegraphics[width=0.19\linewidth]{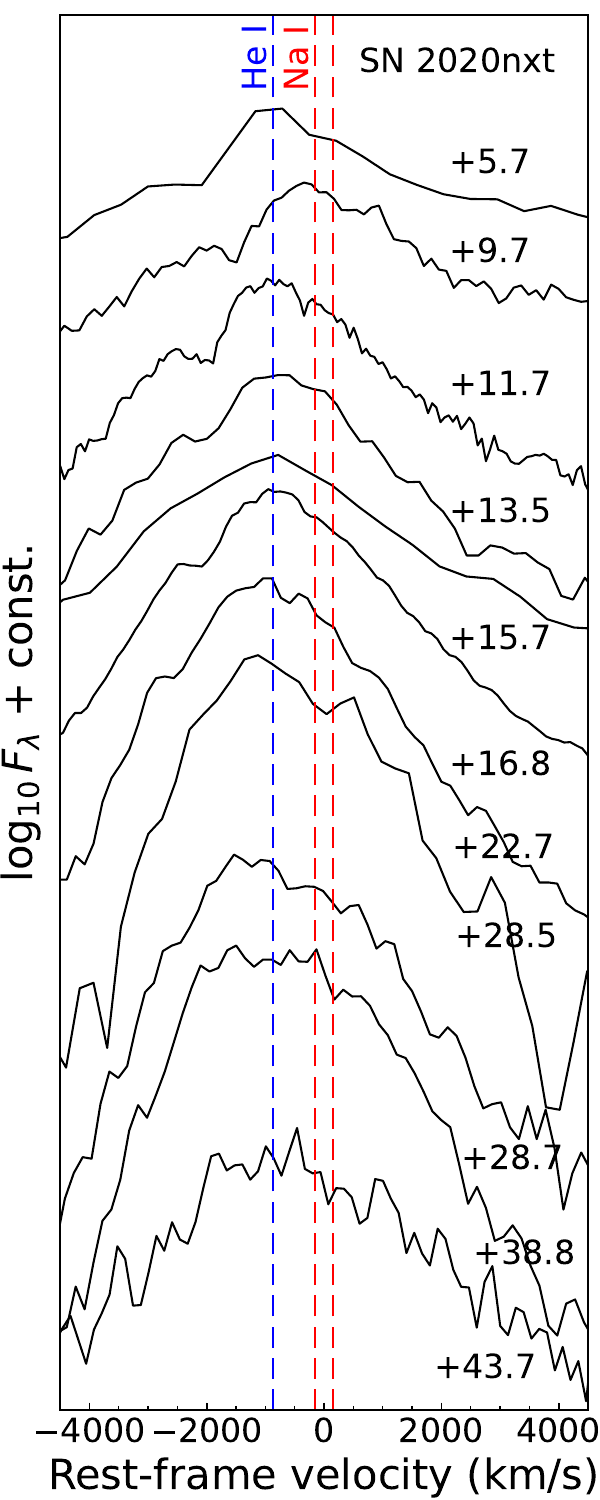}
\includegraphics[width=0.19\linewidth]{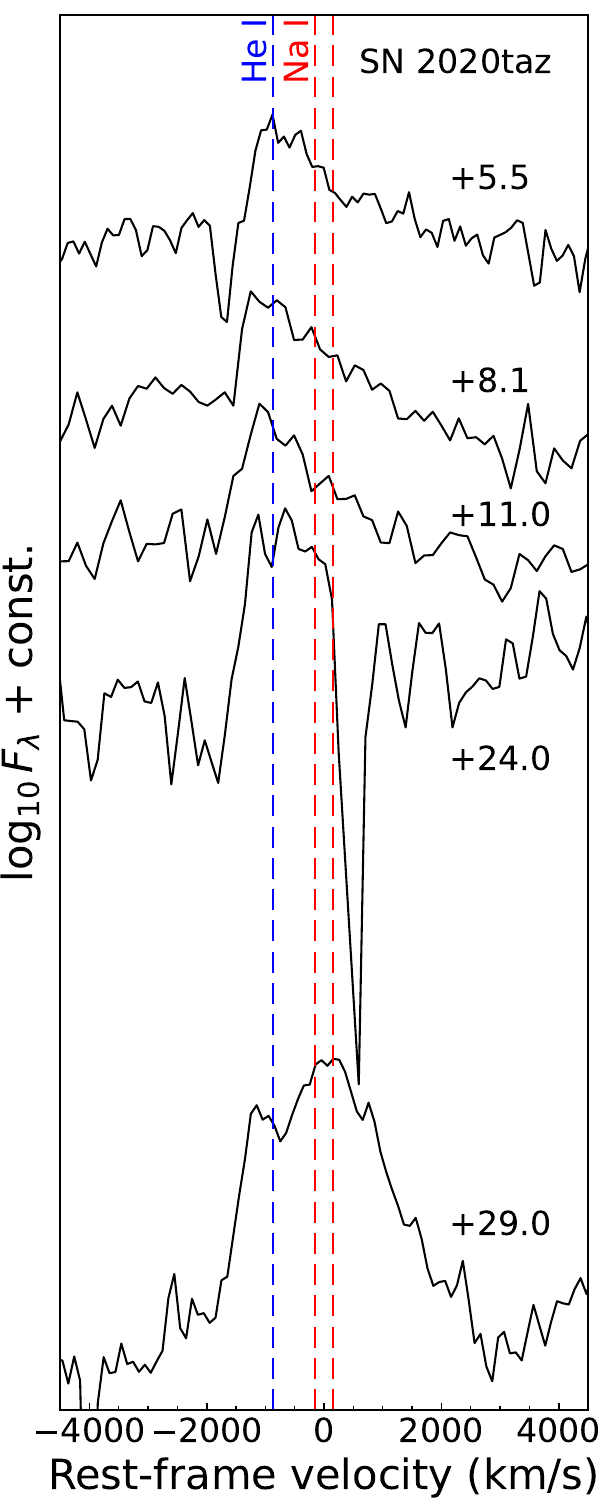}
\includegraphics[width=0.19\linewidth]{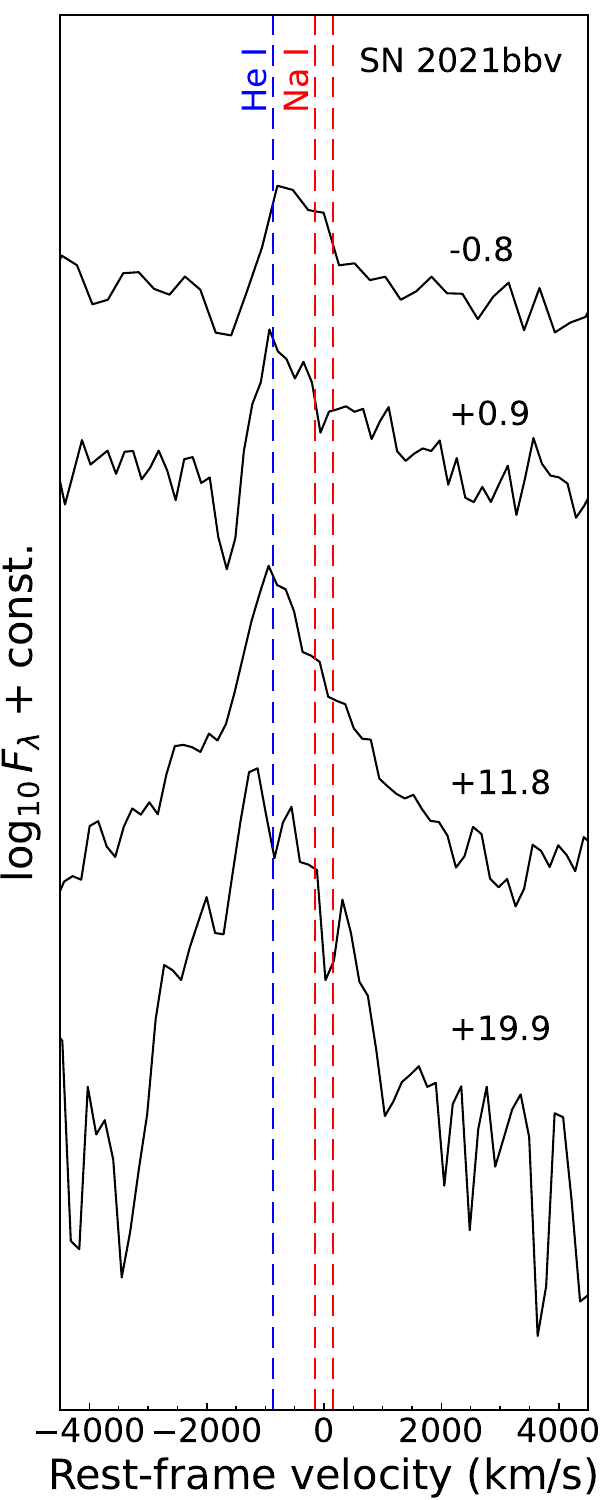}
\includegraphics[width=0.19\linewidth]{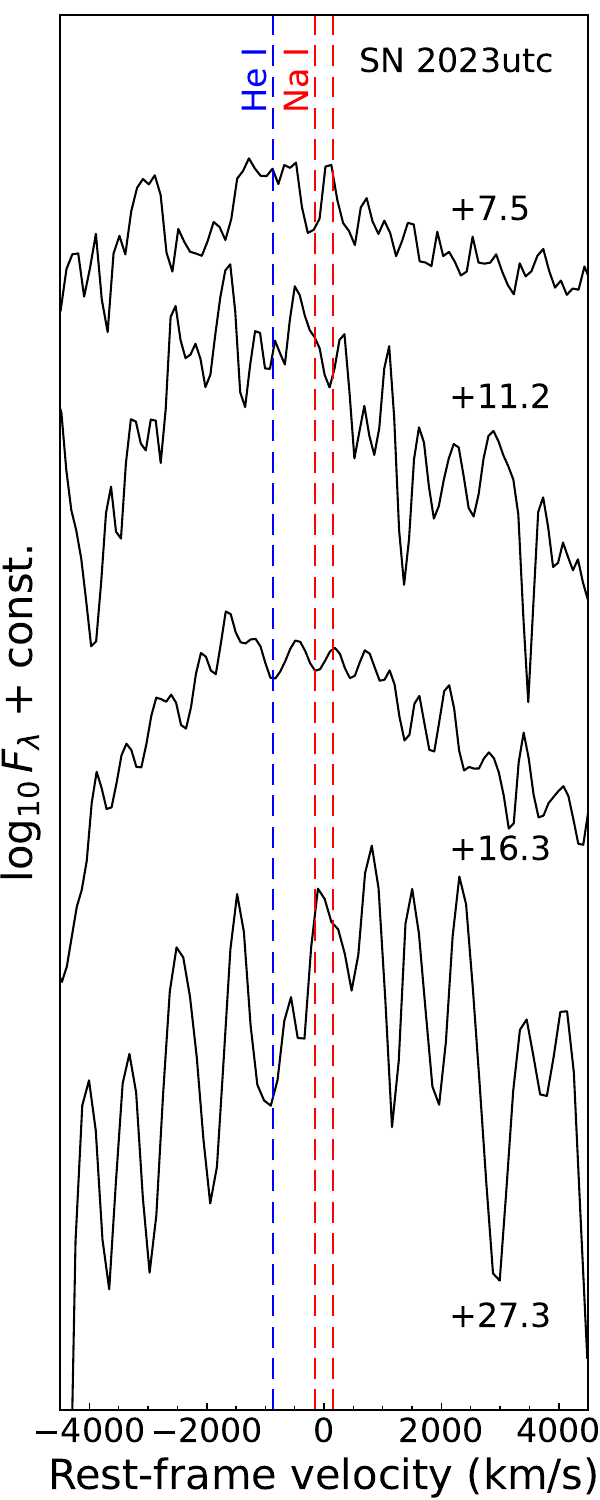}
\includegraphics[width=0.19\linewidth]{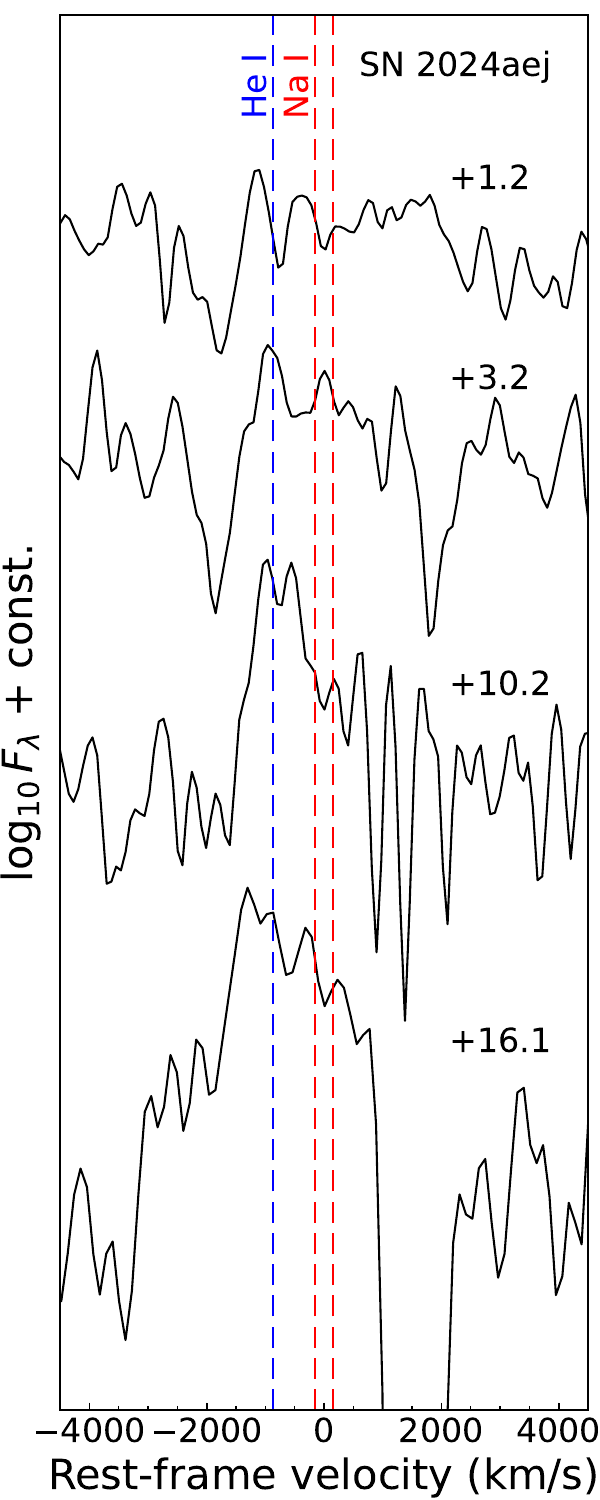}
\caption{Detail of the spectral evolution of SNe~2020nxt, 2020taz, 2021bbv, 2023utc, and 2024aej near the the expected position of the narrow interstellar Na\,\textsc{i} D absorption line (in the velocity space). 
$v = 0$ km s$^{-1}$ corresponds to the rest wavelength of the core of the Na\,\textsc{i} D line.
}
\label{fig:NaID}
\end{figure*}

We investigated the potential reddening contribution of dust within host galaxies by searching for signatures of Na\,\textsc{i}\,D absorption doublet features in the spectra of the five SNe~Ibn of our sample. In Fig.~\ref{fig:NaID}, for each SN we show the evolution of the spectral region where the narrow (interstellar) Na\,\textsc{i}\,D line is expected to be found, in the velocity space. In all cases, the features are not detected securely. For all spectra of each SN, we estimate the EW of the noise patterns close to the expected position of the narrow host-galaxy Na\,\textsc{i}\,D absorption. Using all spectra, we estimate the average value for the EW of these patterns ($\overline{EW}$) and compute the standard deviation $\sigma$. The conservative upper limit $3\sigma$ is then derived as $EW_\mathrm{upper} < \overline{EW} + 3 \times \mathrm{\sigma}$. The upper limits are then used to constrain the host-galaxy reddening limits using the empirical relation of \citet{Poznanski2012MNRAS.426.1465P}. As the empirical relationship of \citet{Poznanski2012MNRAS.426.1465P} tends to saturate for EW $\gtrsim 0.8$\,\AA, we adopt the prescription of \citet{Turatto2003fthp.conf..200T} for cases exceeding this threshold. The resulting upper limits are the following.

\begin{itemize}
    \item SN~2020nxt: EW~$\lesssim 0.69$\,\AA, $E(B-V)_{\mathrm{host}} \lesssim 0.09$\,mag
    \item SN~2020taz: EW~$\lesssim 0.30$\,\AA, $E(B-V)_{\mathrm{host}} \lesssim 0.03$\,mag
    \item SN~2021bbv: EW~$\lesssim 0.40$\,\AA, $E(B-V)_{\mathrm{host}} \lesssim 0.04$\,mag
    \item SN~2023utc: EW~$\lesssim 1.66$\,\AA, $E(B-V)_{\mathrm{host}} \lesssim 0.26$\,mag
    \item SN~2024aej: EW~$\lesssim 1.30$\,\AA, $E(B-V)_{\mathrm{host}} \lesssim 0.20$\,mag
\end{itemize}

The above upper limits to the colour excess due to the host galaxy dust attenuation are not very stringent, implying that - at least for SNe 2023utc and 2024aej - a non-negligible host galaxy extinction cannot be ruled out in principle.
However, most SNe Ibn in our sample (including SNe 2023utc and 2024aej) are hosted in very faint dwarf galaxies or remote locations from the host nucleus, where large line-of-sight reddening values are not expected. Hence, while reliable estimates for $E(B-V)_{\mathrm{host}}$ are not available, the position of the SNe in their host galaxies and/or their morphologic types support the choice of assuming a negligible host galaxy reddening for all objects of our sample.

\newpage
\section[]{Basic information for observational facilities used for the five SNe Ibn}
\label{sec:facilities}
We report the basic information for observational facilities in Table \ref{table_telescope}, which were used for the five SNe Ibn.

\begin{table*}[h]
\centering
\caption{Information on the instrumental setups.}
\label{table_telescope}
\scalebox{0.8}{
\begin{tabular}{@{}lllll@{}}
\hline \hline
Code & Diameter&Telescope & Instrument & Site \\
                & $\mathrm{m}$&            &                   & \\
\hline
Moravian       & 0.67/0.92 & Schmidt Telescope            & Moravian   &  Osservatorio Astronomico di Asiago, Asiago, Italy\\
TNT            & 0.80      &  Tsinghua-NAOC Telescope     & Andor DZ936& Xinglong Observatory, Hebei Province, China \\
fa03$^*$       & 1.00      & LCO (LSC site)               & Sinistro   & LCO node at Cerro Tololo Inter-American Observatory, Cerro Tololo, Chile\\
fa05$^*$       & 1.00      & LCO (ELP site)               & Sinistro   & LCO node at McDonald Observatory, Texas, USA\\
fa06$^*$       & 1.00      & LCO (CPT site)               & Sinistro   & LCO node at South African Astronomical Observatory, Cape Town, South Africa\\
fa07$^*$       & 1.00      & LCO (ELP site)               & Sinistro   & LCO node at McDonald Observatory, Texas, USA\\
fa08$^*$       & 1.00      & LCO (ELP site)               & Sinistro   & LCO node at McDonald Observatory, Texas, USA\\ 
fa11$^*$       & 1.00      & LCO (TFN site)               & Sinistro   & LCO node at Teide Observatory, Tenerife, Spain\\
fa16$^*$       & 1.00      & LCO (ELP site)               & Sinistro   & LCO node at McDonald Observatory, Texas, USA\\
fa19$^*$       & 1.00      & LCO (COJ site)               & Sinistro   & LCO node at Siding Spring Observatory, New South Wales, Australia  \\
fa20$^*$       & 1.00      & LCO (TFN site)               & Sinistro   & LCO node at Teide Observatory, Tenerife, Spain\\
AFOSC          & 1.82      & Copernico Telescope          & AFOSC      & Osservatorio Astronomico di Asiago, Asiago, Italy    \\
IO:O           & 2.00      & Liverpool Telescope          & IO:O       &  Observatorio Roque de Los Muchachos, La Palma, Spain\\
SPRAT          & 2.00      & Liverpool Telescope          & SPRAT      &  Observatorio Roque de Los Muchachos, La Palma, Spain\\
en06$^*$       & 2.00      & Faulkes Telescope North      & FLOYDS     & LCO node at Haleakala Observatory, Maui, USA\\
ALFOSC         & 2.56      & Nordic Optical Telescope     & ALFOSC     &  Observatorio Roque de Los Muchachos, La Palma, Spain\\
DIS            & 3.50      & APO 3.5m telescope           & DIS        & Apache Point Observatory, New Mexico, USA\\
DOLORES        & 3.58      & Telescopio Nazionale Galileo & DOLORES    & Roque de los Muchachos Observatory, La Palma, Spain\\
EFOSC2         & 3.58      & New Technology Telescope     & EFOSC2     & ESO La Silla Observatory, La Silla, Chile\\
KOOLS          & 3.80      & Seimei Telescope             & KOOLS-IFU  & Okayama Astrophysical Observatory, Okayama, Japan\\
OSIRIS         & 10.40     & Gran Telescopio CANARIAS     & OSIRIS     & Observatorio Roque de Los Muchachos, La Palma, Spain\\
\hline \hline
\end{tabular}
}
\medskip
\\ 
\begin{flushleft}
$*$ They are distributed globally at different sites and form part of the LCO global telescope network \citep{Brown2013PASP..125.1031B}. These data come from the Global Supernova Project. \\ 
\end{flushleft}
\end{table*}

\newpage 
\section{Light curve of magnitudes}
\label{tables:LC_mag}
\begin{longtable}{ccccccc}
    \caption{Optical and NIR observed magnitudes of SN\,2020nxt.} 
    \label{table:mag_2020nxt}\\
        \hline \hline
        Date & MJD & Filter & MagType & Magnitude & Error & Instrument/Source \\
        \hline
    \endfirsthead

    \multicolumn{7}{c}%
    {{\tablename\ \thetable{} -- continued from previous page}} \\
        \hline \hline
        Date & MJD & Filter & MagType & Magnitude & Error & Instrument/Source \\
        \hline
    \endhead

        \hline \hline
    \multicolumn{7}{r}{{Continued on next page}} \\
    \endfoot

        \hline \hline
    \endlastfoot

20200605&59005.56&$o$&AB&> 19.4&-&ATLAS\\
20200610&59010.60&$o$&AB&> 20.2&-&ATLAS\\
20200614&59014.57&$o$&AB&> 20.3&-&ATLAS\\
20200617&59017.60&$o$&AB&> 20.6&-&ATLAS\\
20200618&59018.58&$o$&AB&> 19.6&-&ATLAS\\
20200621&59021.54&$o$&AB&> 21.0&-&ATLAS\\
20200626&59026.55&$o$&AB&> 20.6&-&ATLAS\\
20200627&59027.53&$c$&AB&> 20.8&-&ATLAS\\
20200630&59030.56&$o$&AB&> 20.8&-&ATLAS\\
20200701&59031.44&$g$&AB&> 17.3&-&ASAS-SN\\
20200701&59031.58&$c$&AB&20.570&0.230&ATLAS\\
20200702&59032.53&$g$&AB&18.280&0.236&ASAS-SN\\
20200703&59033.56&$o$&AB&17.250&0.020&ATLAS\\
20200704&59034.61&$o$&AB&16.640&0.010&ATLAS\\
20200705&59035.60&$o$&AB&16.280&0.010&ATLAS\\
20200707&59037.48&$g$&AB&15.857&0.057&ASAS-SN\\
20200707&59037.52&$o$&AB&16.000&0.010&ATLAS\\
20200709&59039.50&$o$&AB&15.990&0.010&ATLAS\\
20200709&59039.61&$g$&AB&16.090&0.051&ASAS-SN\\
20200711&59041.16&$r$&AB&16.251&0.033&IO:O\\
20200711&59041.52&$g$&AB&16.469&0.071&ASAS-SN\\
20200712&59042.52&$o$&AB&16.340&0.010&ATLAS\\
20200713&59043.51&$g$&AB&16.545&0.061&ASAS-SN\\
20200715&59045.16&$u$&AB&16.666&0.011&IO:O\\
20200715&59045.16&$g$&AB&16.648&0.009&IO:O\\
20200715&59045.16&$r$&AB&16.811&0.009&IO:O\\
20200715&59045.16&$i$&AB&16.726&0.010&IO:O\\
20200715&59045.16&$z$&AB&16.642&0.019&IO:O\\
20200716&59046.12&$U$&Vega&15.818&0.053&Swift\\
20200716&59046.12&$B$&Vega&16.879&0.069&Swift\\
20200716&59046.12&$V$&Vega&16.772&0.127&Swift\\
20200716&59046.12&$UVW2$&Vega&16.878&0.071&Swift\\
20200716&59046.12&$UVM2$&Vega&16.614&0.124&Swift\\
20200716&59046.12&$UVW1$&Vega&16.306&0.068&Swift\\
20200716&59046.24&$g$&AB&16.934&0.086&ASAS-SN\\
20200716&59046.52&$o$&AB&17.150&0.060&ATLAS\\
20200717&59047.57&$c$&AB&17.040&0.010&ATLAS\\
20200718&59048.55&$g$&AB&17.202&0.073&ASAS-SN\\
20200718&59048.85&$U$&Vega&16.335&0.095&Swift\\
20200718&59048.85&$B$&Vega&17.282&0.128&Swift\\
20200718&59048.85&$V$&Vega&17.111&0.211&Swift\\
20200718&59048.85&$UVW2$&Vega&17.405&0.125&Swift\\
20200718&59048.85&$UVM2$&Vega&17.285&0.113&Swift\\
20200718&59048.85&$UVW1$&Vega&16.902&0.114&Swift\\
20200719&59049.06&$B$&Vega&17.228&0.010&ALFOSC\\
20200719&59049.06&$V$&Vega&17.267&0.020&ALFOSC\\
20200719&59049.06&$u$&AB&17.641&0.015&ALFOSC\\
20200719&59049.06&$g$&AB&17.069&0.004&ALFOSC\\
20200719&59049.06&$r$&AB&17.499&0.028&ALFOSC\\
20200719&59049.06&$i$&AB&17.373&0.011&ALFOSC\\
20200719&59049.06&$z$&AB&17.166&0.013&ALFOSC\\
20200719&59049.97&$B$&Vega&17.215&0.040&Moravian\\
20200719&59049.97&$V$&Vega&17.543&0.029&Moravian\\
20200719&59049.97&$u$&AB&17.787&0.115&Moravian\\
20200719&59049.97&$g$&AB&17.210&0.034&Moravian\\
20200719&59049.97&$r$&AB&17.736&0.061&Moravian\\
20200719&59049.97&$i$&AB&17.637&0.117&Moravian\\
20200720&59050.11&$U$&Vega&16.631&0.085&Swift\\
20200720&59050.11&$B$&Vega&17.540&0.117&Swift\\
20200720&59050.11&$V$&Vega&17.310&0.205&Swift\\
20200720&59050.11&$UVW2$&Vega&18.031&0.155&Swift\\
20200720&59050.11&$UVM2$&Vega&17.734&0.159&Swift\\
20200720&59050.11&$UVW1$&Vega&17.282&0.123&Swift\\
20200720&59050.51&$o$&AB&17.760&0.020&ATLAS\\
20200721&59051.51&$c$&AB&17.650&0.020&ATLAS\\
20200721&59051.51&$g$&AB&17.686&0.091&ASAS-SN\\
20200721&59051.90&$B$&Vega&17.575&0.032&Moravian\\
20200721&59051.90&$V$&Vega&17.806&0.034&Moravian\\
20200722&59052.57&$U$&Vega&17.129&0.120&Swift\\
20200722&59052.57&$B$&Vega&17.845&0.150&Swift\\
20200722&59052.57&$V$&Vega&> 18.0&-&Swift\\
20200722&59052.57&$UVW2$&Vega&18.344&0.198&Swift\\
20200722&59052.57&$UVM2$&Vega&18.048&0.189&Swift\\
20200722&59052.57&$UVW1$&Vega&17.799&0.176&Swift\\
20200722&59052.60&$c$&AB&17.920&0.040&ATLAS\\
20200723&59053.36&$g$&AB&17.840&0.152&ASAS-SN\\
20200723&59053.48&$g$&AB&17.721&0.015&PAN-STARRS\\
20200723&59053.48&$g$&AB&17.720&0.020&PAN-STARRS\\
20200723&59053.49&$i$&AB&20.910&0.130&PAN-STARRS\\
20200723&59053.49&$i$&AB&18.204&0.019&PAN-STARRS\\
20200723&59053.68&$U$&Vega&17.270&0.165&Swift\\
20200723&59053.68&$B$&Vega&18.011&0.212&Swift\\
20200723&59053.68&$V$&Vega&> 17.7&-&Swift\\
20200723&59053.68&$UVW2$&Vega&18.667&0.321&Swift\\
20200723&59053.68&$UVM2$&Vega&> 18.3&-&Swift\\
20200723&59053.68&$UVW1$&Vega&18.021&0.261&Swift\\
20200724&59054.12&$u$&AB&18.249&0.015&IO:O\\
20200724&59054.12&$g$&AB&17.906&0.024&IO:O\\
20200724&59054.12&$r$&AB&18.337&0.018&IO:O\\
20200724&59054.12&$i$&AB&18.307&0.012&IO:O\\
20200724&59054.12&$z$&AB&17.980&0.026&IO:O\\
20200724&59054.58&$o$&AB&18.400&0.050&ATLAS\\
20200725&59055.08&$B$&Vega&18.113&0.008&ALFOSC\\
20200725&59055.08&$V$&Vega&18.173&0.010&ALFOSC\\
20200725&59055.08&$u$&AB&18.462&0.087&ALFOSC\\
20200725&59055.08&$g$&AB&17.908&0.006&ALFOSC\\
20200725&59055.08&$r$&AB&18.460&0.040&ALFOSC\\
20200725&59055.08&$i$&AB&18.439&0.012&ALFOSC\\
20200725&59055.08&$z$&AB&18.074&0.021&ALFOSC\\
20200725&59055.45&$g$&AB&18.532&0.154&ASAS-SN\\
20200725&59055.68&$U$&Vega&17.349&0.143&Swift\\
20200725&59055.68&$B$&Vega&18.235&0.210&Swift\\
20200725&59055.68&$V$&Vega&> 18.0&-&Swift\\
20200725&59055.68&$UVW2$&Vega&> 18.9&-&Swift\\
20200725&59055.68&$UVM2$&Vega&18.693&0.335&Swift\\
20200725&59055.68&$UVW1$&Vega&18.192&0.241&Swift\\
20200725&59055.90&$B$&Vega&18.210&0.051&Moravian\\
20200725&59055.90&$V$&Vega&18.272&0.051&Moravian\\
20200725&59055.90&$u$&AB&18.400&0.197&Moravian\\
20200725&59055.90&$g$&AB&18.029&0.036&Moravian\\
20200725&59055.90&$r$&AB&18.616&0.052&Moravian\\
20200725&59055.90&$i$&AB&18.555&0.137&Moravian\\
20200726&59056.07&$u$&AB&18.458&0.020&IO:O\\
20200726&59056.07&$g$&AB&18.177&0.011&IO:O\\
20200726&59056.07&$r$&AB&18.680&0.018&IO:O\\
20200726&59056.07&$i$&AB&18.680&0.015&IO:O\\
20200726&59056.07&$z$&AB&18.203&0.022&IO:O\\
20200726&59056.62&$U$&Vega&17.482&0.158&Swift\\
20200726&59056.62&$B$&Vega&18.375&0.210&Swift\\
20200726&59056.62&$V$&Vega&18.000&0.360&Swift\\
20200726&59056.62&$UVW2$&Vega&18.775&0.285&Swift\\
20200726&59056.62&$UVM2$&Vega&> 18.8&-&Swift\\
20200726&59056.62&$UVW1$&Vega&18.161&0.234&Swift\\
20200727&59057.14&$u$&AB&18.630&0.018&IO:O\\
20200727&59057.14&$g$&AB&18.085&0.015&IO:O\\
20200727&59057.14&$r$&AB&18.789&0.030&IO:O\\
20200727&59057.14&$i$&AB&18.718&0.021&IO:O\\
20200727&59057.14&$z$&AB&18.384&0.045&IO:O\\
20200727&59057.47&$U$&Vega&17.641&0.171&Swift\\
20200727&59057.47&$B$&Vega&18.480&0.219&Swift\\
20200727&59057.47&$V$&Vega&> 18.0&-&Swift\\
20200727&59057.47&$UVW2$&Vega&> 19.1&-&Swift\\
20200727&59057.47&$UVM2$&Vega&> 18.9&-&Swift\\
20200727&59057.47&$UVW1$&Vega&18.467&0.263&Swift\\
20200728&59058.04&$B$&Vega&18.505&0.087&Moravian\\
20200728&59058.04&$V$&Vega&18.525&0.125&Moravian\\
20200728&59058.05&$u$&AB&18.772&0.096&Moravian\\
20200728&59058.05&$g$&AB&18.210&0.047&Moravian\\
20200728&59058.05&$r$&AB&18.954&0.134&Moravian\\
20200728&59058.05&$i$&AB&18.821&0.127&Moravian\\
20200728&59058.35&$g$&AB&17.610&0.192&ASAS-SN\\
20200729&59059.31&$g$&AB&> 18.4&-&ASAS-SN\\
20200729&59059.74&$U$&Vega&17.743&0.192&Swift\\
20200729&59059.74&$B$&Vega&18.670&0.261&Swift\\
20200729&59059.74&$V$&Vega&> 18.0&-&Swift\\
20200729&59059.74&$UVW2$&Vega&> 19.1&-&Swift\\
20200729&59059.74&$UVM2$&Vega&> 18.8&-&Swift\\
20200730&59060.06&$B$&Vega&18.765&0.026&IO:O\\
20200730&59060.06&$V$&Vega&18.720&0.021&IO:O\\
20200730&59060.06&$u$&AB&19.139&0.029&IO:O\\
20200730&59060.06&$g$&AB&18.458&0.011&IO:O\\
20200730&59060.06&$r$&AB&19.139&0.029&IO:O\\
20200730&59060.06&$i$&AB&19.107&0.020&IO:O\\
20200730&59060.06&$z$&AB&18.646&0.036&IO:O\\
20200730&59060.24&$g$&AB&> 17.7&-&ASAS-SN\\
20200730&59060.45&$o$&AB&19.140&0.120&ATLAS\\
20200730&59060.46&$U$&Vega&17.785&0.199&Swift\\
20200730&59060.46&$V$&Vega&> 18.0&-&Swift\\
20200730&59060.46&$UVW2$&Vega&> 19.1&-&Swift\\
20200730&59060.46&$UVM2$&Vega&> 18.8&-&Swift\\
20200730&59060.46&$UVW1$&Vega&18.565&0.297&Swift\\
20200730&59060.59&$c$&AB&18.830&0.050&ATLAS\\
20200731&59061.45&$o$&AB&19.160&0.090&ATLAS\\
20200731&59061.45&$g$&AB&> 18.2&-&ASAS-SN\\
20200801&59062.57&$g$&AB&> 18.1&-&ASAS-SN\\
20200803&59064.47&$o$&AB&19.460&0.160&ATLAS\\
20200803&59064.51&$i$&AB&19.319&0.131&PAN-STARRS\\
20200803&59064.51&$r$&AB&19.348&0.089&PAN-STARRS\\
20200804&59065.04&$B$&Vega&19.304&0.059&IO:O\\
20200804&59065.04&$V$&Vega&19.208&0.047&IO:O\\
20200804&59065.19&$g$&AB&> 16.6&-&ASAS-SN\\
20200804&59065.55&$o$&AB&19.370&0.150&ATLAS\\
20200805&59066.05&$U$&Vega&18.102&0.257&Swift\\
20200805&59066.05&$B$&Vega&> 18.9&-&Swift\\
20200805&59066.05&$V$&Vega&> 18.0&-&Swift\\
20200805&59066.05&$UVW2$&Vega&> 19.2&-&Swift\\
20200805&59066.05&$UVM2$&Vega&> 19.2&-&Swift\\
20200805&59066.05&$UVW1$&Vega&> 18.9&-&Swift\\
20200805&59066.08&$u$&AB&19.490&0.062&IO:O\\
20200805&59066.08&$g$&AB&18.947&0.016&IO:O\\
20200805&59066.08&$r$&AB&19.527&0.055&IO:O\\
20200805&59066.08&$i$&AB&19.560&0.031&IO:O\\
20200805&59066.08&$z$&AB&19.094&0.061&IO:O\\
20200805&59066.51&$z$&AB&19.086&0.233&PAN-STARRS\\
20200805&59066.98&$B$&Vega&19.375&0.061&Moravian\\
20200805&59066.98&$V$&Vega&19.270&0.095&Moravian\\
20200805&59066.99&$u$&AB&19.632&0.154&Moravian\\
20200805&59066.99&$g$&AB&18.966&0.059&Moravian\\
20200805&59066.99&$r$&AB&19.601&0.110&Moravian\\
20200805&59066.99&$i$&AB&19.638&0.129&Moravian\\
20200806&59067.03&$B$&Vega&19.428&0.061&IO:O\\
20200806&59067.03&$V$&Vega&19.350&0.050&IO:O\\
20200806&59067.22&$g$&AB&> 17.1&-&ASAS-SN\\
20200806&59067.38&$U$&Vega&> 18.3&-&Swift\\
20200806&59067.38&$V$&Vega&> 17.7&-&Swift\\
20200806&59067.38&$UVW2$&Vega&> 18.9&-&Swift\\
20200806&59067.38&$UVM2$&Vega&> 18.7&-&Swift\\
20200806&59067.38&$UVW1$&Vega&> 18.7&-&Swift\\
20200807&59068.96&$B$&Vega&19.509&0.088&Moravian\\
20200808&59069.06&$u$&AB&19.673&0.029&IO:O\\
20200808&59069.06&$g$&AB&19.297&0.016&IO:O\\
20200808&59069.06&$r$&AB&19.799&0.022&IO:O\\
20200808&59069.06&$i$&AB&19.871&0.022&IO:O\\
20200808&59069.06&$z$&AB&19.544&0.070&IO:O\\
20200808&59069.40&$g$&AB&17.788&0.203&ASAS-SN\\
20200808&59069.47&$o$&AB&19.310&0.150&ATLAS\\
20200809&59070.92&$B$&Vega&19.651&0.097&Moravian\\
20200809&59070.92&$V$&Vega&19.533&0.121&Moravian\\
20200810&59071.39&$g$&AB&18.777&0.211&ASAS-SN\\
20200810&59071.47&$o$&AB&19.760&0.140&ATLAS\\
20200810&59071.76&$U$&Vega&18.559&0.342&Swift\\
20200810&59071.76&$B$&Vega&> 19.0&-&Swift\\
20200810&59071.76&$V$&Vega&> 17.9&-&Swift\\
20200810&59071.76&$UVW2$&Vega&> 19.2&-&Swift\\
20200810&59071.76&$UVM2$&Vega&> 19.2&-&Swift\\
20200810&59071.76&$UVW1$&Vega&> 19.0&-&Swift\\
20200811&59072.04&$u$&AB&20.036&0.045&IO:O\\
20200811&59072.04&$g$&AB&19.618&0.047&IO:O\\
20200811&59072.04&$r$&AB&20.035&0.031&IO:O\\
20200811&59072.04&$i$&AB&20.070&0.029&IO:O\\
20200811&59072.04&$z$&AB&19.557&0.079&IO:O\\
20200812&59073.05&$B$&Vega&19.800&0.042&IO:O\\
20200812&59073.05&$V$&Vega&19.749&0.028&IO:O\\
20200812&59073.12&$B$&Vega&19.815&0.206&Moravian\\
20200812&59073.49&$i$&AB&20.078&0.094&PAN-STARRS\\
20200812&59073.50&$g$&AB&19.520&0.075&PAN-STARRS\\
20200812&59073.53&$o$&AB&19.990&0.160&ATLAS\\
20200813&59074.28&$U$&Vega&> 18.5&-&Swift\\
20200813&59074.28&$B$&Vega&> 18.8&-&Swift\\
20200813&59074.28&$V$&Vega&> 17.9&-&Swift\\
20200813&59074.28&$UVW2$&Vega&> 19.1&-&Swift\\
20200813&59074.28&$UVM2$&Vega&> 19.1&-&Swift\\
20200813&59074.28&$UVW1$&Vega&> 18.8&-&Swift\\
20200813&59074.85&$B$&Vega&19.894&0.157&Moravian\\
20200813&59074.85&$V$&Vega&19.786&0.112&Moravian\\
20200814&59075.23&$B$&Vega&19.934&0.032&IO:O\\
20200814&59075.23&$V$&Vega&19.831&0.039&IO:O\\
20200815&59076.14&$U$&Vega&> 18.5&-&Swift\\
20200815&59076.14&$B$&Vega&> 18.9&-&Swift\\
20200815&59076.14&$V$&Vega&> 18.0&-&Swift\\
20200815&59076.14&$UVW2$&Vega&> 19.1&-&Swift\\
20200815&59076.14&$UVM2$&Vega&> 19.2&-&Swift\\
20200815&59076.14&$UVW1$&Vega&> 18.9&-&Swift\\
20200815&59076.48&$g$&AB&19.726&0.046&PAN-STARRS\\
20200815&59076.48&$r$&AB&20.718&0.131&PAN-STARRS\\
20200815&59076.49&$c$&AB&19.690&0.230&ATLAS\\
20200816&59077.08&$B$&Vega&20.001&0.042&IO:O\\
20200816&59077.08&$V$&Vega&19.923&0.043&IO:O\\
20200816&59077.34&$U$&Vega&> 18.4&-&Swift\\
20200816&59077.34&$B$&Vega&> 18.8&-&Swift\\
20200816&59077.34&$V$&Vega&> 17.9&-&Swift\\
20200816&59077.34&$UVW2$&Vega&> 19.1&-&Swift\\
20200816&59077.34&$UVM2$&Vega&> 19.1&-&Swift\\
20200816&59077.34&$UVW1$&Vega&> 18.8&-&Swift\\
20200816&59077.49&$o$&AB&20.420&0.210&ATLAS\\
20200817&59078.04&$u$&AB&20.164&0.043&IO:O\\
20200817&59078.04&$g$&AB&19.865&0.021&IO:O\\
20200817&59078.04&$r$&AB&20.411&0.039&IO:O\\
20200817&59078.04&$i$&AB&20.425&0.021&IO:O\\
20200817&59078.04&$z$&AB&19.885&0.108&IO:O\\
20200818&59079.05&$B$&Vega&20.144&0.029&IO:O\\
20200818&59079.05&$V$&Vega&20.016&0.028&IO:O\\
20200818&59079.47&$c$&AB&19.700&0.090&ATLAS\\
20200819&59080.11&$u$&AB&> 20.1&-&Moravian\\
20200819&59080.42&$g$&AB&19.891&0.057&PAN-STARRS\\
20200819&59080.42&$i$&AB&20.527&0.116&PAN-STARRS\\
20200819&59080.44&$c$&AB&19.870&0.110&ATLAS\\
20200819&59080.82&$B$&Vega&20.171&0.254&Moravian\\
20200819&59080.82&$V$&Vega&20.163&0.199&Moravian\\
20200819&59080.84&$g$&AB&19.813&0.121&Moravian\\
20200819&59080.84&$r$&AB&> 19.9&-&Moravian\\
20200819&59080.84&$i$&AB&> 20.2&-&Moravian\\
20200821&59082.07&$B$&Vega&20.230&0.048&IO:O\\
20200821&59082.07&$V$&Vega&20.305&0.055&IO:O\\
20200821&59082.08&$u$&AB&20.314&0.045&IO:O\\
20200821&59082.08&$g$&AB&19.916&0.028&IO:O\\
20200821&59082.08&$r$&AB&20.473&0.046&IO:O\\
20200821&59082.08&$i$&AB&20.516&0.046&IO:O\\
20200821&59082.08&$z$&AB&19.998&0.124&IO:O\\
20200821&59082.14&$B$&Vega&20.212&0.018&LRS\\
20200821&59082.14&$V$&Vega&20.285&0.027&LRS\\
20200821&59082.15&$u$&AB&20.515&0.044&LRS\\
20200821&59082.15&$g$&AB&19.993&0.014&LRS\\
20200821&59082.15&$r$&AB&20.605&0.047&LRS\\
20200821&59082.15&$i$&AB&20.596&0.045&LRS\\
20200823&59084.48&$c$&AB&20.360&0.160&ATLAS\\
20200824&59085.45&$o$&AB&20.510&0.220&ATLAS\\
20200827&59088.43&$o$&AB&> 20.3&-&ATLAS\\
20200828&59089.03&$B$&Vega&21.095&0.082&ALFOSC\\
20200828&59089.03&$V$&Vega&21.184&0.101&ALFOSC\\
20200828&59089.04&$u$&AB&21.358&0.088&ALFOSC\\
20200828&59089.04&$g$&AB&20.772&0.020&ALFOSC\\
20200828&59089.04&$r$&AB&21.415&0.056&ALFOSC\\
20200828&59089.04&$i$&AB&21.306&0.054&ALFOSC\\
20200828&59089.04&$z$&AB&> 20.2&-&ALFOSC\\
20200829&59090.42&$o$&AB&> 20.5&-&ATLAS\\
20200831&59092.02&$u$&AB&21.395&0.093&ALFOSC\\
20200831&59092.02&$g$&AB&21.750&0.044&ALFOSC\\
20200831&59092.02&$r$&AB&22.307&0.105&ALFOSC\\
20200831&59092.02&$i$&AB&22.008&0.060&ALFOSC\\
20200831&59092.02&$z$&AB&> 20.2&-&ALFOSC\\
20200831&59092.04&$u$&AB&> 21.4&-&IO:O\\
20200831&59092.04&$g$&AB&> 21.0&-&IO:O\\
20200831&59092.04&$r$&AB&> 20.9&-&IO:O\\
20200831&59092.04&$i$&AB&> 21.1&-&IO:O\\
20200831&59092.04&$z$&AB&> 20.1&-&IO:O\\
20200901&59093.41&$o$&AB&> 19.7&-&ATLAS\\
20200902&59094.05&$B$&Vega&22.670&0.086&ALFOSC\\
20200902&59094.05&$V$&Vega&22.353&0.074&ALFOSC\\
20200902&59094.06&$u$&AB&> 21.6&-&ALFOSC\\
20200902&59094.06&$g$&AB&> 22.4&-&ALFOSC\\
20200902&59094.06&$r$&AB&> 22.1&-&ALFOSC\\
20200902&59094.06&$i$&AB&22.046&0.072&ALFOSC\\
20200902&59094.41&$o$&AB&> 20.0&-&ATLAS\\
20200902&59094.43&$z$&AB&21.428&0.387&PAN-STARRS\\
20200903&59095.49&$o$&AB&> 19.4&-&ATLAS\\
20200905&59097.51&$o$&AB&> 18.9&-&ATLAS\\
20200906&59098.44&$o$&AB&> 20.1&-&ATLAS\\
20200906&59098.96&$u$&AB&> 21.2&-&IO:O\\
20200906&59098.96&$g$&AB&> 22.1&-&IO:O\\
20200906&59098.96&$r$&AB&> 21.6&-&IO:O\\
20200906&59098.96&$i$&AB&> 20.8&-&IO:O\\
20200906&59098.96&$z$&AB&> 20.1&-&IO:O\\
20200908&59100.44&$o$&AB&> 20.4&-&ATLAS\\
20200909&59101.47&$o$&AB&> 21.0&-&ATLAS\\
20200910&59102.44&$o$&AB&> 20.9&-&ATLAS\\
20200911&59103.51&$o$&AB&> 19.6&-&ATLAS\\
20200912&59104.44&$c$&AB&> 21.2&-&ATLAS\\
20200913&59105.42&$o$&AB&> 19.8&-&ATLAS\\
20200915&59107.47&$c$&AB&> 20.7&-&ATLAS\\
20200919&59111.44&$c$&AB&> 20.3&-&ATLAS\\
20200922&59114.38&$o$&AB&> 20.0&-&ATLAS\\
20200926&59118.37&$o$&AB&> 20.4&-&ATLAS\\
20200927&59119.40&$o$&AB&> 20.4&-&ATLAS\\
20200928&59120.42&$o$&AB&> 19.6&-&ATLAS\\
20200929&59121.38&$o$&AB&> 19.8&-&ATLAS\\

\end{longtable}
\begin{longtable}{ccccccc}
    \caption{Optical observed magnitudes of SN\,2020taz.} 
    \label{table:mag_2020taz}\\
        \hline \hline
        Date & MJD & Filter & MagType & Magnitude & Error & Instrument/Source \\
        \hline
    \endfirsthead

    \multicolumn{7}{c}%
    {{\tablename\ \thetable{} -- continued from previous page}} \\
        \hline \hline
        Date & MJD & Filter & MagType & Magnitude & Error & Instrument/Source \\
        \hline
    \endhead

        \hline \hline
    \multicolumn{7}{r}{{Continued on next page}} \\
    \endfoot

        \hline \hline
    \endlastfoot

20200721&59051.54&$c$&AB&> 20.7&-&ATLAS\\
20200722&59052.51&$o$&AB&> 20.7&-&ATLAS\\
20200728&59058.45&$c$&AB&> 20.4&-&ATLAS\\
20200730&59060.44&$o$&AB&> 20.1&-&ATLAS\\
20200731&59061.54&$o$&AB&> 20.1&-&ATLAS\\
20200801&59062.43&$o$&AB&> 20.2&-&ATLAS\\
20200803&59064.49&$o$&AB&> 19.9&-&ATLAS\\
20200807&59068.56&$o$&AB&> 19.6&-&ATLAS\\
20200808&59069.43&$o$&AB&> 20.1&-&ATLAS\\
20200810&59071.54&$o$&AB&> 20.4&-&ATLAS\\
20200812&59073.49&$o$&AB&> 20.7&-&ATLAS\\
20200813&59074.51&$o$&AB&> 20.8&-&ATLAS\\
20200816&59077.36&$g$&AB&> 20.8&-&ZTF\\
20200816&59077.54&$o$&AB&> 19.6&-&ATLAS\\
20200817&59078.31&$r$&AB&> 20.6&-&ZTF\\
20200817&59078.34&$g$&AB&> 20.8&-&ZTF\\
20200817&59078.34&$r$&AB&> 20.7&-&ZTF\\
20200817&59078.38&$c$&AB&> 20.9&-&ATLAS\\
20200817&59078.41&$g$&AB&> 20.4&-&ZTF\\
20200818&59079.33&$g$&AB&> 20.7&-&ZTF\\
20200818&59079.33&$r$&AB&> 20.5&-&ZTF\\
20200818&59079.48&$c$&AB&> 21.0&-&ATLAS\\
20200819&59080.33&$g$&AB&> 20.4&-&ZTF\\
20200819&59080.33&$r$&AB&> 20.2&-&ZTF\\
20200819&59080.53&$o$&AB&> 20.6&-&ATLAS\\
20200820&59081.31&$g$&AB&> 20.7&-&ZTF\\
20200820&59081.31&$r$&AB&> 20.5&-&ZTF\\
20200821&59082.35&$c$&AB&> 20.3&-&ATLAS\\
20200821&59082.36&$g$&AB&> 19.0&-&ZTF\\
20200822&59083.50&$c$&AB&> 20.9&-&ATLAS\\
20200823&59084.31&$g$&AB&> 20.6&-&ZTF\\
20200823&59084.31&$r$&AB&> 20.6&-&ZTF\\
20200823&59084.38&$o$&AB&> 20.7&-&ATLAS\\
20200823&59084.40&$w$&AB&> 22.8&-&PAN-STARRS\\
20200824&59085.39&$g$&AB&> 20.7&-&ZTF\\
20200825&59086.33&$r$&AB&> 20.7&-&ZTF\\
20200826&59087.28&$r$&AB&> 20.2&-&ZTF\\
20200826&59087.31&$r$&AB&> 20.5&-&ZTF\\
20200826&59087.35&$g$&AB&> 20.6&-&ZTF\\
20200826&59087.46&$c$&AB&> 20.8&-&ATLAS\\
20200827&59088.29&$g$&AB&> 19.9&-&ZTF\\
20200827&59088.47&$o$&AB&> 20.4&-&ATLAS\\
20200828&59089.28&$g$&AB&> 19.8&-&ZTF\\
20200828&59089.45&$o$&AB&> 20.0&-&ATLAS\\
20200829&59090.34&$g$&AB&> 19.8&-&ZTF\\
20200829&59090.39&$o$&AB&> 20.2&-&ATLAS\\
20200830&59091.30&$g$&AB&> 19.7&-&ZTF\\
20200831&59092.30&$g$&AB&> 19.3&-&ZTF\\
20200903&59095.30&$g$&AB&> 19.1&-&ZTF\\
20200903&59095.30&$r$&AB&> 19.3&-&ZTF\\
20200904&59096.37&$r$&AB&> 19.7&-&ZTF\\
20200904&59096.38&$o$&AB&> 20.3&-&ATLAS\\
20200905&59097.24&$g$&AB&> 17.6&-&ASAS-SN\\
20200905&59097.49&$o$&AB&> 19.0&-&ATLAS\\
20200906&59098.30&$r$&AB&> 20.0&-&ZTF\\
20200906&59098.36&$r$&AB&> 20.0&-&ZTF\\
20200906&59098.46&$o$&AB&> 20.1&-&ATLAS\\
20200906&59098.50&$i$&AB&> 21.4&-&PAN-STARRS\\
20200907&59099.33&$r$&AB&> 20.1&-&ZTF\\
20200908&59100.44&$o$&AB&20.540&0.240&ATLAS\\
20200908&59100.49&$g$&AB&> 18.6&-&ASAS-SN\\
20200909&59101.43&$o$&AB&19.870&0.120&ATLAS\\
20200910&59102.44&$g$&AB&19.510&0.269&ASAS-SN\\
20200911&59103.47&$o$&AB&19.330&0.110&ATLAS\\
20200912&59104.39&$g$&AB&18.689&0.172&ASAS-SN\\
20200913&59105.42&$o$&AB&19.210&0.150&ATLAS\\
20200914&59106.43&$c$&AB&18.770&0.050&ATLAS\\
20200915&59107.28&$g$&AB&18.689&0.109&ZTF\\
20200915&59107.28&$r$&AB&19.018&0.108&ZTF\\
20200916&59108.32&$w$&AB&18.916&0.035&PAN-STARRS\\
20200916&59108.43&$w$&AB&18.885&0.018&PAN-STARRS\\
20200917&59109.43&$o$&AB&18.950&0.050&ATLAS\\
20200918&59110.26&$r$&AB&18.907&0.119&ZTF\\
20200918&59110.29&$r$&AB&18.849&0.119&ZTF\\
20200918&59110.31&$g$&AB&18.671&0.086&ZTF\\
20200918&59110.36&$c$&AB&18.780&0.040&ATLAS\\
20200920&59112.43&$o$&AB&19.070&0.070&ATLAS\\
20200921&59113.19&$g$&AB&18.760&0.094&ZTF\\
20200921&59113.19&$r$&AB&18.903&0.082&ZTF\\
20200921&59113.39&$o$&AB&19.080&0.080&ATLAS\\
20200922&59114.40&$c$&AB&18.770&0.050&ATLAS\\
20200923&59115.40&$g$&AB&18.959&0.208&ASAS-SN\\
20200923&59115.47&$c$&AB&18.950&0.060&ATLAS\\
20200924&59116.25&$g$&AB&18.830&0.079&ZTF\\
20200924&59116.29&$g$&AB&18.847&0.091&ZTF\\
20200924&59116.29&$r$&AB&18.967&0.068&ZTF\\
20200925&59117.27&$g$&AB&18.920&0.090&ZTF\\
20200925&59117.40&$o$&AB&19.110&0.080&ATLAS\\
20200926&59118.25&$r$&AB&18.954&0.091&ZTF\\
20200926&59118.34&$o$&AB&19.080&0.100&ATLAS\\
20200927&59119.22&$g$&AB&19.023&0.109&ZTF\\
20200927&59119.32&$g$&AB&18.827&0.300&ASAS-SN\\
20200927&59119.39&$o$&AB&19.260&0.130&ATLAS\\
20200927&59119.91&$B$&Vega&18.949&0.032&ALFOSC\\
20200927&59119.91&$V$&Vega&18.959&0.028&ALFOSC\\
20200927&59119.91&$u$&AB&19.405&0.091&ALFOSC\\
20200927&59119.91&$g$&AB&18.958&0.019&ALFOSC\\
20200927&59119.91&$r$&AB&19.045&0.029&ALFOSC\\
20200927&59119.91&$i$&AB&19.182&0.040&ALFOSC\\
20200927&59119.91&$z$&AB&19.250&0.064&ALFOSC\\
20200928&59120.28&$g$&AB&19.053&0.157&ZTF\\
20200928&59120.37&$o$&AB&19.570&0.170&ATLAS\\
20200928&59120.90&$g$&AB&> 16.9&-&ASAS-SN\\
20200929&59121.66&$U$&Vega&17.720&0.289&TNT\\
20200929&59121.66&$B$&Vega&19.173&0.393&TNT\\
20200929&59121.66&$V$&Vega&18.914&0.238&TNT\\
20200929&59121.66&$g$&AB&19.079&0.174&TNT\\
20200929&59121.66&$r$&AB&18.999&0.162&TNT\\
20200929&59121.66&$i$&AB&19.107&0.259&TNT\\
20200930&59122.30&$g$&AB&19.079&0.212&ZTF\\
20201001&59123.19&$g$&AB&19.289&0.274&ZTF\\
20201001&59123.19&$r$&AB&19.065&0.173&ZTF\\
20201002&59124.00&$u$&AB&19.819&0.133&ALFOSC\\
20201002&59124.00&$g$&AB&19.198&0.028&ALFOSC\\
20201002&59124.00&$r$&AB&19.246&0.025&ALFOSC\\
20201002&59124.00&$i$&AB&19.361&0.029&ALFOSC\\
20201002&59124.00&$z$&AB&19.465&0.063&ALFOSC\\
20201002&59124.27&$r$&AB&19.384&0.179&ZTF\\
20201003&59125.22&$g$&AB&19.189&0.255&ZTF\\
20201003&59125.38&$o$&AB&19.660&0.180&ATLAS\\
20201003&59125.85&$i$&AB&> 18.5&-&Moravian\\
20201004&59126.23&$g$&AB&19.334&0.178&ZTF\\
20201004&59126.40&$o$&AB&19.210&0.120&ATLAS\\
20201004&59126.57&$g$&AB&19.421&0.096&TNT\\
20201004&59126.57&$r$&AB&19.560&0.104&TNT\\
20201004&59126.57&$i$&AB&19.417&0.110&TNT\\
20201005&59127.24&$g$&AB&19.217&0.198&ZTF\\
20201005&59127.37&$o$&AB&19.650&0.290&ATLAS\\
20201005&59127.65&$g$&AB&19.566&0.052&TNT\\
20201005&59127.65&$r$&AB&19.580&0.035&TNT\\
20201005&59127.65&$i$&AB&19.478&0.045&TNT\\
20201006&59128.41&$g$&AB&> 18.4&-&ASAS-SN\\
20201006&59128.46&$o$&AB&20.000&0.210&ATLAS\\
20201007&59129.23&$g$&AB&19.889&0.210&ZTF\\
20201007&59129.23&$r$&AB&19.582&0.137&ZTF\\
20201007&59129.28&$g$&AB&> 17.9&-&ASAS-SN\\
20201007&59129.42&$o$&AB&19.690&0.140&ATLAS\\
20201007&59129.56&$g$&AB&19.778&0.090&TNT\\
20201007&59129.56&$r$&AB&19.774&0.059&TNT\\
20201007&59129.56&$i$&AB&19.595&0.054&TNT\\
20201007&59130.00&$B$&Vega&19.796&0.186&Moravian\\
20201007&59130.00&$V$&Vega&19.809&0.179&Moravian\\
20201008&59130.02&$g$&AB&> 19.2&-&Moravian\\
20201008&59130.02&$r$&AB&> 19.0&-&Moravian\\
20201008&59130.39&$g$&AB&> 18.8&-&ASAS-SN\\
20201008&59130.40&$o$&AB&19.760&0.140&ATLAS\\
20201009&59131.00&$u$&AB&21.103&0.062&ALFOSC\\
20201009&59131.00&$i$&AB&19.763&0.026&ALFOSC\\
20201009&59131.00&$z$&AB&19.824&0.045&ALFOSC\\
20201009&59131.21&$g$&AB&20.037&0.145&ZTF\\
20201009&59131.41&$o$&AB&20.080&0.190&ATLAS\\
20201010&59132.35&$o$&AB&> 19.9&-&ATLAS\\
20201010&59132.39&$g$&AB&> 19.2&-&ASAS-SN\\
20201011&59133.36&$o$&AB&20.560&0.300&ATLAS\\
20201012&59134.88&$u$&AB&> 22.2&-&ALFOSC\\
20201012&59134.88&$g$&AB&20.836&0.039&ALFOSC\\
20201012&59134.88&$r$&AB&20.425&0.030&ALFOSC\\
20201012&59134.88&$i$&AB&20.415&0.040&ALFOSC\\
20201012&59134.88&$z$&AB&20.298&0.100&ALFOSC\\
20201012&59134.88&$B$&Vega&21.172&0.037&ALFOSC\\
20201012&59134.88&$V$&Vega&20.627&0.037&ALFOSC\\
20201014&59136.37&$o$&AB&> 20.7&-&ATLAS\\
20201016&59138.88&$g$&AB&22.163&0.140&IO:O\\
20201016&59138.88&$r$&AB&22.047&0.188&IO:O\\
20201016&59138.88&$i$&AB&21.833&0.223&IO:O\\
20201016&59138.88&$z$&AB&21.987&0.389&IO:O\\
20201016&59138.89&$B$&Vega&22.343&0.191&IO:O\\
20201016&59138.89&$V$&Vega&22.168&0.261&IO:O\\
20201017&59139.34&$c$&AB&> 20.9&-&ATLAS\\
20201018&59140.31&$g$&AB&> 18.6&-&ASAS-SN\\
20201018&59140.34&$o$&AB&> 19.2&-&ATLAS\\
20201021&59143.23&$w$&AB&> 21.8&-&PAN-STARRS\\
20201021&59143.29&$o$&AB&> 20.6&-&ATLAS\\
20201023&59145.35&$o$&AB&> 20.6&-&ATLAS\\
20201024&59146.36&$o$&AB&> 19.9&-&ATLAS\\
20201024&59146.90&$g$&AB&> 20.9&-&IO:O\\
20201024&59146.90&$r$&AB&> 21.3&-&IO:O\\
20201024&59146.90&$i$&AB&> 21.0&-&IO:O\\
20201024&59146.90&$z$&AB&> 20.6&-&IO:O\\
20201025&59147.33&$o$&AB&> 20.0&-&ATLAS\\
20201026&59148.32&$o$&AB&> 18.3&-&ATLAS\\
20201031&59153.37&$o$&AB&> 19.9&-&ATLAS\\
20201101&59154.38&$o$&AB&> 18.8&-&ATLAS\\
20201102&59155.39&$o$&AB&> 17.5&-&ATLAS\\
20201103&59156.31&$o$&AB&> 20.3&-&ATLAS\\
20201105&59158.29&$o$&AB&> 20.5&-&ATLAS\\
20201106&59159.30&$c$&AB&> 18.2&-&ATLAS\\
20201110&59163.35&$c$&AB&> 21.0&-&ATLAS\\
20201112&59165.31&$o$&AB&> 20.9&-&ATLAS\\
20201116&59169.29&$o$&AB&> 20.8&-&ATLAS\\
20201117&59170.27&$c$&AB&> 20.8&-&ATLAS\\
20201119&59172.27&$o$&AB&> 19.7&-&ATLAS\\
20201126&59179.28&$o$&AB&> 19.9&-&ATLAS\\
20201127&59180.24&$o$&AB&> 19.8&-&ATLAS\\
20201128&59181.29&$o$&AB&> 19.1&-&ATLAS\\
20201129&59182.26&$o$&AB&> 20.2&-&ATLAS\\
20201202&59185.28&$o$&AB&> 20.4&-&ATLAS\\
20201203&59186.28&$o$&AB&> 20.0&-&ATLAS\\
20201204&59187.30&$o$&AB&> 20.1&-&ATLAS\\
20201206&59189.26&$o$&AB&> 20.1&-&ATLAS\\
20201207&59190.23&$c$&AB&> 20.8&-&ATLAS\\
20201212&59195.27&$c$&AB&> 20.5&-&ATLAS\\
20201221&59204.21&$o$&AB&> 19.4&-&ATLAS\\
20201224&59207.20&$o$&AB&> 19.4&-&ATLAS\\
20201226&59209.24&$o$&AB&> 19.8&-&ATLAS\\
20201228&59211.24&$o$&AB&> 19.4&-&ATLAS\\
20210103&59217.24&$o$&AB&> 20.2&-&ATLAS\\
20210104&59218.24&$c$&AB&> 20.3&-&ATLAS\\
20210108&59222.24&$c$&AB&> 20.6&-&ATLAS\\
20210113&59227.22&$c$&AB&> 20.3&-&ATLAS\\

\end{longtable}
\begin{longtable}{ccccccc}
    \caption{Optical observed magnitudes of SN\,2021bbv.} 
    \label{table:mag_2021bbv}\\
        \hline \hline
        Date & MJD & Filter & MagType & Magnitude & Error & Instrument/Source \\
        \hline
    \endfirsthead

    \multicolumn{7}{c}%
    {{\tablename\ \thetable{} -- continued from previous page}} \\
        \hline \hline
        Date & MJD & Filter & MagType & Magnitude & Error & Instrument/Source \\
        \hline
    \endhead

        \hline \hline
    \multicolumn{7}{r}{{Continued on next page}} \\
    \endfoot

        \hline \hline
    \endlastfoot

20201217&59200.48&$c$&AB&> 20.1&-&ATLAS\\
20201221&59204.49&$c$&AB&> 20.4&-&ATLAS\\
20201223&59206.64&$c$&AB&> 20.3&-&ATLAS\\
20201224&59207.57&$c$&AB&> 20.2&-&ATLAS\\
20201226&59209.60&$o$&AB&> 20.7&-&ATLAS\\
20210106&59220.53&$o$&AB&> 19.9&-&ATLAS\\
20210107&59221.43&$r$&AB&> 19.9&-&ZTF\\
20210107&59221.46&$g$&AB&> 19.5&-&ZTF\\
20210109&59223.40&$g$&AB&> 20.3&-&ZTF\\
20210109&59223.46&$r$&AB&> 20.1&-&ZTF\\
20210111&59225.41&$r$&AB&> 19.9&-&ZTF\\
20210111&59225.44&$g$&AB&> 20.0&-&ZTF\\
20210112&59226.56&$c$&AB&> 20.3&-&ATLAS\\
20210113&59227.46&$g$&AB&> 20.4&-&ZTF\\
20210115&59229.44&$r$&AB&> 19.8&-&ZTF\\
20210115&59229.64&$c$&AB&> 21.0&-&ATLAS\\
20210116&59230.49&$c$&AB&> 20.6&-&ATLAS\\
20210117&59231.42&$g$&AB&> 20.5&-&ZTF\\
20210117&59231.46&$r$&AB&> 20.6&-&ZTF\\
20210118&59232.52&$c$&AB&> 20.5&-&ATLAS\\
20210120&59234.45&$g$&AB&20.451&0.333&ZTF\\
20210120&59234.54&$c$&AB&> 19.8&-&ATLAS\\
20210128&59242.18&$u$&AB&> 18.9&-&Moravian\\
20210128&59242.18&$g$&AB&18.358&0.058&Moravian\\
20210128&59242.18&$r$&AB&18.592&0.070&Moravian\\
20210128&59242.20&$B$&Vega&18.627&0.132&Moravian\\
20210128&59242.20&$V$&Vega&18.526&0.080&Moravian\\
20210129&59243.08&$u$&AB&18.958&0.018&ALFOSC\\
20210129&59243.08&$g$&AB&18.522&0.050&ALFOSC\\
20210129&59243.08&$r$&AB&18.619&0.033&ALFOSC\\
20210129&59243.08&$i$&AB&18.767&0.045&ALFOSC\\
20210129&59243.09&$B$&Vega&18.885&0.020&ALFOSC\\
20210129&59243.09&$V$&Vega&18.642&0.034&ALFOSC\\
20210131&59245.04&$u$&AB&19.288&0.059&ALFOSC\\
20210131&59245.04&$g$&AB&18.841&0.039&ALFOSC\\
20210131&59245.04&$r$&AB&18.851&0.057&ALFOSC\\
20210131&59245.04&$i$&AB&18.804&0.056&ALFOSC\\
20210131&59245.04&$z$&AB&18.913&0.055&ALFOSC\\
20210131&59245.05&$B$&Vega&19.120&0.047&ALFOSC\\
20210131&59245.05&$V$&Vega&18.894&0.047&ALFOSC\\
20210203&59248.05&$U$&Vega&18.639&0.063&fa16\\
20210203&59248.05&$B$&Vega&19.310&0.050&fa16\\
20210203&59248.05&$V$&Vega&19.184&0.056&fa16\\
20210203&59248.08&$g$&AB&19.286&0.046&fa16\\
20210203&59248.08&$r$&AB&19.124&0.061&fa16\\
20210203&59248.08&$i$&AB&19.096&0.070&fa16\\
20210203&59248.12&$u$&AB&19.358&0.043&IO:O\\
20210203&59248.12&$g$&AB&19.195&0.023&IO:O\\
20210203&59248.12&$r$&AB&19.309&0.029&IO:O\\
20210203&59248.12&$i$&AB&19.281&0.032&IO:O\\
20210203&59248.12&$z$&AB&19.260&0.056&IO:O\\
20210203&59248.12&$B$&Vega&19.509&0.036&IO:O\\
20210203&59248.12&$V$&Vega&19.242&0.032&IO:O\\
20210203&59248.59&$o$&AB&19.380&0.110&ATLAS\\
20210204&59249.04&$U$&Vega&18.742&0.089&fa06\\
20210204&59249.04&$B$&Vega&19.408&0.052&fa06\\
20210204&59249.04&$V$&Vega&19.406&0.053&fa06\\
20210204&59249.06&$g$&AB&19.284&0.035&fa06\\
20210204&59249.06&$r$&AB&19.353&0.043&fa06\\
20210204&59249.06&$i$&AB&19.436&0.061&fa06\\
20210204&59249.38&$r$&AB&19.444&0.231&ZTF\\
20210204&59249.42&$g$&AB&19.378&0.200&ZTF\\
20210206&59251.40&$g$&AB&19.444&0.177&ZTF\\
20210206&59251.44&$r$&AB&19.698&0.151&ZTF\\
20210206&59251.56&$o$&AB&19.580&0.110&ATLAS\\
20210207&59252.21&$U$&Vega&19.175&0.091&fa03\\
20210207&59252.21&$B$&Vega&19.842&0.031&fa03\\
20210207&59252.21&$V$&Vega&19.612&0.047&fa03\\
20210207&59252.22&$g$&AB&19.833&0.032&fa03\\
20210207&59252.22&$r$&AB&19.659&0.042&fa03\\
20210207&59252.22&$i$&AB&19.538&0.032&fa03\\
20210207&59252.54&$c$&AB&19.820&0.110&ATLAS\\
20210208&59253.35&$r$&AB&19.873&0.132&ZTF\\
20210208&59253.38&$g$&AB&19.754&0.220&ZTF\\
20210208&59253.60&$c$&AB&19.870&0.120&ATLAS\\
20210210&59255.38&$g$&AB&19.944&0.170&ZTF\\
20210210&59255.40&$r$&AB&19.991&0.145&ZTF\\
20210210&59255.48&$c$&AB&20.240&0.180&ATLAS\\
20210211&59256.49&$c$&AB&> 20.6&-&ATLAS\\
20210211&59256.74&$B$&Vega&20.296&0.032&fa19\\
20210211&59256.74&$V$&Vega&19.999&0.045&fa19\\
20210211&59256.76&$g$&AB&20.277&0.043&fa19\\
20210211&59256.76&$r$&AB&20.219&0.067&fa19\\
20210211&59256.77&$i$&AB&> 19.7&-&fa19\\
20210212&59257.38&$g$&AB&20.102&0.282&ZTF\\
20210212&59257.55&$c$&AB&20.520&0.190&ATLAS\\
20210212&59257.61&$w$&AB&20.199&0.043&PAN-STARRS\\
20210213&59258.46&$c$&AB&20.350&0.190&ATLAS\\
20210215&59260.44&$c$&AB&20.340&0.250&ATLAS\\
20210216&59261.47&$c$&AB&20.480&0.180&ATLAS\\
20210217&59262.04&$u$&AB&21.248&0.078&ALFOSC\\
20210217&59262.04&$g$&AB&20.592&0.030&ALFOSC\\
20210217&59262.04&$r$&AB&20.483&0.062&ALFOSC\\
20210217&59262.04&$i$&AB&20.364&0.055&ALFOSC\\
20210217&59262.04&$z$&AB&> 19.7&-&ALFOSC\\
20210217&59262.05&$B$&Vega&20.805&0.041&ALFOSC\\
20210217&59262.05&$V$&Vega&20.567&0.041&ALFOSC\\
20210217&59262.34&$r$&AB&20.445&0.193&ZTF\\
20210217&59262.36&$B$&Vega&20.712&0.053&fa03\\
20210217&59262.36&$V$&Vega&20.446&0.071&fa03\\
20210217&59262.38&$g$&AB&20.714&0.045&fa03\\
20210217&59262.38&$r$&AB&20.720&0.094&fa03\\
20210217&59262.38&$i$&AB&20.430&0.083&fa03\\
20210217&59262.53&$c$&AB&20.570&0.220&ATLAS\\
20210217&59262.60&$w$&AB&20.531&0.068&PAN-STARRS\\
20210221&59266.65&$c$&AB&> 17.1&-&ATLAS\\
20210222&59267.51&$o$&AB&20.370&0.250&ATLAS\\
20210223&59268.42&$o$&AB&> 20.5&-&ATLAS\\
20210224&59269.28&$B$&Vega&21.116&0.148&fa03\\
20210224&59269.28&$V$&Vega&21.223&0.205&fa03\\
20210224&59269.31&$g$&AB&21.195&0.088&fa03\\
20210224&59269.31&$r$&AB&20.990&0.105&fa03\\
20210224&59269.31&$i$&AB&20.616&0.079&fa03\\
20210224&59269.38&$o$&AB&> 20.1&-&ATLAS\\
20210303&59276.50&$o$&AB&> 20.8&-&ATLAS\\
20210304&59277.50&$o$&AB&> 20.2&-&ATLAS\\
20210305&59278.52&$o$&AB&> 19.1&-&ATLAS\\
20210307&59280.48&$o$&AB&> 20.9&-&ATLAS\\
20210309&59282.05&$B$&Vega&22.018&0.076&ALFOSC\\
20210315&59288.98&$B$&Vega&22.262&0.138&IO:O\\
20210315&59288.98&$V$&Vega&21.856&0.106&IO:O\\
20210318&59291.92&$g$&AB&22.343&0.084&ALFOSC\\
20210318&59291.92&$r$&AB&21.957&0.110&ALFOSC\\
20210318&59291.92&$i$&AB&21.568&0.107&ALFOSC\\
20210318&59291.92&$z$&AB&> 21.1&-&ALFOSC\\
20210319&59292.39&$o$&AB&> 20.7&-&ATLAS\\
20210320&59293.43&$o$&AB&> 20.7&-&ATLAS\\
20210321&59294.39&$o$&AB&> 19.8&-&ATLAS\\
20210324&59297.46&$o$&AB&> 20.1&-&ATLAS\\
20210411&59315.28&$w$&AB&> 22.7&-&PAN-STARRS\\

\end{longtable}
\begin{longtable}{ccccccc}
    \caption{Optical observed magnitudes of SN\,2023utc.} 
    \label{table:mag_2023utc}\\
        \hline \hline
        Date & MJD & Filter & MagType & Magnitude & Error & Instrument/Source \\
        \hline
    \endfirsthead

    \multicolumn{7}{c}%
    {{\tablename\ \thetable{} -- continued from previous page}} \\
        \hline \hline
        Date & MJD & Filter & MagType & Magnitude & Error & Instrument/Source \\
        \hline
    \endhead

        \hline \hline
    \multicolumn{7}{r}{{Continued on next page}} \\
    \endfoot

        \hline \hline
    \endlastfoot

20230605&60100.26&$o$&AB&> 20.1&-&ATLAS\\
20230913&60200.63&$g$&AB&> 17.7&-&ASAS-SN\\
20230918&60205.63&$g$&AB&> 18.1&-&ASAS-SN\\
20230925&60212.63&$c$&AB&> 20.1&-&ATLAS\\
20230927&60214.62&$g$&AB&> 18.5&-&ASAS-SN\\
20231003&60220.44&$g$&AB&> 19.1&-&ZTF\\
20231003&60220.49&$r$&AB&> 19.6&-&ZTF\\
20231003&60220.58&$g$&AB&> 17.3&-&ASAS-SN\\
20231003&60220.63&$o$&AB&> 20.1&-&ATLAS\\
20231005&60222.48&$g$&AB&> 19.5&-&ZTF\\
20231005&60222.49&$r$&AB&> 19.9&-&ZTF\\
20231006&60223.63&$g$&AB&> 18.3&-&ASAS-SN\\
20231007&60224.44&$g$&AB&> 19.8&-&ZTF\\
20231012&60229.45&$g$&AB&17.961&0.086&ZTF\\
20231012&60229.51&$r$&AB&18.254&0.083&ZTF\\
20231013&60230.56&$g$&AB&17.807&0.127&ASAS-SN\\
20231015&60232.45&$g$&AB&17.274&0.042&ZTF\\
20231015&60232.49&$r$&AB&17.505&0.035&ZTF\\
20231015&60232.55&$g$&AB&17.463&0.110&ASAS-SN\\
20231017&60234.42&$g$&AB&17.674&0.145&ASAS-SN\\
20231017&60234.43&$r$&AB&17.518&0.054&ZTF\\
20231017&60234.47&$g$&AB&17.405&0.044&ZTF\\
20231019&60236.42&$g$&AB&17.505&0.115&ASAS-SN\\
20231019&60236.43&$g$&AB&17.635&0.047&ZTF\\
20231019&60236.49&$r$&AB&17.651&0.047&ZTF\\
20231019&60236.59&$c$&AB&17.840&0.030&ATLAS\\
20231020&60237.43&$g$&AB&17.875&0.149&ASAS-SN\\
20231021&60238.43&$g$&AB&17.893&0.045&ZTF\\
20231021&60238.51&$r$&AB&17.864&0.046&ZTF\\
20231022&60239.56&$g$&AB&17.975&0.150&ASAS-SN\\
20231023&60240.61&$c$&AB&18.060&0.030&ATLAS\\
20231025&60242.43&$g$&AB&18.426&0.205&ASAS-SN\\
20231025&60242.49&$g$&AB&18.425&0.055&ZTF\\
20231025&60242.52&$r$&AB&18.242&0.048&ZTF\\
20231025&60242.53&$o$&AB&18.160&0.060&ATLAS\\
20231026&60243.54&$g$&AB&> 17.5&-&ASAS-SN\\
20231027&60244.18&$U$&Vega&18.499&0.024&fa20\\
20231027&60244.18&$B$&Vega&18.590&0.035&fa20\\
20231027&60244.18&$V$&Vega&18.269&0.038&fa20\\
20231027&60244.20&$g$&AB&18.497&0.027&fa20\\
20231027&60244.20&$r$&AB&18.283&0.031&fa20\\
20231027&60244.20&$i$&AB&18.189&0.035&fa20\\
20231027&60244.41&$g$&AB&> 17.0&-&ASAS-SN\\
20231027&60244.60&$o$&AB&18.440&0.140&ATLAS\\
20231029&60246.45&$g$&AB&18.575&0.138&ZTF\\
20231029&60246.50&$r$&AB&18.411&0.079&ZTF\\
20231029&60246.63&$o$&AB&18.380&0.080&ATLAS\\
20231030&60247.18&$U$&Vega&18.803&0.137&fa20\\
20231030&60247.18&$B$&Vega&18.699&0.087&fa20\\
20231030&60247.18&$V$&Vega&18.430&0.138&fa20\\
20231030&60247.20&$g$&AB&18.636&0.153&fa20\\
20231030&60247.20&$r$&AB&18.325&0.095&fa20\\
20231030&60247.20&$i$&AB&18.304&0.173&fa20\\
20231031&60248.53&$g$&AB&> 17.3&-&ASAS-SN\\
20231031&60248.57&$o$&AB&> 18.1&-&ATLAS\\
20231101&60249.45&$g$&AB&> 17.3&-&ASAS-SN\\
20231102&60250.42&$r$&AB&18.670&0.104&ZTF\\
20231102&60250.44&$g$&AB&> 17.5&-&ASAS-SN\\
20231102&60250.47&$g$&AB&19.075&0.109&ZTF\\
20231102&60250.48&$U$&Vega&19.308&0.053&fa16\\
20231102&60250.48&$B$&Vega&19.212&0.046&fa16\\
20231102&60250.48&$V$&Vega&18.773&0.039&fa16\\
20231102&60250.50&$g$&AB&19.149&0.037&fa16\\
20231102&60250.50&$r$&AB&18.709&0.027&fa16\\
20231102&60250.50&$i$&AB&19.043&0.069&fa16\\
20231103&60251.38&$g$&AB&> 17.6&-&ASAS-SN\\
20231104&60252.38&$g$&AB&> 17.8&-&ASAS-SN\\
20231104&60252.43&$r$&AB&18.871&0.078&ZTF\\
20231104&60252.48&$g$&AB&19.504&0.151&ZTF\\
20231104&60252.57&$o$&AB&18.920&0.090&ATLAS\\
20231105&60253.38&$g$&AB&> 16.7&-&ASAS-SN\\
20231105&60253.46&$U$&Vega&20.064&0.189&fa05\\
20231105&60253.46&$B$&Vega&20.036&0.090&fa05\\
20231105&60253.46&$V$&Vega&19.297&0.064&fa05\\
20231105&60253.47&$g$&AB&19.582&0.038&fa05\\
20231106&60254.38&$g$&AB&> 17.4&-&ASAS-SN\\
20231106&60254.47&$g$&AB&19.985&0.152&ZTF\\
20231106&60254.52&$r$&AB&19.209&0.106&ZTF\\
20231107&60255.39&$g$&AB&> 18.2&-&ASAS-SN\\
20231108&60256.37&$g$&AB&> 18.4&-&ASAS-SN\\
20231108&60256.58&$o$&AB&19.450&0.130&ATLAS\\
20231110&60258.45&$r$&AB&19.586&0.110&ZTF\\
20231110&60258.51&$g$&AB&> 20.5&-&ZTF\\
20231110&60258.52&$c$&AB&> 18.8&-&ATLAS\\
20231112&60260.34&$g$&AB&> 20.0&-&ZTF\\
20231112&60260.38&$r$&AB&19.909&0.134&ZTF\\
20231112&60260.48&$g$&AB&> 18.5&-&ASAS-SN\\
20231112&60260.58&$c$&AB&20.150&0.220&ATLAS\\
20231114&60262.45&$g$&AB&> 20.6&-&ZTF\\
20231114&60262.49&$r$&AB&20.025&0.147&ZTF\\
20231115&60263.49&$g$&AB&> 18.5&-&ASAS-SN\\
20231116&60264.62&$c$&AB&20.220&0.290&ATLAS\\
20231118&60266.46&$g$&AB&> 18.7&-&ASAS-SN\\
20231118&60266.59&$c$&AB&20.670&0.330&ATLAS\\
20231120&60268.48&$g$&AB&> 18.5&-&ASAS-SN\\
20231122&60270.48&$g$&AB&> 18.1&-&ASAS-SN\\
20231122&60270.57&$c$&AB&20.780&0.280&ATLAS\\
20231124&60272.32&$g$&AB&> 17.3&-&ASAS-SN\\
20231124&60272.53&$i$&AB&20.290&0.091&PAN-STARRS\\
20231124&60272.59&$o$&AB&20.240&0.190&ATLAS\\
20231126&60274.53&$o$&AB&> 19.3&-&ATLAS\\
20231204&60282.57&$o$&AB&> 20.4&-&ATLAS\\
20231205&60283.56&$o$&AB&> 20.6&-&ATLAS\\
20231206&60284.56&$o$&AB&20.610&0.330&ATLAS\\
20231207&60285.28&$g$&AB&> 18.3&-&ASAS-SN\\
20231207&60285.51&$o$&AB&> 20.6&-&ATLAS\\
20231208&60286.27&$g$&AB&> 18.6&-&ASAS-SN\\
20231208&60286.56&$o$&AB&> 20.9&-&ATLAS\\
20231209&60287.27&$g$&AB&> 18.4&-&ASAS-SN\\
20231211&60289.26&$g$&AB&> 18.4&-&ASAS-SN\\
20231211&60289.57&$o$&AB&> 20.7&-&ATLAS\\
20231213&60291.42&$g$&AB&> 18.6&-&ASAS-SN\\
20231213&60291.50&$c$&AB&> 20.5&-&ATLAS\\
20231215&60293.53&$o$&AB&> 20.6&-&ATLAS\\
20231215&60293.61&$w$&AB&21.291&0.230&PAN-STARRS\\
20231217&60295.52&$c$&AB&> 20.8&-&ATLAS\\
20231224&60302.36&$g$&AB&> 17.2&-&ASAS-SN\\
20231225&60303.49&$i$&AB&21.224&0.204&PAN-STARRS\\
20231225&60303.53&$g$&AB&> 17.5&-&ASAS-SN\\
20231227&60305.51&$g$&AB&> 17.2&-&ASAS-SN\\
20231230&60308.30&$g$&AB&> 17.3&-&ASAS-SN\\
20231231&60309.38&$g$&AB&> 18.0&-&ASAS-SN\\
20240102&60311.34&$g$&AB&> 18.4&-&ASAS-SN\\
20240103&60312.42&$g$&AB&> 18.3&-&ASAS-SN\\
20240106&60315.50&$g$&AB&> 18.8&-&ASAS-SN\\
20240107&60316.39&$g$&AB&> 18.9&-&ASAS-SN\\
20240110&60319.51&$g$&AB&> 18.1&-&ASAS-SN\\
20240111&60320.57&$w$&AB&20.732&0.147&PAN-STARRS\\
20240201&60341.40&$g$&AB&> 17.0&-&ASAS-SN\\
20240202&60342.39&$g$&AB&> 17.1&-&ASAS-SN\\
20240205&60345.24&$g$&AB&> 18.8&-&ASAS-SN\\
20240206&60346.27&$g$&AB&> 18.5&-&ASAS-SN\\
20240208&60348.30&$g$&AB&> 18.5&-&ASAS-SN\\
20240213&60353.28&$g$&AB&> 18.4&-&ASAS-SN\\
20240214&60354.30&$g$&AB&> 18.3&-&ASAS-SN\\
20240217&60357.22&$g$&AB&> 18.1&-&ASAS-SN\\
20240219&60359.18&$g$&AB&> 17.7&-&ASAS-SN\\
20240220&60360.21&$g$&AB&> 17.2&-&ASAS-SN\\
20240221&60361.20&$g$&AB&> 16.6&-&ASAS-SN\\
20240222&60362.28&$g$&AB&> 16.4&-&ASAS-SN\\
20240223&60363.24&$g$&AB&> 16.6&-&ASAS-SN\\
20240226&60366.37&$g$&AB&> 16.7&-&ASAS-SN\\
20240301&60370.30&$g$&AB&> 17.7&-&ASAS-SN\\
20240302&60371.35&$g$&AB&> 17.8&-&ASAS-SN\\
20240304&60373.34&$g$&AB&> 18.6&-&ASAS-SN\\
20240305&60374.34&$g$&AB&> 18.5&-&ASAS-SN\\
20240306&60375.32&$g$&AB&> 18.1&-&ASAS-SN\\
20240311&60380.29&$g$&AB&> 18.3&-&ASAS-SN\\
20240313&60382.29&$g$&AB&> 18.5&-&ASAS-SN\\
20240325&60394.26&$g$&AB&> 17.3&-&ASAS-SN\\
20240330&60399.28&$g$&AB&> 16.8&-&ASAS-SN\\

\end{longtable}
\begin{longtable}{ccccccc}
    \caption{Optical observed magnitudes of SN\,2024aej.} 
    \label{table:mag_2024aej}\\
        \hline \hline
        Date & MJD & Filter & MagType & Magnitude & Error & Instrument/Source \\
        \hline
    \endfirsthead

    \multicolumn{7}{c}%
    {{\tablename\ \thetable{} -- continued from previous page}} \\
        \hline \hline
        Date & MJD & Filter & MagType & Magnitude & Error & Instrument/Source \\
        \hline
    \endhead

        \hline \hline
    \multicolumn{7}{r}{{Continued on next page}} \\
    \endfoot

        \hline \hline
    \endlastfoot

20231215&60293.23&$g$&AB&> 20.8&-&ZTF\\
20231216&60294.08&$r$&AB&> 20.5&-&ZTF\\
20231216&60294.09&$g$&AB&> 20.6&-&ZTF\\
20231216&60294.14&$g$&AB&> 20.8&-&ZTF\\
20231216&60294.21&$g$&AB&> 20.6&-&ZTF\\
20231224&60302.37&$o$&AB&> 19.8&-&ATLAS\\
20231225&60303.36&$o$&AB&> 19.9&-&ATLAS\\
20231227&60305.19&$g$&AB&> 18.8&-&ZTF\\
20231227&60305.38&$o$&AB&> 19.6&-&ATLAS\\
20231228&60306.36&$o$&AB&> 20.2&-&ATLAS\\
20231229&60307.15&$g$&AB&> 20.1&-&ZTF\\
20231229&60307.18&$r$&AB&> 20.2&-&ZTF\\
20231229&60307.38&$o$&AB&> 20.3&-&ATLAS\\
20231230&60308.37&$o$&AB&> 20.3&-&ATLAS\\
20240101&60310.34&$o$&AB&> 20.8&-&ATLAS\\
20240102&60311.34&$o$&AB&> 20.7&-&ATLAS\\
20240103&60312.33&$o$&AB&> 20.8&-&ATLAS\\
20240104&60313.34&$o$&AB&> 20.3&-&ATLAS\\
20240105&60314.12&$g$&AB&> 20.4&-&ZTF\\
20240105&60314.16&$r$&AB&> 20.3&-&ZTF\\
20240106&60315.29&$o$&AB&> 20.7&-&ATLAS\\
20240107&60316.33&$o$&AB&> 20.8&-&ATLAS\\
20240111&60320.32&$o$&AB&> 20.8&-&ATLAS\\
20240114&60323.22&$g$&AB&18.854&0.128&ZTF\\
20240114&60323.26&$r$&AB&18.969&0.175&ZTF\\
20240114&60323.33&$o$&AB&19.180&0.080&ATLAS\\
20240115&60324.33&$o$&AB&18.750&0.040&ATLAS\\
20240116&60325.21&$r$&AB&18.412&0.078&ZTF\\
20240116&60325.21&$g$&AB&18.071&0.066&ZTF\\
20240116&60325.29&$o$&AB&18.450&0.030&ATLAS\\
20240117&60326.96&$U$&Vega&17.328&0.016&fa11\\
20240117&60326.96&$B$&Vega&17.860&0.012&fa11\\
20240117&60326.96&$V$&Vega&17.970&0.021&fa11\\
20240117&60326.98&$g$&AB&18.067&0.028&fa11\\
20240117&60326.98&$r$&AB&18.091&0.032&fa11\\
20240117&60326.98&$i$&AB&> 17.6&-&fa11\\
20240118&60327.15&$r$&AB&18.149&0.059&ZTF\\
20240118&60327.20&$g$&AB&17.938&0.059&ZTF\\
20240118&60327.28&$o$&AB&18.180&0.050&ATLAS\\
20240119&60328.29&$o$&AB&18.150&0.050&ATLAS\\
20240119&60328.94&$U$&Vega&17.462&0.021&fa20\\
20240119&60328.94&$B$&Vega&17.968&0.016&fa20\\
20240119&60328.94&$V$&Vega&18.028&0.023&fa20\\
20240119&60328.94&$g$&AB&18.086&0.016&fa20\\
20240119&60328.94&$r$&AB&18.125&0.027&fa20\\
20240119&60328.94&$i$&AB&18.328&0.041&fa20\\
20240120&60329.29&$o$&AB&18.200&0.060&ATLAS\\
20240121&60330.31&$o$&AB&18.160&0.070&ATLAS\\
20240122&60331.34&$o$&AB&18.320&0.080&ATLAS\\
20240123&60332.32&$o$&AB&18.490&0.090&ATLAS\\
20240123&60332.84&$U$&Vega&18.057&0.038&fa11\\
20240123&60332.84&$B$&Vega&18.499&0.030&fa11\\
20240123&60332.84&$V$&Vega&18.531&0.036&fa11\\
20240123&60332.86&$g$&AB&18.570&0.028&fa11\\
20240123&60332.86&$r$&AB&18.582&0.033&fa11\\
20240123&60332.86&$i$&AB&18.648&0.049&fa11\\
20240124&60333.33&$o$&AB&18.710&0.120&ATLAS\\
20240125&60334.18&$U$&Vega&> 18.0&-&fa08\\
20240125&60334.18&$B$&Vega&18.966&0.053&fa08\\
20240125&60334.18&$V$&Vega&18.729&0.034&fa08\\
20240125&60334.18&$g$&AB&18.950&0.059&fa08\\
20240125&60334.18&$r$&AB&18.689&0.050&fa08\\
20240125&60334.18&$i$&AB&18.754&0.065&fa08\\
20240125&60334.32&$o$&AB&19.060&0.280&ATLAS\\
20240126&60335.31&$o$&AB&18.880&0.200&ATLAS\\
20240127&60336.22&$g$&AB&19.021&0.177&ZTF\\
20240128&60337.16&$U$&Vega&19.054&0.124&fa16\\
20240128&60337.16&$B$&Vega&19.154&0.082&fa16\\
20240128&60337.16&$V$&Vega&19.127&0.109&fa16\\
20240128&60337.18&$g$&AB&19.169&0.066&fa16\\
20240128&60337.18&$r$&AB&19.134&0.096&fa16\\
20240128&60337.18&$i$&AB&19.154&0.115&fa16\\
20240128&60337.31&$o$&AB&19.330&0.150&ATLAS\\
20240129&60338.30&$o$&AB&19.410&0.180&ATLAS\\
20240131&60340.12&$g$&AB&19.700&0.129&ZTF\\
20240201&60341.28&$c$&AB&19.830&0.140&ATLAS\\
20240203&60343.27&$o$&AB&> 20.7&-&ATLAS\\
20240204&60344.27&$o$&AB&20.210&0.220&ATLAS\\
20240205&60345.16&$B$&Vega&20.841&0.138&fa08\\
20240205&60345.16&$V$&Vega&20.041&0.228&fa08\\
20240205&60345.17&$g$&AB&20.484&0.122&fa08\\
20240205&60345.17&$r$&AB&20.264&0.195&fa08\\
20240206&60346.10&$U$&Vega&> 19.1&-&fa08\\
20240206&60346.10&$B$&Vega&> 20.2&-&fa08\\
20240206&60346.10&$V$&Vega&> 19.6&-&fa08\\
20240206&60346.10&$g$&AB&20.670&0.118&fa08\\
20240206&60346.10&$r$&AB&20.545&0.210&fa08\\
20240206&60346.10&$i$&AB&20.249&0.187&fa08\\
20240207&60347.24&$c$&AB&20.540&0.340&ATLAS\\
20240207&60347.90&$U$&Vega&20.892&0.219&fa20\\
20240207&60347.90&$B$&Vega&21.131&0.094&fa20\\
20240207&60347.90&$V$&Vega&20.582&0.096&fa20\\
20240207&60347.90&$g$&AB&20.845&0.066&fa20\\
20240207&60347.90&$r$&AB&20.746&0.119&fa20\\
20240207&60347.90&$i$&AB&20.500&0.110&fa20\\
20240210&60350.26&$c$&AB&> 20.4&-&ATLAS\\
20240218&60358.29&$o$&AB&> 19.9&-&ATLAS\\
20240219&60359.24&$o$&AB&> 20.2&-&ATLAS\\
20240221&60361.30&$o$&AB&> 19.9&-&ATLAS\\
20240223&60363.29&$o$&AB&> 18.7&-&ATLAS\\
20240301&60370.23&$o$&AB&> 19.2&-&ATLAS\\
\end{longtable}

\newpage 
\section{Peak time of our sample of SNe Ibn}\label{appendix:peak_time}
\begin{figure*}[h]
\includegraphics[width=0.5\linewidth]{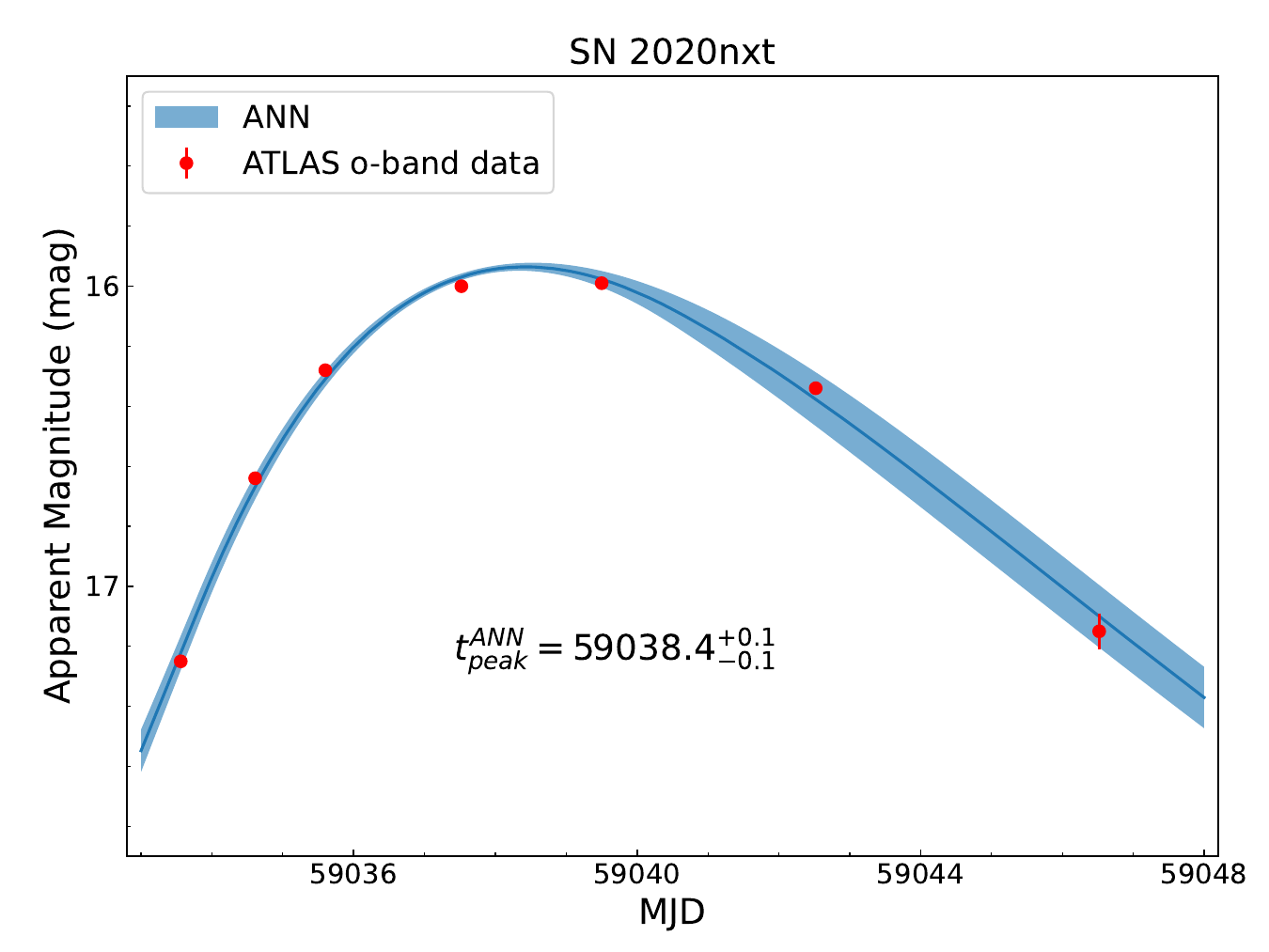}
\includegraphics[width=0.5\linewidth]{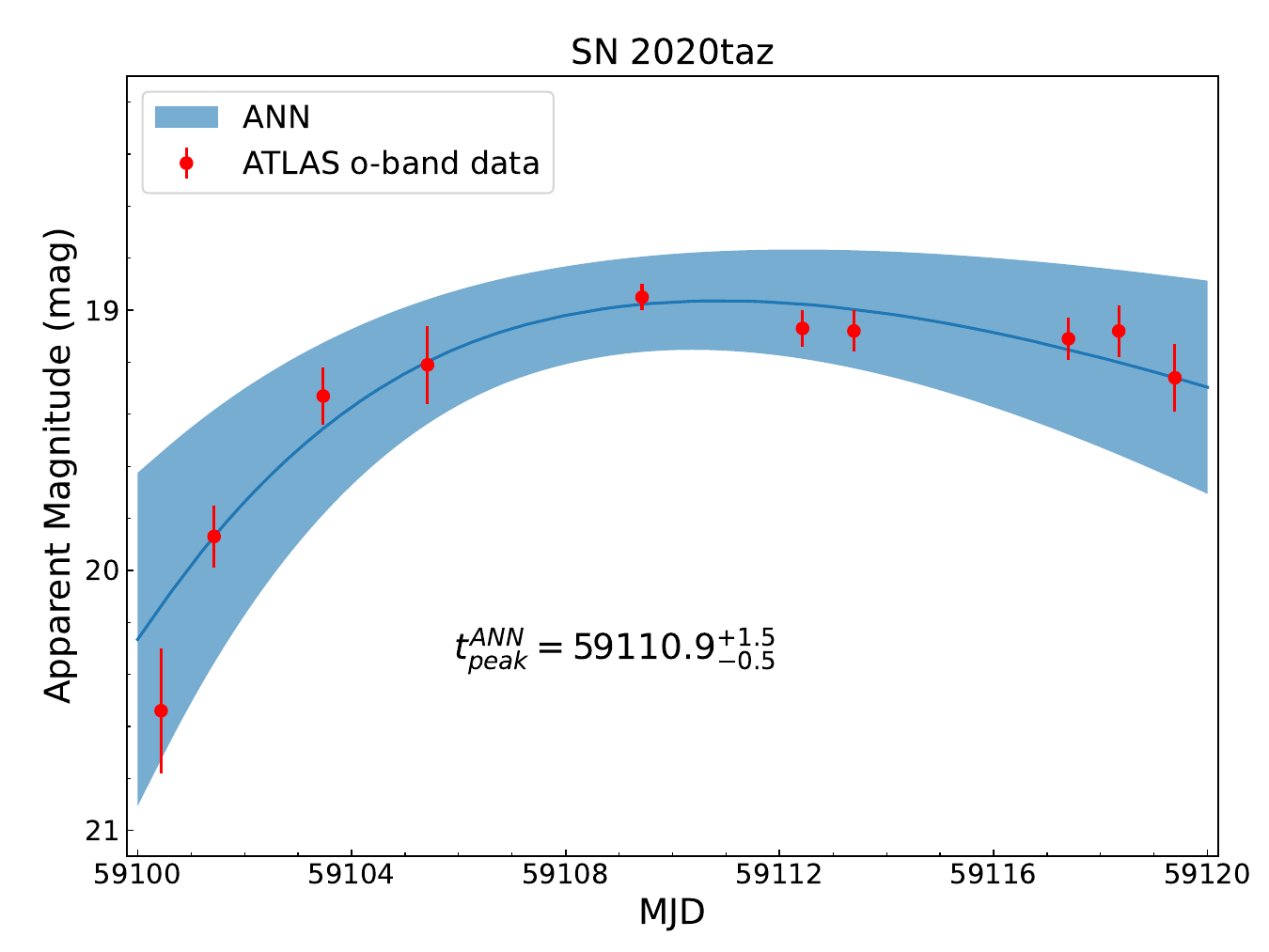}
\includegraphics[width=0.5\linewidth]{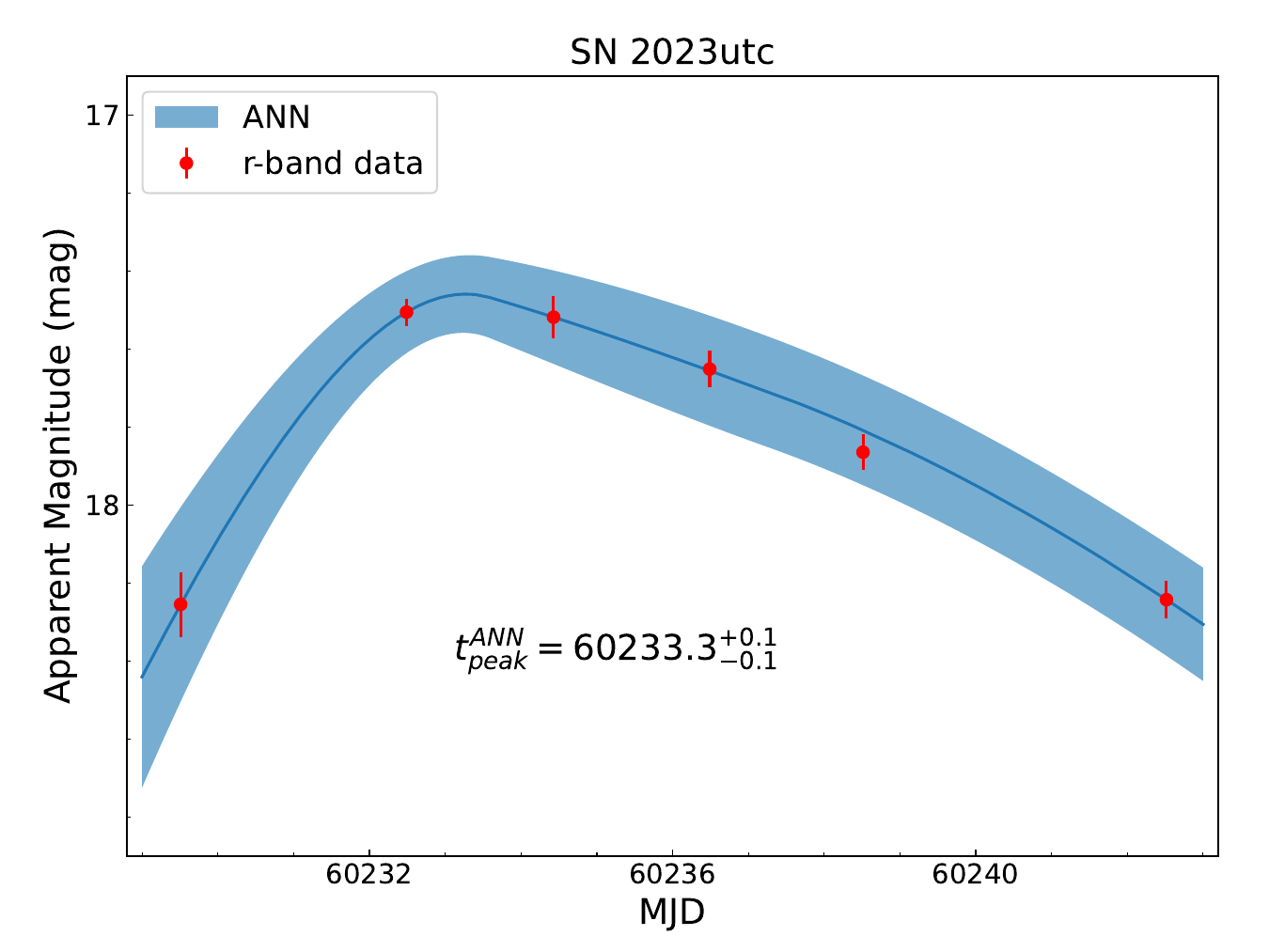}
\includegraphics[width=0.5\linewidth]{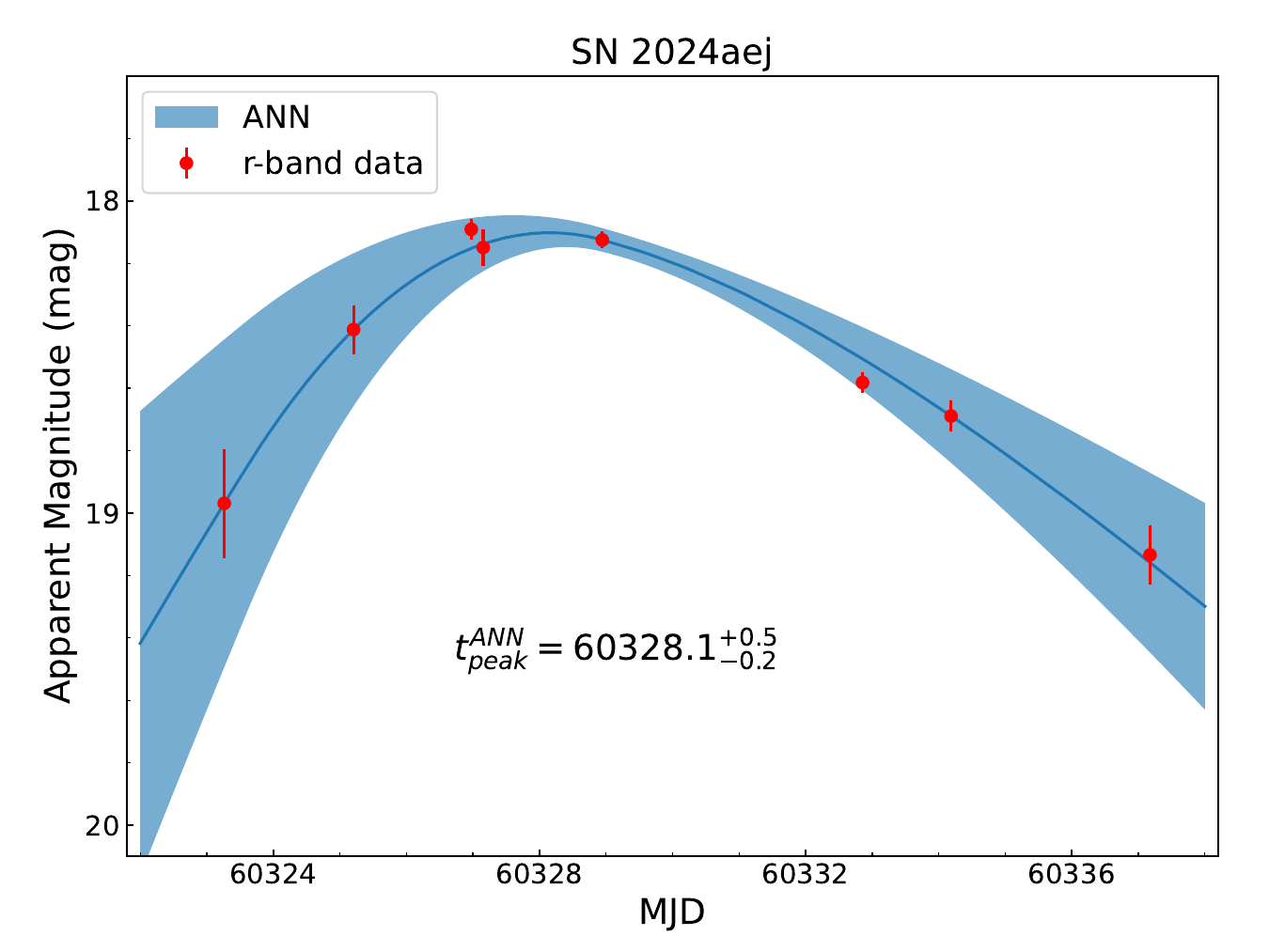}
\caption{Peak time of SNe\,2020nxt, 2020taz, 2023utc, 2024aej. 
}
\label{fig:peak_time}
\end{figure*}
\newpage 

\section{Decline rates of light curves of five SNe Ibn}\label{appendix:decline_rate}
\begin{longtable}{lcccccccccc}
    \caption{Decline rates of the light curves of individual SNe, along with their uncertainties in the units of mag (100 d)$^{-1}$. 
    }
    \label{decline_rate}\\
    \hline \hline
        \multicolumn{4}{c}{SN\,2020nxt} \\
        \hline
        Filter & $\gamma_{0-25}$ & $\gamma_{25-45}$ & $\gamma_{45-60}$\\ \hline
        $UVW2$ & 23.66$\pm$2.38 &$\cdots$&$\cdots$\\
        $UVM2$ & 21.33$\pm$1.66 &$\cdots$&$\cdots$\\
        $UVW1$ & 18.21$\pm$1.46 &$\cdots$&$\cdots$\\
        $u$ & 14.95$\pm$0.94 & 7.15$\pm$0.66 &$\cdots$\\
        $U$ & 14.05$\pm$1.03 &$\cdots$&$\cdots$\\
        $B$ & 13.74$\pm$0.61 & 5.56$\pm$0.19 & 32.45$\pm$3.76 \\
        $g$ & 11.70$\pm$0.34 & 5.92$\pm$0.59 & 32.82$\pm$2.15 \\
        $c$ & 13.53$\pm$1.14 & 8.83$\pm$2.52 &$\cdots$\\
        $V$ & 14.11$\pm$0.73 & 6.27$\pm$0.19 & 21.85$\pm$3.38 \\
        $r$ & 15.86$\pm$0.32 & 6.55$\pm$0.45 & 29.93$\pm$5.40 \\
        $o$ & 14.93$\pm$0.56 & 7.84$\pm$1.93 &$\cdots$\\
        $i$ & 16.20$\pm$0.53 & 6.56$\pm$0.43 & 15.41$\pm$5.97 \\
        $z$ & 13.97$\pm$0.50 & 5.60$\pm$0.91 &$\cdots$\\
        \hline
        \multicolumn{4}{c}{SN\,2020taz} \\
        \hline
        Filter & $\gamma_{0-10}$ & $\gamma_{10-20}$ & $\gamma_{20-30}$\\ \hline
        $u$ &$\cdots$& 10.12$\pm$5.48 &$\cdots$\\
        $B$ &$\cdots$& 5.96$\pm$1.13 & 30.97$\pm$0.87 \\
        $g$ & 2.77$\pm$1.15 & 9.75$\pm$1.44 & 27.78$\pm$3.25 \\
        $c$ & 2.34$\pm$2.93 &$\cdots$&$\cdots$\\
        $V$ &$\cdots$& 9.72$\pm$2.02 & 25.51$\pm$7.15 \\
        $r$ & 1.82$\pm$0.65 & 8.95$\pm$1.20 & 23.74$\pm$7.93 \\
        $o$ & 1.70$\pm$1.12 & 6.57$\pm$5.96 & 26.70$\pm$1.77 \\
        $i$ &$\cdots$& 6.24$\pm$0.70 & 26.32$\pm$5.34\\
        $z$ &$\cdots$& 5.17$\pm$0.03 & 27.54$\pm$8.62\\
        \hline
        \multicolumn{4}{c}{SN\,2021bbv} \\
        \hline
        Filter & $\gamma_{0-25}$ & $\gamma_{25-50}$ & $\cdots$\\ \hline
        $u$ & 12.02$\pm$1.05 &$\cdots$& $\cdots$\\
        $B$ & 10.15$\pm$0.42 & 5.97$\pm$0.91 & $\cdots$\\
        $g$ & 11.10$\pm$0.46 & 5.08$\pm$0.76 & $\cdots$\\
        $c$ & 6.90$\pm$1.48 &$\cdots$& $\cdots$\\
        $V$ & 9.62$\pm$0.35 & 3.32$\pm$1.58 & $\cdots$\\
        $r$ & 9.93$\pm$0.44 & 4.28$\pm$0.95 & $\cdots$\\
        $o$ & 5.14$\pm$0.24 &$\cdots$& $\cdots$\\
        $i$ & 8.35$\pm$0.62 & 4.21$\pm$0.82 & $\cdots$\\
        $z$ & 11.3$\pm$3.62 &$\cdots$& $\cdots$\\
        \hline
        \multicolumn{4}{c}{SN\,2023utc} \\
        \hline
        Filter & $\gamma_{0-30}$ & $\gamma_{30-70}$ & $\cdots$\\ \hline
        $U$ & 16.69$\pm$2.34 &$\cdots$  & $\cdots$\\
        $B$ & 15.58$\pm$3.65 &$\cdots$ & $\cdots$\\
        $g$ & 11.12$\pm$0.54 &$\cdots$  & $\cdots$\\
        $c$ & 9.90$\pm$0.70  & 8.45$\pm$4.95  & $\cdots$\\
        $V$ & 11.00$\pm$1.86 &$\cdots$  & $\cdots$\\
        $r$ & 8.88$\pm$0.41  &$\cdots$ & $\cdots$\\
        $o$ & 8.67$\pm$1.05  & 3.09$\pm$4.34  & $\cdots$ \\
        $i$ & 13.69$\pm$5.33 & 3.02$\pm$0.96 & $\cdots$ \\
        \hline
        \multicolumn{4}{c}{SN\,2024aej} \\
        \hline
        Filter & $\gamma_{0-20}$ & $\cdots$ & $\cdots$ \\ \hline
        $U$ & 18.33$\pm$0.74 & $\cdots$ & $\cdots$\\
        $B$ & 17.23$\pm$0.80 & $\cdots$ & $\cdots$\\
        $g$ & 14.96$\pm$0.45 & $\cdots$ & $\cdots$\\
        $V$ & 13.05$\pm$0.41 & $\cdots$ & $\cdots$\\
        $r$ & 14.18$\pm$0.44 & $\cdots$ & $\cdots$\\
        $o$ & 13.83$\pm$0.79 & $\cdots$ & $\cdots$\\
        $i$ & 11.80$\pm$0.45 & $\cdots$ & $\cdots$\\
        \hline
\end{longtable}

\newpage
\section{MOSFiT corner plots of five SNe Ibn} \label{appendix:corner}
\begin{figure}[h]
    \centering
    \includegraphics[width=1\linewidth]{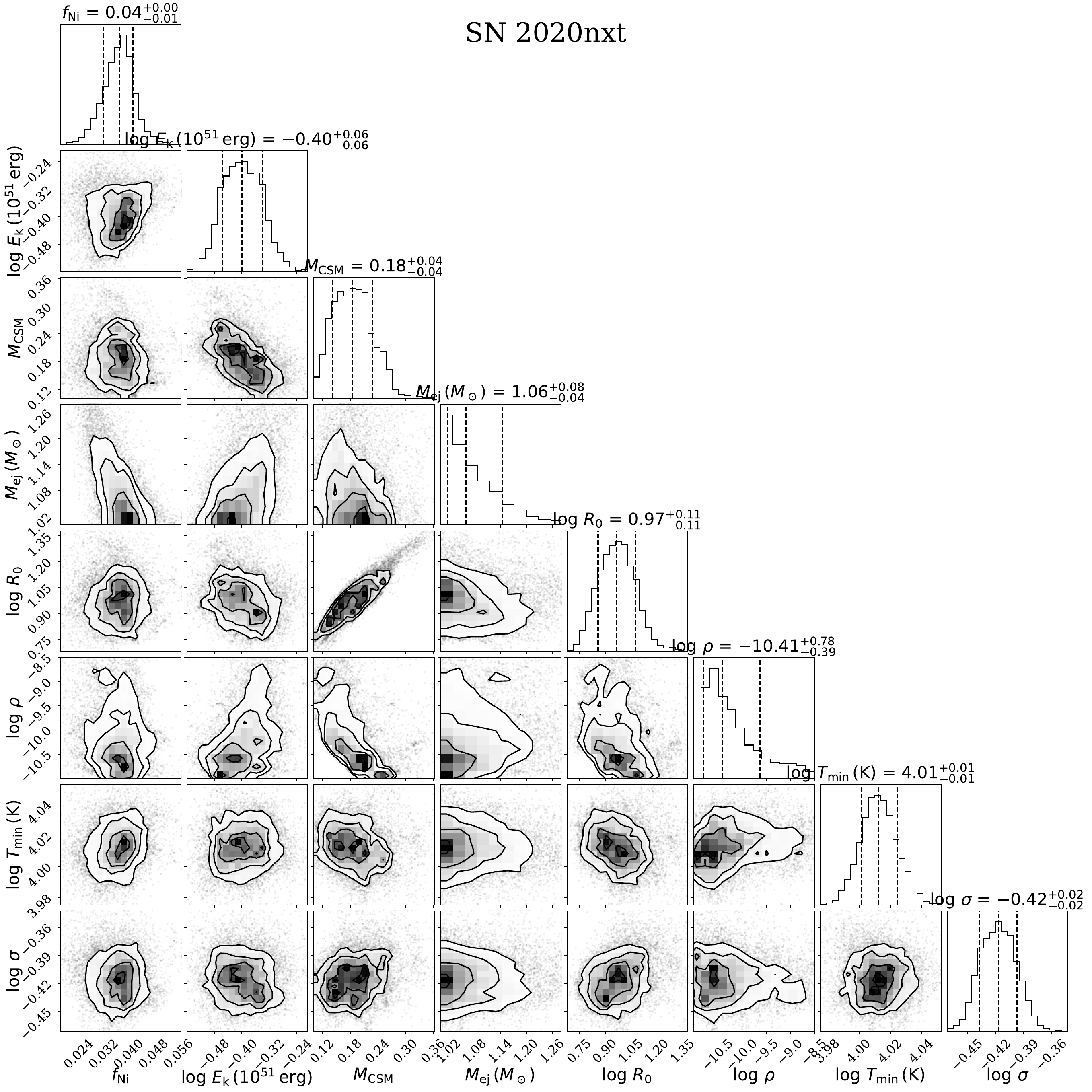}
    \caption{Corner plot showing the posterior distributions of the fitted parameters for SN~2020nxt, based on the \texttt{RD+CSI} model using \texttt{MOSFiT}. Median values are indicated by vertical lines, with the shaded regions representing the 68\% confidence intervals.}
    \label{fig:modfit_corner1}
\end{figure}

\newpage 

\begin{figure}[h]
    \centering
    \includegraphics[width=1\linewidth]{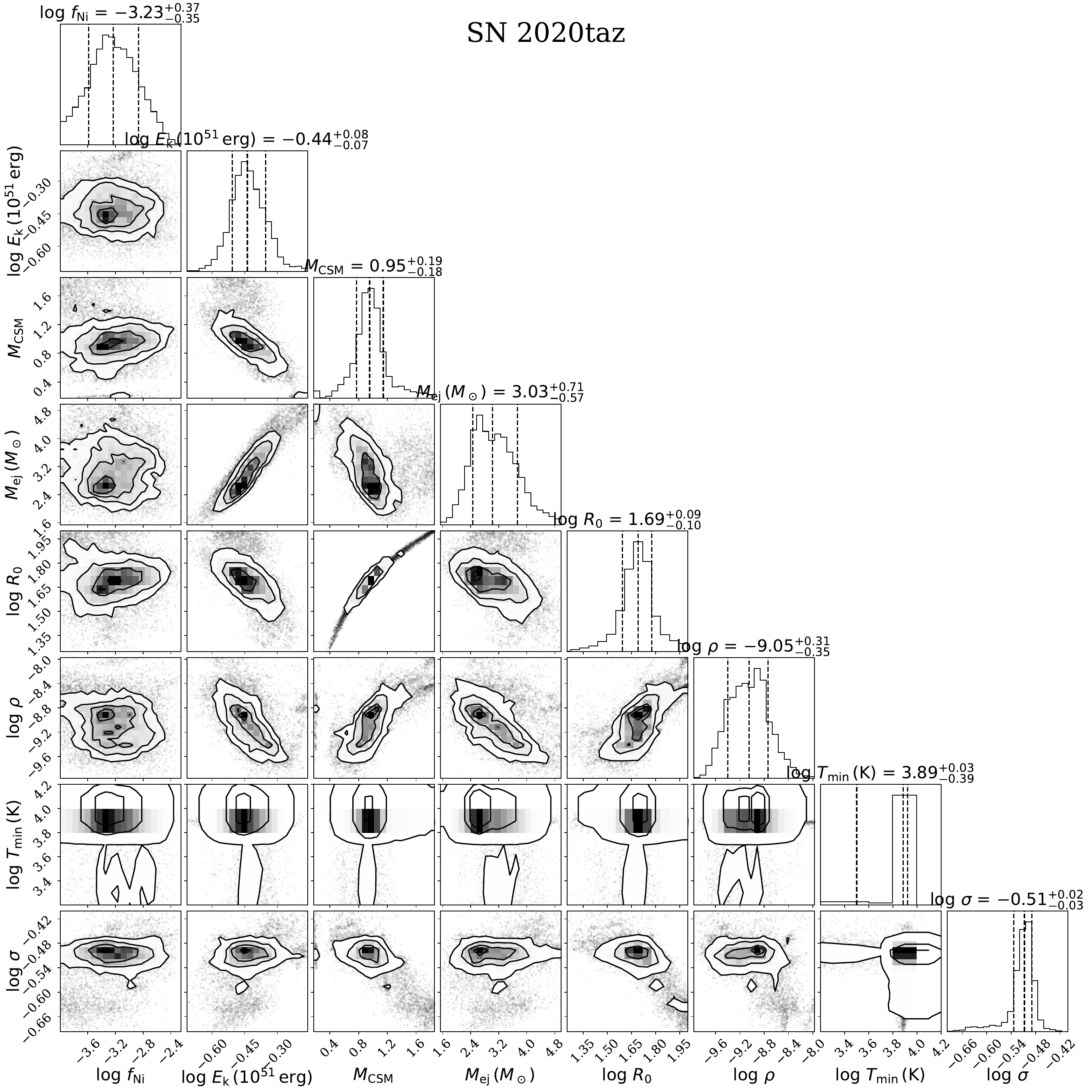}
    \caption{Corner plot showing the posterior distributions of the fitted parameters for SN~2020taz, based on the \texttt{RD+CSI} model using \texttt{MOSFiT}. Median values are indicated by vertical lines, with the shaded regions representing the 68\% confidence intervals.}
    \label{fig:modfit_corner2}
\end{figure}

\newpage 

\begin{figure}[h]
    \centering
    \includegraphics[width=1\linewidth]{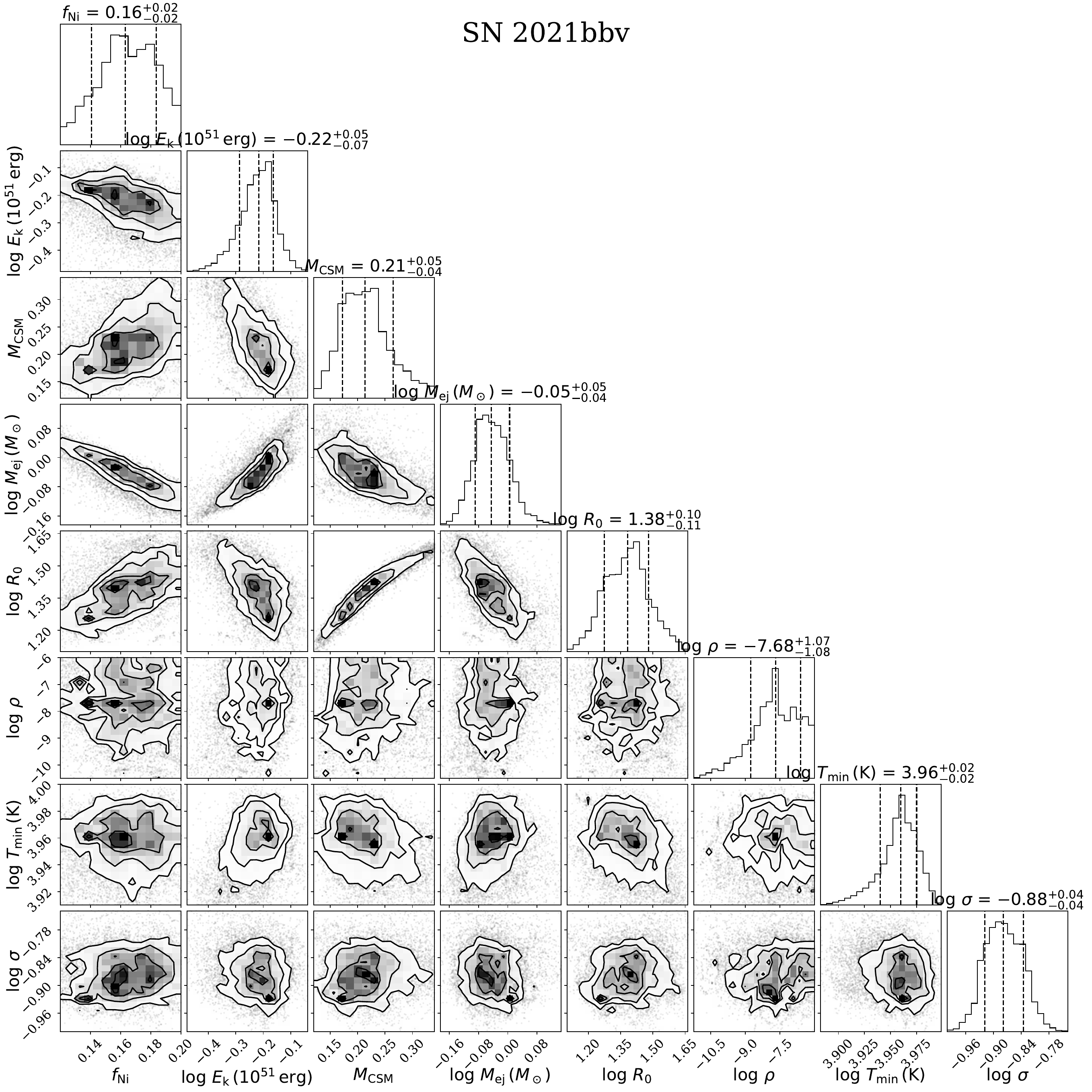}
    \caption{Corner plot showing the posterior distributions of the fitted parameters for SN~2021bbv, based on the \texttt{RD+CSI} model using \texttt{MOSFiT}. Median values are indicated by vertical lines, with the shaded regions representing the 68\% confidence intervals.}
    \label{fig:modfit_corner3}
\end{figure}

\newpage 

\begin{figure}[h]
    \centering
    \includegraphics[width=1\linewidth]{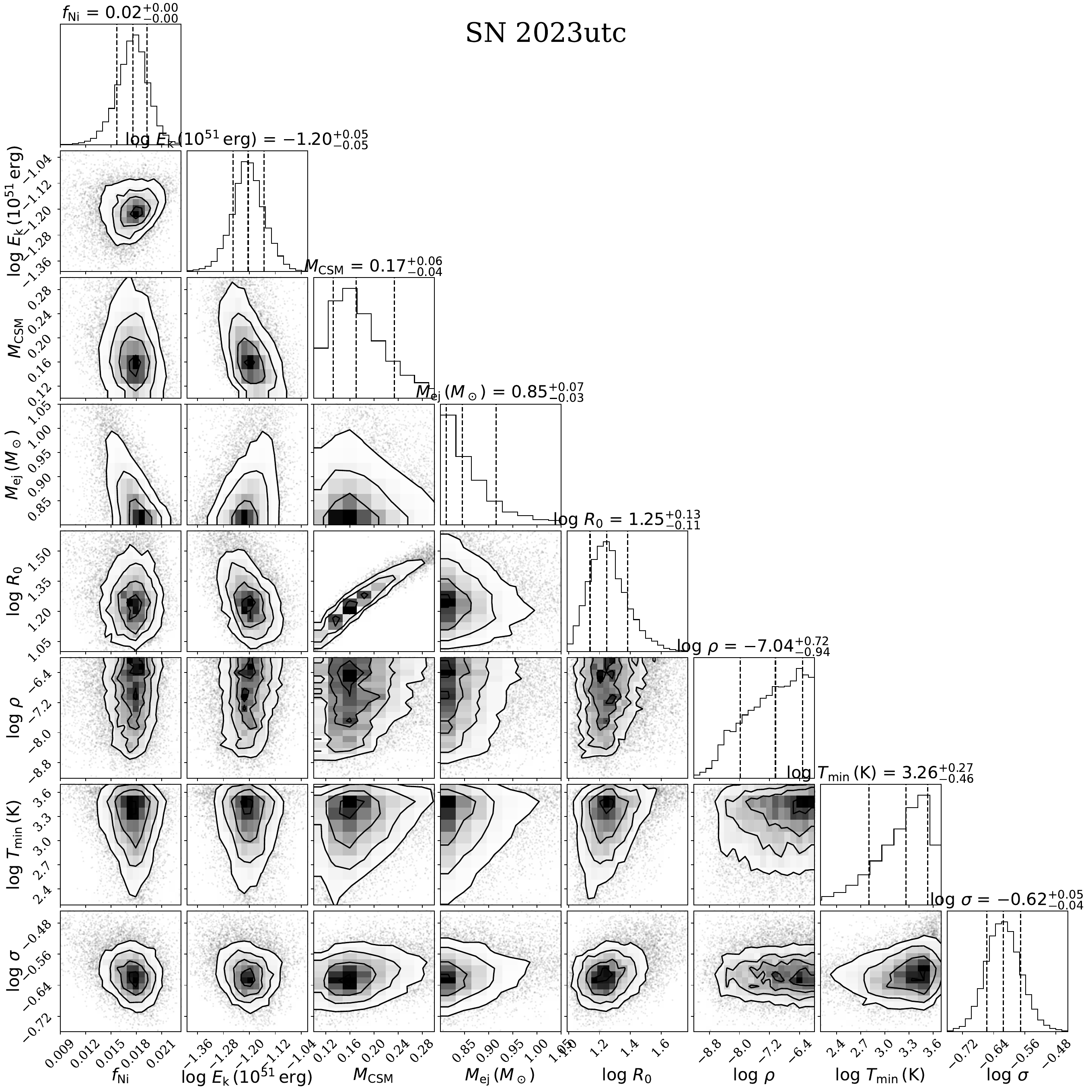}
    \caption{Corner plot showing the posterior distributions of the fitted parameters for SN~2023utc, based on the \texttt{RD+CSI} model using \texttt{MOSFiT}. Median values are indicated by vertical lines, with the shaded regions representing the 68\% confidence intervals.}
    \label{fig:modfit_corner4}
\end{figure}

\newpage 

\begin{figure}[h]
    \centering
    \includegraphics[width=1\linewidth]{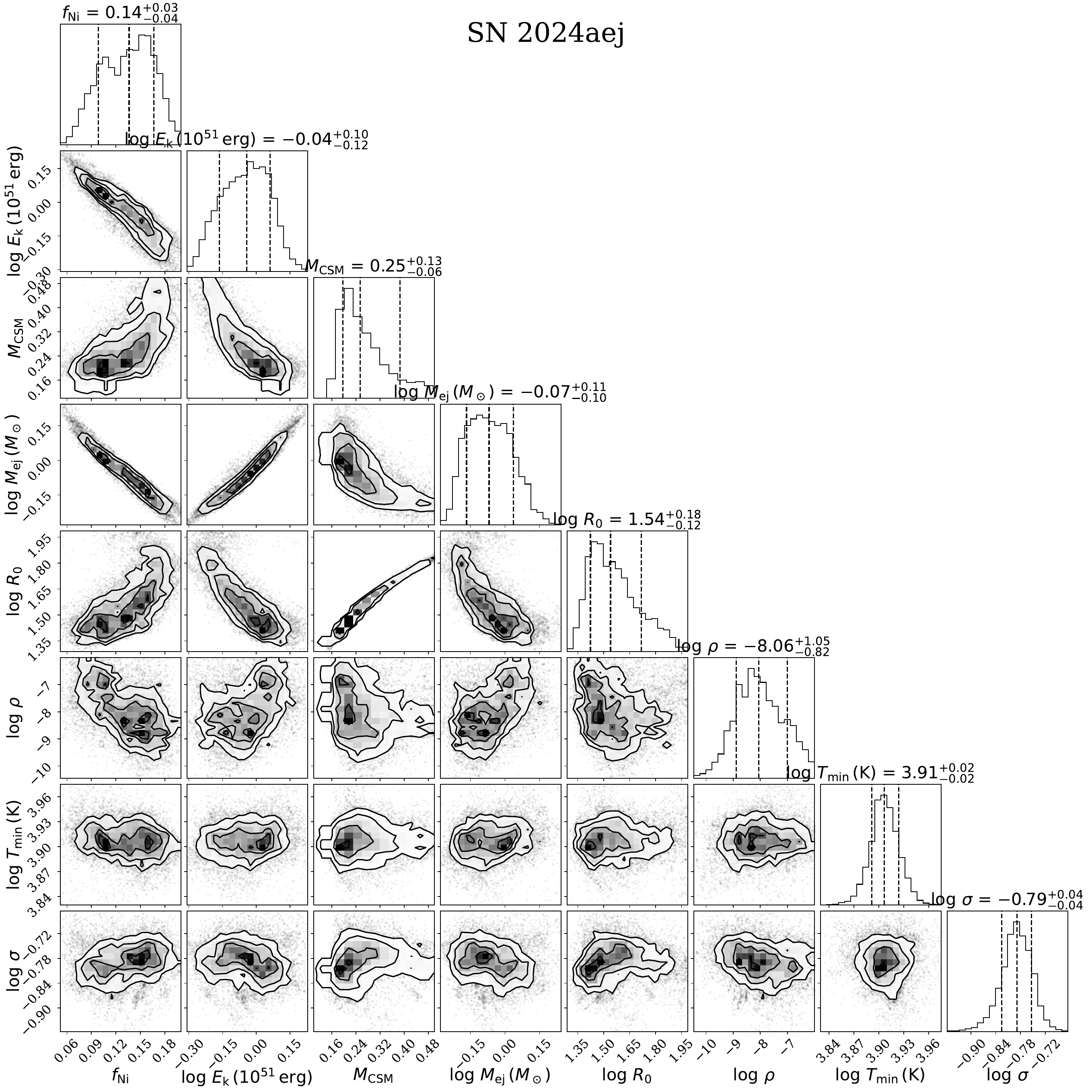}
    \caption{Corner plot showing the posterior distributions of the fitted parameters for SN~2024aej, based on the \texttt{RD+CSI} model using \texttt{MOSFiT}. Median values are indicated by vertical lines, with the shaded regions representing the 68\% confidence intervals.}
    \label{fig:modfit_corner5}
\end{figure}

\newpage 

\section{Log of the spectroscopic observations}
\label{Spec_log}
\begin{longtable}{cccccccc}
    \caption{Log of the spectroscopic observations of SN~2020nxt.}
    \label{table:speclog_2020nxt}\\
        \hline \hline
        Date     & MJD         & Phase$^{a}$ & Instrumental setup & Grism/Grating & Spectral range & Exposure time &  Resolution \\
        	     &             & (d)         &                    &               & (\AA)          & (s)           &  (\AA)     \\
        \hline
        20200713 & 59044.13    & $+5.7$      & LT+SPRAT           & VPH600        & $4000-8000$  & $1200$          & 18 \\
        20200717 & 59048.11    & $+9.7$      & TNG+DOLORES        & LR-B          & $3300-8000$    & $1800$        & 10.8 \\
        20200719 & 59050.11    & $+11.7$     & NOT+ALFOSC         & gm8           & $5700-8600$    & $2500$        & 10.5\\
        20200721 & 59051.93    & $+13.5$     & Ekar1.82m+AFOSC    & VPH6+VPH7     & $3400-9300$    & $2400+2400$   & 17.4,14.6\\
        20200723 & 59054.09    & $+15.7$     & LT+SPRAT           & VPH600        & $4100-8000$  & $2100$          & 18\\
        20170724 & 59055.16    & $+16.8$     & GTC+OSIRIS         & R1000B+R1000R & $3600-10400$   & $1500+1500$   & 7,8 \\
        20200731 & 59061.10    & $+22.7$     & NOT+ALFOSC         & gm4           & $3400-9700$    & $2700$        & 14.0\\
        20200805 & 59066.92    & $+28.5$     & Ekar1.82m+AFOSC    & gm4           & $3500-8200$    & $2700\times2$ & 13.4\\
        20200805 & 59067.14    & $+28.7$     & NOT+ALFOSC         & gm4           & $3400-9700$    & $3000$        & 13.6\\
        20200816 & 59077.15    & $+38.8$     & NOT+ALFOSC         & gm4           & $3400-9700$    & $3600$        & 14.2\\
        20200820 & 59082.09    & $+43.7$     & TNG+DOLORES        & LR-B          & $3500-8000$   & $1800\times2$ & 10.7 \\
        \hline
        \multicolumn{7}{l}{{$^a$Phases are relative to $o$-band maximum light (MJD =  59038.4~$^{+~0.1}_{-~0.1}$; 2020-07-08) in observer frame.}} \\
\end{longtable}

\begin{longtable}{cccccccc}
    \caption{Log of the spectroscopic observations of SN~2020taz.}
    \label{table:speclog_2020taz}\\
        \hline \hline
        Date     & MJD         & Phase$^{a}$ & Instrumental setup & Grism/Grating & Spectral range & Exposure time  &  Resolution \\
                 &             & (d)         &                    &               & (\AA)          & (s)            &  (\AA)      \\
        \hline
        20200924 & 59116.39    & $+5.5$        & APO 3.5m+DIS       & R300          & $4000-9850$    & 3600           &    9        \\ 
        20200927 & 59119.01    & $+8.1$        & NOT+ALFOSC         & gm4           & $3600-9680$    & 2700           &  14.0       \\
        20200929 & 59121.85    & $+11.0$       & NOT+ALFOSC         & gm4           & $3400-9680$    & 3600           &  14.0       \\
        20201012 & 59134.87    & $+24.0$       & TNG+DOLORES        & LR-B          & $3600-7990$    & 1800           &  14.5 \\
        20201017 & 59139.89    & $+29.0$       & GTC+OSIRIS         & R1000B        & $3650-7850$    & 1800           &  7    \\
        \hline
        \multicolumn{7}{l}{{$^a$Phases are relative to $o$-band maximum light (MJD = 59110.9~$^{+~1.5}_{-~0.5}$; 2020-09-18) in observer frame.}} \\
\end{longtable}

\begin{longtable}{cccccccc}
    \caption{Log of the spectroscopic observations of SN~2021bbv.}
    \label{table:speclog_2021bbv}\\
        \hline \hline
        Date     & MJD         & Phase$^{a}$ & Instrumental setup & Grism/Grating & Spectral range & Exposure time  &  Resolution   \\
                 &             & (d)         &                    &               & (\AA)          & (s)            &  (\AA)        \\
        \hline
        20210127 & 59241.37    & $-0.8$         & NTT+EFOSC2         & gm13          & 3650$-$9240      & 900          &  21           \\
        20210129 & 59243.10    & $+0.9$         & NOT+ALFOSC         & gm4           & 3480$-$9680      & 2700         &  12.5         \\
        20210208 & 59253.98    & $+11.8$         & NOT+ALFOSC         & gm4           & 3500$-$9670      & 3600         &  13.7         \\
        20210217 & 59262.06    & $+19.9$         & NOT+ALFOSC         & gm4           & 3600$-$9650      & 2560         &  17.6         \\
        \hline
        \multicolumn{7}{l}{{$^a$Phases are relative to $r$-band maximum light (MJD = 59242.2; 2021-01-28) in observer frame.}} \\
\end{longtable}

\begin{longtable}{cccccccc}
    \caption{Log of the spectroscopic observations of SN~2023utc.}
    \label{table:speclog_2023utc}\\
    \hline \hline
        Date     & MJD         & Phase$^{a}$ & Instrumental setup & Grism/Grating & Spectral range & Exposure time  &  Resolution   \\
                 &             & (d)         &                    &               & (\AA)          & (s)            &  (\AA)        \\
        \hline
        20231023 & 60240.82    & $+7.5$          & ST+KOOLS-IFU       & VPH-blue      & 4000$-$8550    & 900            &  12       \\   
        20231027 & 60244.54    & $+11.2$         & FTN+FLOYDS         & red/blu       & 3500$-$10000   & 3600           &  16       \\
        20231101 & 60249.57    & $+16.3$         & FTN+FLOYDS         & red/blu       & 3500$-$10000   & 3600           &  16       \\
        20231112 & 60260.60    & $+27.3$         & FTN+FLOYDS         & red/blu       & 3500$-$10000   & 3600           &  16       \\
        \hline
        \multicolumn{7}{l}{{$^a$Phases are relative to $r$-band maximum light (MJD = 60233.3~$^{+~0.1}_{-~0.1}$; 2023-10-16) in observer frame.}} \\
\end{longtable}

\begin{longtable}{cccccccc}
    \caption{Log of the spectroscopic observations of SN~2024aej.}
    \label{table:speclog_2024aej}\\
    \hline \hline
        Date     & MJD         & Phase$^{a}$ & Instrumental setup & Grism/Grating & Spectral range & Exposure time  &  Resolution   \\
                 &             & (d)         &                    &               & (\AA)          & (s)            &  (\AA)        \\
        \hline
        20240120 & 60329.25    & $+1.2$         & FTN+FLOYDS         & red/blu       & 3500$-$10000   & 2700           & 16            \\
        20240122 & 60331.28    & $+3.2$         & FTN+FLOYDS         & red/blu       & 3500$-$10000   & 2700           & 16            \\
        20240129 & 60338.26    & $+10.2$         & FTN+FLOYDS         & red/blu       & 3300$-$10000   & 2700           & 16            \\
        20240204 & 60344.21    & $+16.1$         & FTN+FLOYDS         & red/blu       & 3500$-$10000   & 3600           & 16            \\ 
        \hline
        \multicolumn{7}{l}{{$^a$Phases are relative to $r$-band maximum light (MJD = 60328.1~$^{+~0.5}_{-~0.2}$; 2024-01-19) in observer frame.}} \\
\end{longtable}

\newpage

\section{Modeling the Spectra}
\label{appendix:model_spectra}
\begin{figure*}[htp]
\includegraphics[width=1\linewidth]{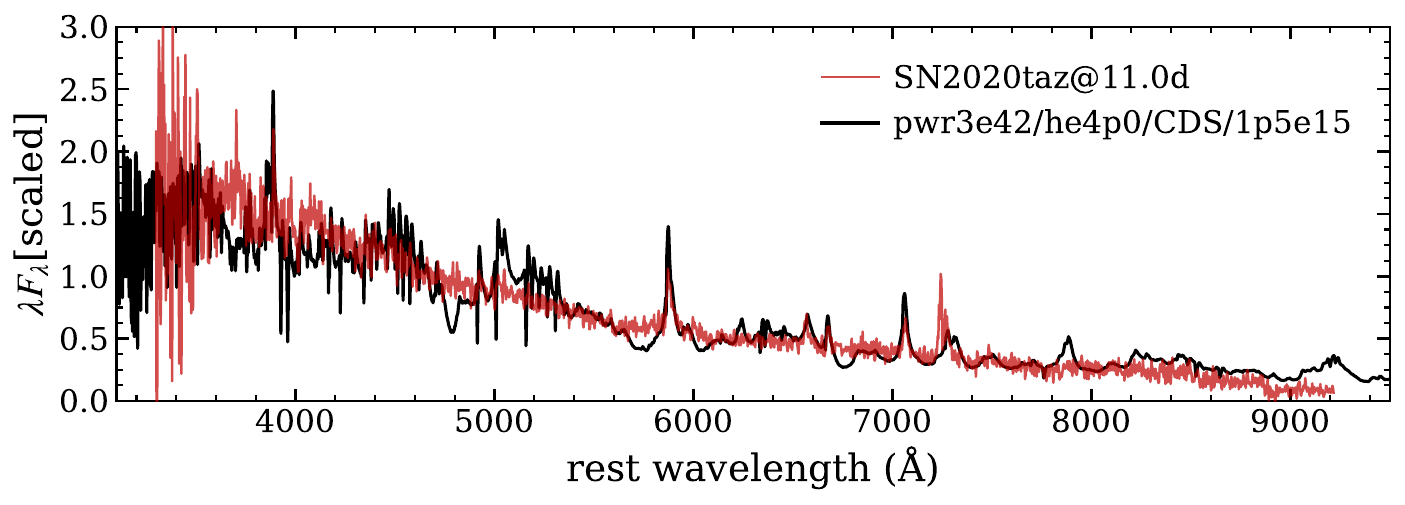}
\caption{Comparison between the observed spectrum of SN~2020taz obtained on 29 September 2020, corresponding to 11.0 days after the estimated time of maximum light (MJD = 59121.85), and a synthetic spectrum from a CDS model based on \texttt{he4p0}. The model assumes a shell located at a radius of $1.5 \times 10^{15}$\,cm, expanding at a velocity of 500\,km\,s$^{-1}$, and powered at a rate of $3 \times 10^{42}$\,erg\,s$^{-1}$. 
No smoothing has been applied to either the observed or synthetic spectra. The model spectra were provided by L.~Dessart (priv. comm.). 
}
\label{fig:model_spectra_low}
\end{figure*}

\twocolumn

\section{Acknowledgements}
We gratefully thank the anonymous referee for his/her insightful comments and suggestions that improved the paper.
We thank Luc Dessart for kindly providing the light curve and spectral models for Type Ibn supernovae, as well as for his valuable guidance and support to this work. We are also grateful to Qinan Wang for sharing data and engaging in helpful discussions. We thank Qiliang Fang for his helpful discussions. This study is supported by the National Natural Science Foundation of China (Nos 12303054,12225304), the National Key Research and Development Program of China (Grant Nos. 2024YFA1611603, 202501AS070078), the Yunnan Fundamental Research Projects (Grant No. 202401AU070063), and the International Centre of Supernovae, Yunnan Key Laboratory (No. 202302AN360001).
AP, AR, SB, EC, NER, LT, GV acknowledge support from the PRIN-INAF 2022, "Shedding light on the nature of gap transients: from the observations to the models".
AR also acknowledges financial support from the GRAWITA Large Program Grant (PI P. D'Avanzo).
WLL is supported by the National Natural Science Foundation of China (NSFC grants 12473045, 12033003 and 12494572) and the Natural Science Foundation of Xiamen city (grant 3502Z202471015).
EC acknowledges support from MIUR, PRIN 2020 (METE, grant 2020KB33TP).
N.E.R. acknowledges support from the Spanish Ministerio de Ciencia e Innovaci\'on (MCIN) and the Agencia Estatal de Investigaci\'on (AEI) 10.13039/501100011033 under the program Unidad de Excelencia Mar\'ia de Maeztu CEX2020-001058-M.
JRF is supported by the National Science Foundation Graduate Research Fellowship Program under Grant No. (NSF 2139319).
A.F. acknowledges the support by the State of Hesse within the Research Cluster ELEMENTS (Project ID 500/10.006).
Shane Moran is funded by Leverhulme Trust grant RPG-2023-240.
M.D. Stritzinger is funded by the Independent Research Fund Denmark (IRFD, grant number 10.46540/2032-00022B).
GJW acknowledges the Africa Europe Cluster of Research Excellence (CoRE-AI) fellowship.
XFW is supported by the National Natural Science Foundation of China (NSFC grants 12288102, 12033003, and 11633002) and the Tencent Xplorer Prize.
CPG acknowledges financial support from the Secretary of Universities and Research (Government of Catalonia) and by the Horizon 2020 Research and Innovation Programme of the European Union under the Marie Sk\l{}odowska-Curie and the Beatriu de Pin\'os 2021 BP 00168 programme, from the Spanish Ministerio de Ciencia e Innovaci\'on (MCIN) and the Agencia Estatal de Investigaci\'on (AEI) 10.13039/501100011033 under the PID2023-151307NB-I00 SNNEXT project, from Centro Superior de Investigaciones Cient\'ificas (CSIC) under the PIE project 20215AT016 and the program Unidad de Excelencia Mar\'ia de Maeztu CEX2020-001058-M, and from the Departament de Recerca i Universitats de la Generalitat de Catalunya through the 2021-SGR-01270 grant.
T.K. acknowledges support from the Research Council of Finland project 360274.
Song-Peng Pei is supported by the Science and Technology Foundation of Guizhou Province (QKHJC-ZK[2023]442).
T.M.R is part of the Cosmic Dawn Center (DAWN), which is funded by the Danish National Research Foundation under grant DNRF140. T.M.R acknowledges support from the Research Council of Finland project 350458.
C.-Y. Wu is supported by the National Natural Science Foundation of China (NSFC, Grant No.12473032), the Yunnan Revitalization Talent Support Program—Young Talent project, and the International Centre of Supernovae, Yunnan Key Laboratory (No. 202302AN360001).
J.Z. is supported by the National Key R\&D Program of China with No. 2021YFA1600404, the National Natural Science Foundation of China (12173082, 12333008), the Yunnan Fundamental Research Projects (grants 202401BC070007 and 202201AT070069), the Top-notch Young Talents Program of Yunnan Province, the Light of West China Program provided by the Chinese Academy of Sciences, and the International Centre of Supernovae, Yunnan Key Laboratory (No. 202302AN360001).
XJZ is supported by the National Natural Science Foundation of China (Grant No.~12203004) and by the Fundamental Research Funds for the Central Universities.
We acknowledge the support of the staffs of the various observatories at which data were obtained.
We thank Melissa L. Graham for providing an APO-3.5m spectrum of SN 2020taz, and Kenta Taguchi for sharing a spectrum of SN 2023utc obtained through the 3.8 m Seimei Telescope.
We acknowledge the support of the staff of the Xinglong 80 cm telescope (TNT). 
Based on observations made with the Nordic Optical Telescope, owned in collaboration by the University of Turku and Aarhus University, and operated jointly by Aarhus University, the University of Turku, and the University of Oslo, representing Denmark, Finland, and Norway, the University of Iceland, and Stockholm University at the Observatorio del Roque de los Muchachos, La Palma, Spain, of the Instituto de Astrofisica de Canarias.
Observations from the NOT were obtained through the NUTS2 collaboration which is supported in part by the Instrument Centre for Danish Astrophysics (IDA), and the Finnish Centre for Astronomy with ESO (FINCA) via Academy of Finland grant nr 306531. The data presented here were obtained in part with ALFOSC, which is provided by the Instituto de Astrofisica de Andalucia (IAA) under a joint agreement with the University of Copenhagen and NOTSA.
The Liverpool Telescope is operated on the island of La Palma by Liverpool John Moores University in the Spanish Observatorio del Roque de los Muchachos of the Instituto de Astrofisica de Canarias with financial support from the UK Science and Technology Facilities Council.
The Italian Telescopio Nazionale Galileo (TNG) operated on the island of La Palma by the Fundaci\'on Galileo Galilei of the INAF (Istituto Nazionale di Astrofisica) at the Spanish Observatorio del Roque de los Muchachos of the Instituto de Astrofísica de Canarias.
Based on observations collected at Copernico and Schmidt telescopes (Asiago, Italy) of the INAF -- Osservatorio Astronomico di Padova.
Based on observations obtained with the Gran Telescopio Canarias (GTC), installed in the Spanish Observatorio del Roque de los Muchachos of the Instituto de Astrofisica de Canarias, on the island of La Palma.
Based on data products from observations made with ESO Telescopes at the La Silla Paranal Observatory under programmes 106.216C.004/010 (D/J): ePESSTO+ (the advanced Public ESO Spectroscopic Survey for Transient Objects).
This work has made use of data from the Asteroid Terrestrial-impact Last Alert System (ATLAS) project. The Asteroid Terrestrial-impact Last Alert System (ATLAS) project is primarily funded to search for near earth asteroids through NASA grants NN12AR55G, 80NSSC18K0284, and 80NSSC18K1575; byproducts of the NEO search include images and catalogs from the survey area. This work was partially funded by Kepler/K2 grant J1944/80NSSC19K0112 and HST GO-15889, and STFC grants ST/T000198/1 and ST/S006109/1. The ATLAS science products have been made possible through the contributions of the University of Hawaii Institute for Astronomy, the Queen's University Belfast, the Space Telescope Science Institute, the South African Astronomical Observatory, and The Millennium Institute of Astrophysics (MAS), Chile.
This work makes use of data from the Las Cumbres Observatory Network and the Global Supernova Project. The LCO team is supported by U.S. NSF grants AST-1911225 and AST-1911151, and NASA.
We thank Las Cumbres Observatory and its staff for their continued support of ASAS-SN. ASAS-SN is funded in part by the Gordon and Betty Moore Foundation through grants GBMF5490 and GBMF10501 to the Ohio State University, and also funded in part by the Alfred P. Sloan Foundation grant G-2021-14192. Development of ASAS-SN has been supported by NSF grant AST-0908816, the Mt. Cuba Astronomical Foundation, the Center for Cosmology and AstroParticle Physics at the Ohio State University, the Chinese Academy of Sciences South America Center for Astronomy (CAS-SACA), and the Villum Foundation.
Pan-STARRS is a project of the Institute for Astronomy of the University of Hawaii, and is supported by the NASA SSO Near Earth Observation Program under grants 80NSSC18K0971, NNX14AM74G, NNX12AR65G, NNX13AQ47G, NNX08AR22G, 80NSSC21K1572 and by the State of Hawaii. The Pan-STARRS1 Surveys (PS1) and the PS1 public science archive have been made possible through contributions by the Institute for Astronomy, the University of Hawaii, the Pan-STARRS Project Office, the Max-Planck Society and its participating institutes, the Max Planck Institute for Astronomy, Heidelberg and the Max Planck Institute for Extraterrestrial Physics, Garching, The Johns Hopkins University, Durham University, the University of Edinburgh, the Queen's University Belfast, the Harvard-Smithsonian Center for Astrophysics, the Las Cumbres Observatory Global Telescope Network Incorporated, the National Central University of Taiwan, STScI, NASA under grant NNX08AR22G issued through the Planetary Science Division of the NASA Science Mission Directorate, NSF grant AST-1238877, the University of Maryland, Eotvos Lorand University (ELTE), the Los Alamos National Laboratory, and the Gordon and Betty Moore Foundation.
The Zwicky Transient Facility (ZTF) is supported by the National Science Foundation under Grants No. AST-1440341 and AST-2034437 and involves a collaboration that includes current partners such as Caltech, IPAC, the Oskar Klein Center at Stockholm University, the University of Maryland, the University of California, Berkeley, the University of Wisconsin-Milwaukee, the University of Warwick, Ruhr University, Cornell University, Northwestern University, and Drexel University. Operations are conducted by COO, IPAC, and UW.
We acknowledge the use of public data from the Swift data archive.
SDSS is managed by the Astrophysical Research Consortium for the Participating Institutions of the SDSS Collaboration including the Brazilian Participation Group, the Carnegie Institution for Science, Carnegie Mellon University, Center for Astrophysics | Harvard \& Smithsonian (CfA), the Chilean Participation Group, the French Participation Group, Instituto de Astrof\'isica de Canarias, The Johns Hopkins University, Kavli Institute for the Physics and Mathematics of the Universe (IPMU) / University of Tokyo, the Korean Participation Group, Lawrence Berkeley National Laboratory, Leibniz Institut f\"{u}r Astrophysik Potsdam (AIP), Max-Planck-Institut f\"{u}r Astronomie (MPIA Heidelberg), Max-Planck-Institut f\"{u}r Astrophysik (MPA Garching), Max-Planck-Institut f\"{u}r Extraterrestrische Physik (MPE), National Astronomical Observatories of China, New Mexico State University, New York University, University of Notre Dame, Observat\'orio Nacional / MCTI, The Ohio State University, Pennsylvania State University, Shanghai Astronomical Observatory, United Kingdom Participation Group, Universidad Nacional Aut\'onoma de M\'exico, University of Arizona, University of Colorado Boulder, University of Oxford, University of Portsmouth, University of Utah, University of Virginia, University of Washington, University of Wisconsin, Vanderbilt University, and Yale University.
This research has made use of the NASA/IPAC Extragalactic Database (NED), which is operated by the Jet Propulsion Laboratory, California Institute of Technology, under contract with the National Aeronautics and Space Administration.
{\sc iraf} was distributed by the National Optical Astronomy Observatory, which was managed by the Association of Universities for Research in Astronomy (AURA), Inc., under a cooperative agreement with the U.S. NSF.

\end{appendix}
\end{document}